\documentclass[]{jfm}
\usepackage{graphicx}
\usepackage{amssymb}
\usepackage{amsmath}
\usepackage{times}
\usepackage{colordvi}

\newcommand{\dxi}[1]{\frac{\mathrm{d}#1}{\mathrm{d}\xi}} 

\begin{document}

\title[Exact solution of the Riemann problem in relativistic magnetohydrodynamics]
{The exact solution of the Riemann problem in relativistic magnetohydrodynamics}

\author
[B. Giacomazzo \& L. Rezzolla]
{B\ls R\ls U\ls N\ls O\ls
\ns G\ls I\ls A\ls C\ls O\ls M\ls A\ls Z\ls Z\ls O$^{1}$\ns \and
L\ls U\ls C\ls I\ls A\ls N\ls O\ls \ns R\ls E\ls Z\ls Z\ls
O\ls L\ls L\ls A$^{1,2}$ }

\affiliation{$^1$ SISSA, International School for Advanced Studies and INFN,
		Trieste, Italy\\[\affilskip]
	$^2$ Department of Physics, Louisiana State University, Baton
        Rouge, USA} 

\pubyear{2006}
\volume{562}
\pagerange{223-259}
\date{15 July 2005 and in revised form 26 January 2006}

\maketitle
\begin{abstract}
	We discuss the procedure for the exact solution of the Riemann
	problem in special relativistic magnetohydrodynamics (MHD).
	We consider both initial states leading to a set of only three
	waves analogous to the ones in relativistic hydrodynamics, as
	well as generic initial states leading to the full set of
	seven MHD waves.  Because of its generality, the solution
	presented here could serve as an important test for those
	numerical codes solving the MHD equations in relativistic
	regimes\footnote{The numerical code computing the exact
	solution is available from the authors upon request.}.
\end{abstract}

\section{Introduction}
\label{intro}

        As first formulated by Riemann more than a hundred years ago, the
solution of the one-dimensional Riemann problem in hydrodynamics consists
of determining the temporal evolution of a fluid which, at some initial
time, has two adjacent uniform states characterized by different values
of uniform velocity, pressure and density. These initial conditions
completely determine the way in which the discontinuity will decay after
removal of the barrier separating the initial ``left'' and ``right''
states.

	The Riemann problem has ceased to be merely academic and has
gained enormous importance when it was realized that its numerical
solution can serve as the building block of hydrodynamical codes based
on Godunov-type finite difference methods (\cite{God59}). In such
methods, the computational domain is discretized and each interface
between two adjacent grid-zones is used to construct the initial left
and right states of a ``local'' Riemann problem. The evolution of the
hydrodynamical equations is then obtained through the solution across
the computational grid of the sequence of local Riemann problems set
up at the interfaces between successive grid-zones (see \cite{God59},
but also \cite{MM03} and \cite{F03} for the use of the Riemann problem
in relativistic regimes).

	In general, the Riemann problem requires the solution of a
nonlinear algebraic system of equations written as a function of a single
unknown quantity ({\it e.g.} the total pressure at the contact
discontinuity in purely hydrodynamical problems). With the exception of
few trivial initial configurations, the solution of the Riemann problem
cannot be obtained analytically but requires a numerical approach. The
solution found in this way is referred to as the {\it ``exact''} solution
of the Riemann problem, to distinguish it from the {\it ``approximate''}
solution of the Riemann problem, which is instead obtained when the
system of equations is reduced to a {locally-linear} form (an exhaustive
discussion of approximate Riemann solvers can be found in
\cite{toro99}). It is therefore useful to stress that although named
``exact'', the solution of the Riemann problem is necessarily obtained
with a small but nonzero truncation error.

	The exact solution of the Riemann problem in relativistic
hydrodynamics has been obtained only rather recently and was proposed
by~\cite{marti94} for flows that are purely along the direction normal
to the initial discontinuity. This work has then been extended to the
case in which tangential velocities are present (\cite{pons00}) and
improved in efficiency by exploiting the relativistic invariant
relative velocity between the two states to predict the wave pattern
produced (\cite{rezzolla01} and~\cite{rezzolla03}). The relevance of
these calculations has not been restricted to fundamental issues of
relativistic hydrodynamics. Quite the opposite, these solutions have
been of great importance for the testing of complex multidimensional
codes implementing High Resolution Shock Capturing (HRSC) methods, and
that are based on the approximate or exact solution of Riemann
problems at the interfaces between the numerical cells
(\cite{leveque92}). These codes have then been used in various
simulations in either fixed (\cite{aloyetal99}, \cite{fd02a},
\cite{zrf02}) or dynamical spacetimes (\cite{duez04}, \cite{ss05},
\cite{baiottieta05}).

	This intense and recent development of numerical codes for the
solution of the relativistic hydrodynamic equations has been accompanied
by an equally intense development of codes solving the equations of
magnetohydrodynamics (MHD) in relativistic regimes. The reason behind
this activity is the widespread expectation that strong magnetic fields
are crucial in the study and explanation of several puzzling
astrophysical phenomena such as relativistic jets or $\gamma$-ray
bursts. As a result, and in the hope of clarifying issues in relativistic
astrophysics which cannot be described satisfactorily through analytic
techniques, several groups have recently constructed numerical codes
solving the equations of relativistic MHD on either fixed spacetimes
(see, for example, \cite{delzanna03} and \cite{KO99} for a flat
background and \cite{gammie03}, \cite{devilliers03}, \cite{KO04},
\cite{mizunoetal04}, \cite{fragile05}, \cite{antonetal05} for a
black-hole background) or in fully dynamical spacetimes (\cite{duez05}).

	Just like their hydrodynamical counterparts, some of these
codes are based on the solution of a local Riemann problem suitably
formulated for a magnetized fluid, and all are meant to be used for
ultrarelativistic flows.  However, unlike their hydrodynamical
counterparts, these codes cannot benefit from the comparison with the
exact solution of the Riemann problem in relativistic MHD. The
literature on the Riemann problem in MHD is, in fact, much more
limited and a general exact solution was found rather recently and for
a Newtonian fluid only (\cite{Ryu95,falle98}). The background
knowledge in this area is even more scarce for a relativistic fluid
and while no general exact solution has been proposed yet, recent work
has been made to derive an exact solution in the particular case in
which the magnetic field of the initial states is tangential to the
discontinuity and orthogonal to the fluid velocity
(\cite{romeroetal05}). {Besides having a larger set of equations when
compared to the corresponding problem in relativistic hydrodynamics, a
considerable addition to the complexity of the Riemann problem in
relativistic MHD is represented by the fact that the mathematical
structure of the problem itself is modified and the system of
equations is no longer strictly hyperbolic~(\cite{lich67})\footnote{We
recall that a systems of $m$ quasi-linear partial differential
equations is said to be {\em hyperbolic} if the matrix of coefficients
has $m$ real eigenvalues; furthermore, the system is said to be {\em
totally} or {\em strictly hyperbolic} if the eigenvalues are real and
also all distinct.}. The possibility of having coincident eigenvalues
poses the question of the uniqueness of the solutions and this
represents then a problem within the problem. As we will comment also
later on, a lively debate on these issues is presently ongoing and
progress is starting to be made, although first results are known in
Newtonian MHD only (see~\cite{torrilhon03b}). Because the focus of
this work is the exact solution of the Riemann problem in relativistic
MHD as an aid to the development of numerical codes, hereafter we will
adopt the {\em working assumption} that the Riemann problems
considered here have a solution and that this solution is
unique. Clearly, this hypothesis avoids the issue rather than solving
it, but it allows for a marked progress at least in those cases in
which compound waves are not found in numerically approximate
solutions.}

	{A direct and important consequence of the scarcity of works
in this area of fundamental relativistic MHD is that the validation of
modern complex MHD codes for the most elementary and yet demanding tests
has not been made in a quantitative manner for generic initial
conditions. Rather, it has taken place through the qualitative comparison
with the large set of test-problems in relativistic MHD meticulously
collected over the years (see, for instance,~\cite{KO99}
and~\cite{BA01}). It should be recognized, however, that for non-generic
initial states it is sufficient to have exact solutions for MHD shocks
and rarefactions as this covers all types of basic hyperbolic waves of
the system and indeed exact solutions of this type were used by
Komissarov (1999) for quantitative testing.}

	{On the other hand, the purpose of this paper is to present
the procedure for the exact solution of the Riemann problem in
relativistic MHD with generic initial conditions.} Our approach considers
both initial states with a zero component of the magnetic field along the
flow and leading to a set of only three waves analogous to the ones in
relativistic hydrodynamics, as well as generic initial states leading to
the full set of seven MHD waves. The approach discussed for the numerical
solution is based on a ``hybrid'' approach which adopts different sets of
equations according to the values of the normal magnetic field and that
has turned out to be crucial for a successful solution.

	The paper is organized as follows: Section \ref{equations}
contains the basic equations of relativistic MHD, while Section
\ref{strategy} describes the strategy used to solve the Riemann problem
numerically and which combines the methods discussed in
Section~\ref{pmethod} and Section~\ref{btmethod}. Section \ref{numerical}
focuses on the details of the numerical implementation and discusses
the solution of a number of tests that have become standard
references. Finally, the conclusions are collected in Section
\ref{conclusions}.

        We use a spacelike signature $(-,+,+,+)$ and a system of units in
which $c=1$. Greek indices are taken to run from 0 to 3, Latin indices
from 1 to 3 and we adopt the standard convention for the summation over
repeated indices. Finally we indicate 3-vectors with an arrow and use
bold letters to denote 4-vectors and tensors.


\section{Equations of Relativistic MHD}
\label{equations}

	Consider an ideal but magnetized relativistic fluid with an
energy-momentum tensor given by
\begin{equation}
T^{\mu\nu}=\left(\rho + \rho\epsilon + p_{\rm g} + 
	2p_{\rm m}\right)u^{\mu}u^{\nu}+
	\left(p_{\rm g} + p_{\rm m}\right)
	\eta^{\mu\nu}-b^{\mu}b^{\nu} \ ,
\end{equation}
where $\rho$ is the rest mass density, $\epsilon$ the {specific
internal energy}, $p_{\rm g}$ the gas pressure, $p_{\rm m}$ the
magnetic pressure, $u^{\mu}\equiv W(1,v^x,v^y,v^z)$ the four-velocity,
$W\equiv {1}/{\sqrt{1-v^i v_i}}={1}/{\sqrt{1-v^2}}$ the Lorentz factor
and the 4-vector ${\boldsymbol b}$ has components
\begin{equation}
b^{\alpha}\equiv\left\{ W(\vec{v}\cdot\vec{B}), \frac{\vec{B}}{W}+
	W(\vec{v}\cdot\vec{B})\vec{v}\right\} \ .
\label{defb}
\end{equation}
Here $\vec{B}$ is the magnetic field 3-vector and 
\begin{equation}
\qquad b^2\equiv b^i b_i =\frac{B^2}{W^2} + (\vec{v}\cdot\vec{B})^2 
	= 2 p_{\rm m}\ .
\end{equation}

The general relativistic equations of MHD are then simply obtained after
requiring the conservation of baryon number
\begin{equation}
\label{barno}
\nabla_\mu (\rho u^\mu) = 0 \ ,
\end{equation}
where $\nabla$ represents a covariant derivative, the conservation of
energy and momentum
\begin{equation}
\label{en-mom}
\nabla_\mu T^{\mu\nu} = 0 \ ,
\end{equation}
together with the relevant pair of Maxwell equations. If the fluid is
assumed to have an infinite electrical conductivity ({\it i.e.} ideal
MHD limit), the Maxwell equations reduce to $\partial_{[\alpha}
F_{\beta\gamma]} = 0$, where ${\boldsymbol F}$ is the Faraday tensor and
the square brackets refer to antisymmetrised indices. Using the
definition (\ref{defb}), the Maxwell equations can be simply written as
\begin{equation}
\label{maxwell}
\nabla_{\mu}(b^\mu u^\nu-u^\mu b^\nu)=0 \ .
\end{equation}
The system of equations (\ref{barno})--(\ref{maxwell}) is completed
with an equation of state (EOS) relating the pressure to the rest-mass
density and/or to the energy density. Although hereafter we will use
{an ideal-gas EOS}: $p_{\rm g}=\rho \epsilon (\Gamma-1)$, where
$\Gamma$ is the polytropic index, the procedure described for the
solution of the Riemann problem is valid for a generic EOS.

	We next assume that the system has a planar-symmetry, {\it i.e.}
that in a Cartesian coordinate system $(t,x,y,z)$ all the variables
depend only on $t$ and $x$, and that the spacetime is flat so that
covariant derivatives in equations~(\ref{barno})--(\ref{maxwell}) can be
replaced by partial derivatives and $A^i = A_i$ for any 3-vector ${\vec
A}$. In this case, the complete set of MHD equations can be written as a
set of first-order partial differential equations in a flux-conservative
form
\begin{equation}
\frac{\partial \mathbf{U}}{\partial t}+\frac{\partial\mathbf{F}}
	{\partial x} = 0 \ ,
\end{equation}
where $\mathbf{U}$ and $\mathbf{F}$ are respectively the vectors of
conserved quantities and fluxes, defined as
\begin{equation}
\mathbf{U} \equiv \left(
\begin{array}{c} 
D		\\ 
\tau-b^0b^0	\\ 
S^x-b^0b^x	\\ 
S^y-b^0b^y	\\ 
S^z-b^0b^z	\\ 
B^y		\\ 
B^z  
\end{array}
\right) \ ,
\hskip 1.5cm
\mathbf{F} \equiv \left( 
\begin{array}{c}
D v^x		\\ 
S^x-b^0b^x-Dv^x	\\ 
S^xv^x+p-b^xb^x	\\ 
S^yv^x-b^xb^y	\\ 
S^zv^x-b^xb^z	\\ 
B^yv^x-B^xv^y	\\ 
B^zv^x-B^xv^z 
\end{array}
\right) \ ,
\end{equation}
and where the following definitions have been used 
\begin{eqnarray}
\tau &\equiv& wW^2-p-D \ ,
\\ 
D &\equiv& \rho W \ ,
\\ 
S^j &\equiv& \rho h W^2 v^j \ ,
\\ 
p &\equiv& p_{\rm g}+p_{\rm m} = p_{\rm g}+\frac{1}{2}b^2 \ ,
\\ 
w &\equiv& \rho h \ ,
\\ 
h &\equiv& h_{\rm g}+ \frac{b^2}{\rho} = 1 + \epsilon + 
	\frac{p_{\rm g}}{\rho} + \frac{b^2}{\rho} \ ,
\end{eqnarray}
where $h$  is the total specific enthalpy and $h_{\rm g}$ the one of the
gas only.

	Note that the divergence-free condition for the magnetic field
and the Maxwell equation for the evolution of the $x$-component of the
magnetic field imply that $\partial_t B^x = 0 = \partial_x B^x$, {\it
i.e.} $B^x$ is uniform in space, constant in time and thus always
maintains its initial values.

\section{Strategy of Solution}
\label{strategy}

	The general Riemann problem in relativistic MHD consists of a set
of seven nonlinear waves: two {\it fast-waves} (FW), two {\it slow-waves}
(SW), two {\it Alfv\`en-waves} (AW), and a {\it contact discontinuity}
(CD) at which only the density may be discontinuous. The fast and slow
nonlinear waves can be either shocks or rarefaction waves, depending on
the change in the pressure and in the norm of the magnetic field across
the wave.

	Building on the experience with relativistic hydrodynamics, our
general strategy in the search for the solution consists of expressing
all of the variables after each wave as functions of the values of the
same variables ahead of the wave and of an unknown variable behind the
wave. When considering the Riemann problem in relativistic hydrodynamics,
in fact, the solution is found after expressing all of the quantities
behind the wave as functions of the value of the pressure at the contact
discontinuity. In this way, the problem is reduced to the search for the
value of the pressure that satisfies the jump conditions at the contact
discontinuity. 

\begin{figure}
  \begin{center}
    \includegraphics[height=6.5cm]{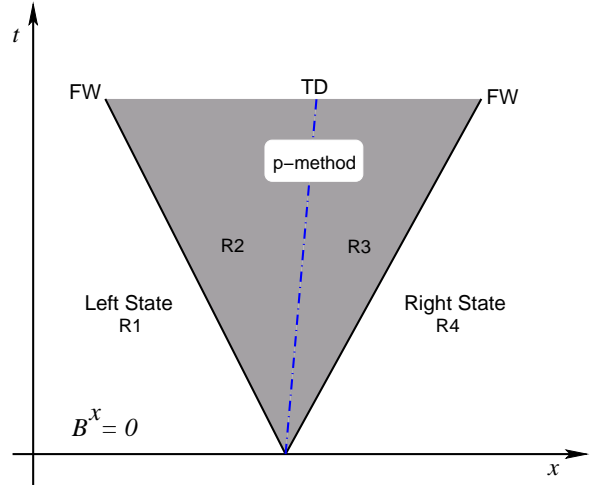}
  \end{center}
  \caption{\label{mhd_zones_nullBx} Spacetime structure of the MHD
    Riemann problem in the case in which the magnetic field has
    tangential components only, {\it i.e.}  $B^x=0$. The ``Riemann-fan''
    in this case is composed of only two fast-waves (FW) and of a central
    tangential discontinuity (TD), thus resembling structure of the
    Riemann problem in pure hydrodynamics. Indicated with R1--R4 are the
    4 different regions into which the Riemann problem can be decomposed,
    each representing a different state.}
\end{figure}

	When considering the Riemann problem in relativistic MHD, on the
other hand, two different cases need to be distinguished. Assuming the
initial discontinuity to have normal along the $x$-axis, the initial
magnetic field in this direction can either be zero ({\it i.e.} $B^x=0$)
or not ({\it i.e.} $B^x \ne 0$). In the first case, the structure of the
solution is very similar to the hydrodynamical one, with only two fast
waves and a {\it tangential discontinuity} along which only the total
pressure and the $x$ component of the velocity are continuous. The
spacetime structure of the Riemann problem in this case is sketched in
Figure~\ref{mhd_zones_nullBx}, where the ``Riemann-fan'' is shown to be
composed of only two fast-waves (FW) and of a central tangential
discontinuity (TD). Because of this analogy, the numerical solution of
the Riemann problem when $B^x=0$ follows the same procedure implemented
in relativistic hydrodynamics. We refer to this as the ``total-pressure
approach'' or simply, the {\it ``p-method''}.

	A detailed investigation of the exact solution of the Riemann
problem with tangential magnetic fields and when the additional
condition ${\vec v}\cdot {\vec B}=0$ is imposed, has been recently
proposed by Romero et al. (2005). Among the many points discussed,
this work has shown that when $B^x = 0 = {\vec v}\cdot {\vec B}$ the
Riemann problem in relativistic MHD can be assimilated to the one in
relativistic hydrodynamics and that all of the corrections introduced
by the magnetic field can be incorporated in the definition of a new,
effective EOS.

\begin{figure}
  \begin{center}
    \includegraphics[height=6.5cm]{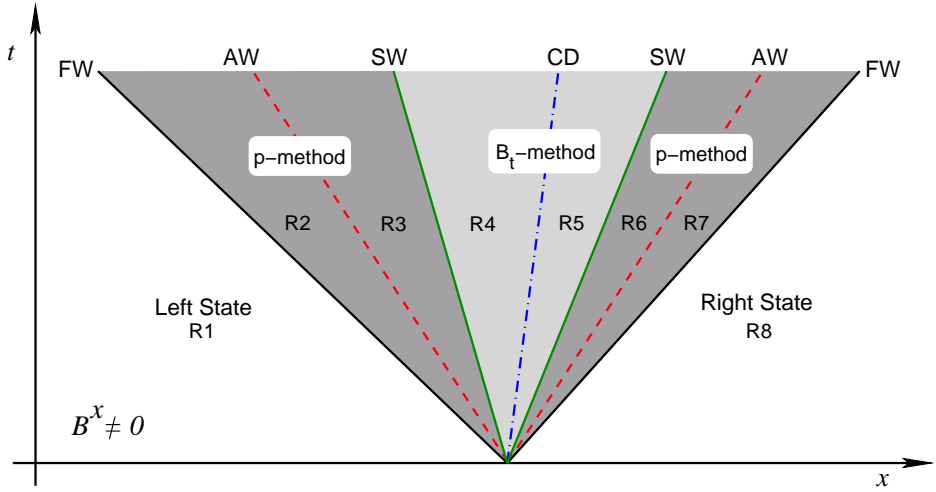}
  \end{center}
  \caption{\label{mhd_zones}Spacetime structure of the MHD Riemann
    problem in the general case in which the magnetic field has also a
    normal component, {\it i.e.} $B^x \ne 0$. The ``Riemann-fan'' is here
    composed of two fast-waves (FW), of two Alfv\`en waves (AW), of two
    slow-waves (SW) and of a central contact discontinuity (CD). Indicated
    with R1--R8 are the 8 different regions into which the Riemann problem
    can be decomposed, each representing a different state. Indicated are
    also the different methods used to compute the solutions in the
    different regions ({\it i.e.} $B_t$-method in regions R4 and R5 and
    $p$-method in regions R2-R3 and R6-R7).}
\end{figure}

	In the second case, on the other hand, the Riemann problem is
considerably more complex and all of the seven waves are allowed to form
when the initial discontinuity is removed. The spacetime structure of the
Riemann problem in this case is sketched in Figure~\ref{mhd_zones}, where
the ``Riemann-fan'' is shown to be composed of two fast-waves (FW), of
two Alfv\`en waves (AW), of two slow-waves (SW) and of a central contact
discontinuity (CD).

	It is important to bear in mind that across the Alfv\`en
discontinuities only the total pressure, the gas pressure and the density
are continuous, while there could be jumps in the other quantities. As a
result, if the total pressure is used as unknown, there would be three
different values for the total pressure (two between the fast and the
slow-waves and one between the two slow-waves) but five conditions to be
satisfied at the contact discontinuity: the continuity of the three
components of the velocity and the continuity of the tangential
components of the magnetic field. The resulting system of five equations
in three unknowns is over-constrained and there is no guarantee that a
global convergent solution is found at the contact discontinuity. Indeed,
experience has shown that small numerical imprecisions at the level of
round-off errors are in general sufficient to prevent the simultaneous
solution of the five constraints.

	To circumvent this difficulty and inspired by the procedure
followed in the exact solution of the corresponding Riemann solver in
nonrelativistic MHD (\cite{Ryu95}), when $B^x \neq 0$ we have implemented
a ``hybrid'' approach in which the total pressure is used as the unknown
variable between the fast and the slow waves ({\it i.e.} in regions
R2-R3, and R6-R7 of Figure~\ref{mhd_zones}), while the tangential
components of the magnetic fields ($B^y$ and $B^z$) is used between the
slow waves ({\it i.e.} in regions R4-R5 of Figure~\ref{mhd_zones}). In
this way, the continuity of the tangential components of the magnetic
field $B^y$ and $B^z$ is automatically guaranteed through the contact
discontinuity and only the continuity of the total pressure and of the
three components of the velocity needs to be satisfied. The resulting
system consists of four equations in four unknowns and, being closed, it
can be solved numerically through root-finding techniques for nonlinear
system of equations ({\it e.g.}  using a Newton-Raphson method). We refer
to this as the ``tangential magnetic field approach'' or simply, the {\it
``$B_t$-method''}.

	{As mentioned in the Introduction, hereafter we will assume
that the Riemann problem has a solution and that this is unique. As a
result, we will not discuss in any detail compound waves which seem to
develop in the numerical solution of some special initial states (one of
these is shown in Section~\ref{num_bxneq0_tests}) and whose admissibility as
solution of the Riemann problem is still debated.}

\section{Total-Pressure Approach: ``${\boldsymbol p}$-method''}
\label{pmethod}

	In the following Sections we describe in detail the approach in
which we calculate all of the variables in the Riemann fan using as
unknown the total pressure, {\it i.e.} the $p$-method. Different set of
equations will be derived according to whether the solution is across a
shock or a rarefaction wave.

\subsection{Solution across a shock front}
\label{sasf_pm}

	Consider $\Sigma$ to be a hypersurface in flat spacetime across
which $\rho$, ${\boldsymbol u}$ and ${\boldsymbol T}$ are
discontinuous. Let also ${\boldsymbol n}$ be the unit 4-vector normal to
$\Sigma$ so that the Rankine-Hugoniot conditions for relativistic MHD
can be expressed as
\begin{eqnarray}
[[\rho u^\alpha]]n_\alpha &=& 0 \ ,
	\label{S1}
\\ 
\left[\left[T^{\alpha\beta}\right]\right]n_\alpha &=& 0 \ ,
\label{S2}
\\ 
\left[\left[b^\alpha u^\beta - u^\alpha b^\beta\right]
	\right]n_\alpha &=& 0 \ ,
\label{S3}
\end{eqnarray}
where we use the double-bracket notation to express the jump of a
quantity $F$ across the hypersurface $\Sigma$, {\it
i.e.}
\[
[[F]]\equiv F_a - F_b\ .
\]
where $F_a$ and $F_b$ are respectively the values ahead $(a)$ and
behind $(b)$ the shock.

In particular, if $\Sigma$ is the 4-dimensional hypersurface
describing the evolution of a shock wave normal to the $x$-axis, the
unitary condition on ${\boldsymbol n}$ can be used to derive the
components
\begin{equation}
n^\alpha=W_{s}(V_{s},1,0,0) \ ,
\end{equation}
where $V_{s}$ is the coordinate velocity of the shock, $W_{s} \equiv
(1-V^2_s)^{-1/2}$ its Lorentz factor, and we can rewrite
equations~(\ref{S1})--(\ref{S3}) explicitly as
\begin{eqnarray}
\label{S24}
[[J]]\equiv[[\rho W(V_{s}-v^x)W_{s}]] &=& 0 \ ,
\\ 
\label{S25}
\left[\left[b^0b^0-\tau\right]\right]V_{s}+\left[\left[S^x-b^0b^x-
	Dv^x\right]\right] &=& 0\ , 
\\ 
\label{S26}
\left[\left[b^0b^x-S^x\right]\right]V_{s}+\left[\left[S^x
	v^x+p-b^xb^x\right]\right] &=& 0 \ , 
\\ 
\label{S27}
\left[\left[b^0b^y-S^y\right]\right]V_{s}+\left[\left[S^y
    v^x-b^xb^y\right]\right] &=& 0 \ , 
\\ 
\label{S28}
\left[\left[b^0b^z-S^z\right]\right]V_{s}+\left[\left[S^z
    v^x-b^xb^z\right]\right] &=& 0 \ , 
\\ 
\label{S29}
\left[\left[B^x\right]\right] &=& 0 \ , 
\\ 
\label{S30}
\left[\left[B^y\right]\right]V_{s}+\left[\left[B^x
	v^y-v^xB^y\right]\right] &=& 0 \ , 
\\ 
\label{S31}
\left[\left[B^z\right]\right]V_{s}+\left[\left[B^x
    v^z-v^xB^z\right]\right] &=& 0 \ , 
\end{eqnarray}
where $J$ is the (rest) mass flux across the shock.

	After a number of tedious but otherwise straightforward algebraic
manipulations, equations {(\ref{S24})--(\ref{S31})} can be recast as
\begin{eqnarray}
\left[\left[v^x\right]\right]+
	\frac{J}{W_{s}}\left[\left[\frac{1}{D}\right]\right] 
	&=& 0 \ , \hskip 1.0cm \label{eqs1}\\
\frac{J}{W_{s}}\left[\left[\frac{W^2\eta^2}{D}\right]\right]-
	B^x\left[\left[\eta\right]\right]-
	\frac{J}{W_{s}}\left[\left[\frac{\tau}{D}\right]\right]+
	\left[\left[pv^x\right]\right] &=& 0\ , \hskip 1.0cm  \label{eqs2}\\
\frac{JB^x}{W_{s}}\left[\left[\frac{\eta}{D}\right]\right]-B^x[\eta
	v^x]+\frac{J}{W_{s}}\left[\left[\frac{W^2\eta^2
        v^x}{D}\right]\right]-\left[\left[\frac{{B^2_x}}{W^2}\right]\right]-
	\frac{J}{W_{s}}\left[\left[\frac{S^x}{D}\right]\right]+
	\left[\left[p\right]\right] &=& 0 \ , \hskip 1.0cm \label{eqs3}\\ 
\frac{J}{W_{s}}\left[\left[\frac{\eta
	B^y}{D}\right]\right]+\frac{J}{W_{s}}\left[\left[
	\frac{W^2\eta^2v^y}{D}\right]\right]-
	B^x\left[\left[\frac{B^y}{W^2}\right]\right]-B^x\left[\left[\eta
	v^y\right]\right]-\frac{J}{W_{s}}\left[\left[
	\frac{S^y}{D}\right]\right] &=& 0 \ , \hskip 1.0cm \label{eqs4}\\ 
\frac{J}{W_{s}}\left[\left[\frac{\eta
	B^z}{D}\right]\right]+\frac{J}{W_{s}}\left[\left[\frac{W^2\eta^2
        v^z}{D}\right]\right]-B^x\left[\left[\frac{B^z}{W^2}\right]\right]-
	B^x\left[\left[\eta v^z\right]\right]-
	\frac{J}{W_{s}}\left[\left[\frac{S^z}{D}\right]\right]
	&=& 0 \ , \label{eqs5}\\ 
\left[\left[B^x\right]\right] &=& 0 \ , \hskip 1.0cm \label{eqs6}\\
\frac{J}{W_{s}}\left[\left[\frac{B^y}{D}\right]\right]+
	B^x\left[\left[v^y\right]\right] &=& 0 \ , \hskip 1.0cm \label{eqs7}\\
\frac{J}{W_{s}}\left[\left[\frac{B^z}{D}\right]\right]+
	B^x\left[\left[v^z\right]\right] &=& 0\ , \hskip 1.0cm 
\label{eqs8}
\end{eqnarray}
where we have defined $\eta \equiv \vec{v} \cdot \vec{B}$ and exploited
the property
\[
[[F(V_{s}-v^x)]]=\frac{J}{W_{s}}\left[\left[\frac{F}{D}\right]\right] \ ,
\]
valid for any scalar quantity $F$.

	The next step to take is to express all of the variables as
functions of $J$ and $p_b$ only. We start by using equation~(\ref{eqs1}) to
obtain
\begin{equation}
\frac{1}{D_b}=\left(v^x_a-v^x_b\right)\frac{W_{s}}{J}+\frac{1}{D_a} \ ,
\end{equation}
so that equation~(\ref{eqs2}) yields
\begin{equation}
\frac{\tau_b}{D_b}=-\frac{W_a^2 \eta_a^2}{D_a}+\frac{W_b^2
	\eta_b^2}{D_b}+\frac{W_{s}}{J}\left[B^x \left(\eta_a-
	\eta_b\right)-p_a v^x_a+p_b
	v^x_b\right]+\frac{\tau_a}{D_a} \ ,
\end{equation}
which depends on $v^x_b$, $p_b$ but also on $B^y_b$, $B^z_b$, $v^y_b$,
$v^z_b$. To remove the dependence from these latter quantities we employ
equations~(\ref{eqs7}) and (\ref{eqs8}) to obtain $B^y_b$ and $B^z_b$ as
functions of $v^x_b$, $v^y_b$ $v^z_b$ and $p_b$, {\it i.e.}
\begin{eqnarray}
B^y_b &=& D_b\left(\frac{B^y_a}{D_a}+\frac{W_{s}}{J}B^x
	v^y_a-\frac{W_{s}}{J}B^x v^y_b\right)\ , \\ \nonumber \\
B^z_b &=& D_b\left(\frac{B^z_a}{D_a}+\frac{W_{s}}{J}B^x v^z_a-
	\frac{W_{s}}{J}B^x v^z_b\right) \ .
\end{eqnarray}
We can now solve equation~(\ref{eqs3}) and finally obtain $v^x_b$ as a
function of $v^y_b$, $v^z_b$, $p_b$ and $J$
\begin{eqnarray}
\label{vxb}
v^x_b &=& \frac{D_a\left\{{B^2_x}
	W_{s}+W_a^2\left[W_{s}(p_b - p_a) - {B^2_x}W_{s}(1 - v^y_a v^y_b - 
	v^z_a v^z_b)+v^x_a(J+B^x W_{s}\eta_a)\right]\right\}}
	{W_a^2\left\{D_a\left[J-W_{s}({B^2_x}+p_a-p_b)v^x_a+B^x
	W_{s}\eta_a\right]-J({B^2_x}-p_b+W_a^2\eta_a^2-\tau_a)\right\}} +
\nonumber\\ \nonumber \\ 
&& \hskip 1.0cm \frac{J \left[B^x(B^y_a v^y_b+B^z_a
	v^z_b-\eta_a)+v^x_a(p_a-W^2_a\eta_a^2
	+\tau_a)\right]}{D_a[J-W_{s}({B^2_x}+p_a-p_b)v^x_a+B^x
	W_{s}\eta_a] - J({B^2_x}-p_b + W_a^2\eta_a^2-\tau_a)} \ ,
\nonumber \\ 
\end{eqnarray}
where it should be noted that equation~(\ref{vxb}) reduces to the
corresponding hydrodynamical expression in the limit of $\vec{B}=0$ [{\it
cf.} equation (4.12) of Pons et al. (2000) or equation (3.13) of Rezzolla
et al. (2003)]. Note also that using equations (\ref{eqs4}) and
(\ref{eqs5}) it is possible to obtain expressions for $v^y_b$ and $v^z_b$
in terms of the post-shock quantities $p$ and $J$; the corresponding
expressions are rather lengthy and uninspiring; for this reason we report
them in Appendix~\ref{appendixa}. 

	When all of the post-shock quantities are expressed as functions
of only $p_b$ and $J$ ({\it i.e.}  $V_{s}$), it is still necessary to
express $V_{s}$ as function of the post-shock pressure $p_b$. To do this
we follow Pons et al. (2000) and use the original jump conditions
(\ref{S1}), (\ref{S2}) and (\ref{S3}) to obtain
\begin{eqnarray}
\label{vs1}
\left[\left[p\right]\right]+J^2\left[\left[\frac{h_{\rm g}}{\rho}\right]\right]
	&=& 0 \ , \\ \nonumber \\
\label{vs2}
\left[\left[h_{\rm g}^2\right]\right]-
	\left( \hskip -0.1cm \left(\frac{h_{\rm g}}{\rho}\right) 
	\hskip -0.1cm \right)
	\left[\left[p\right]\right]-
	H\left[\left[b^2\right]\right]+2\left[\left[b^2\frac{h_{\rm g}}{\rho}
	\right]\right]-2J^2H\left[\left[\frac{h_{\rm g}}{\rho}\right]\right]
	&=& 0 \ ,
\end{eqnarray}
where $((F))\equiv F_a + F_b$ and $H \equiv {B_n^2}/{J^2}-{b^2}/{\rho^2}$
is a shock invariant quantity ({\it i.e.} $[[H]]=0$, \cite{anile89}). Note
that $B_n$ is not just the normal component of the magnetic field but,
rather, the projection of ${\boldsymbol b}$ along ${\boldsymbol n}$, {\it
i.e.}
\begin{equation}
B_n \equiv b^\mu n_\mu = -\frac{\eta}{\rho}J+\frac{W_{s}}{W}B^x \ .
\end{equation}
Equation~(\ref{vs2}) is also known as the {\it Lichnerowicz adiabat},
and represents the relativistic MHD counterpart of the {\it Hugoniot
adiabat}. 

	A couple of remarks should be made. Firstly, equations
(\ref{vs1})--(\ref{vs2}) can be used for fast and slow shocks but not
for an Alfv\`en discontinuity. In this case, in fact,
$[[{h}/{\rho}]]=0$, equations (\ref{vs1})--(\ref{vs2}) are simple
identities and the shock velocity $V_{s}$ is trivially given by the
local Alfv\`en velocity $V_{_{A}}$. Secondly, for purely
hydrodynamical shocks it is possible to find an analytic expression
for $V_{s}$ as a function of the post-shock pressure [{\it cf.}
equation (4.14) of Pons et al. (2000)]. In relativistic MHD, however,
the corresponding analytic expression has not been found and equation
(\ref{vs1}) needs to be solved numerically using a standard
root-finding algorithm, but also increasing the computational costs
considerably. To guarantee that we are using the right shock
velocity, the root is searched in the approriate physical interval,
i.e. $|V_s| \in \left(|V_A|,1\right)$ for fast shocks and $|V_s| \in
\left(|v^x|,|V_A|\right)$ for slow shocks.

\subsection{Solution across a rarefaction wave}
\label{sarw_pm}

	Rarefaction waves are self-similar solutions of the flow
equations, {\it i.e.} equations in which all of the fluid quantities
depend on $x$ and $t$ through the combination $\xi\equiv x/t$. Using this
as the independent variable, the set of partial differential MHD
equations can be rewritten as the following set of ordinary differential
equations (ODEs)
\begin{eqnarray}
\xi\frac{\mathrm{d}D}{\mathrm{d}\xi}-
	\frac{\mathrm{d}(Dv^x)}{\mathrm{d}\xi} &=& 0\ , 
\label{R1}\\ 
\xi\frac{\mathrm{d}(\tau-b^0b^0)}{\mathrm{d}\xi}-
	\frac{\mathrm{d}(S^x-b^0b^x-Dv^x)}{\mathrm {d}\xi} &=& 0\ , 
\label{R2}\\ 
\xi\frac{\mathrm{d}(S^x-b^0b^x)}{\mathrm{d}\xi}-
	\frac{\mathrm{d}(S^xv^x+p-b^xb^x)}{\mathrm{d}\xi} &=& 0\ ,
\label{R3}\\ 
\xi\frac{\mathrm{d}(S^y-b^0b^y)}{\mathrm{d}\xi}-
	\frac{\mathrm{d}(S^yv^x-b^xb^y)}{\mathrm{d} \xi} &=& 0 \ ,
\label{R4}\\ 
\xi\frac{\mathrm{d}(S^z-b^0b^z)}{\mathrm{d}\xi}-
	\frac{\mathrm{d}(S^zv^x-b^xb^z)}{\mathrm{d}\xi} &=& 0\ , 
\label{R5}\\ 
\xi\frac{\mathrm{d}B^x}{\mathrm{d}\xi} &=& 0\ ,
\label{R6}\\ 
\xi\frac{\mathrm{d}B^y}{\mathrm{d}\xi}-
	\frac{\mathrm{d}(B^yv^x-B^xv^y)}{\mathrm{d}\xi} &=& 0 \ ,
\label{R7}\\ 
\xi\frac{\mathrm{d}B^z}{\mathrm{d}\xi}-
	\frac{\mathrm{d}(B^zv^x-B^xv^z)}{\mathrm{d}\xi} &= & 0\ . 
\label{R8}
\end{eqnarray}
Equation (\ref{R1}) can be further decomposed as 
\begin{equation}
(v^x-\xi)\frac{\mathrm{d}\rho}{\mathrm{d}\xi}+
\rho\left[(v^x-\xi)W^2v^x+1\right]\frac{\mathrm{d}v^x}
	{\mathrm{d}\xi}+(v^x-\xi)\rho
	W^2v^y\frac{\mathrm{d}v^y}{\mathrm{d}\xi}+(v^x-\xi)\rho
	W^2v^z\frac{\mathrm{d}v^z}{\mathrm{d}\xi}=0 \ ,
\end{equation}
while combining equation (\ref{R2}) with equations (\ref{R3})--(\ref{R5})
provides us with the relations
\begin{eqnarray}
wW^2(v^x-\xi)\dxi{v^x}+(1-\xi v^x)\dxi{p}-v^x\xi\dxi{(b^0b^0)}+
	(v^x+\xi)\dxi{(b^0b^x)}-\dxi{(b^xb^x)} &=& 0 \ ,  \nonumber \\
&&\\
wW^2(v^x-\xi)\dxi{v^y}-\xi v^y\dxi{p}-
	\xi v^y\dxi{(b^0b^0)}+v^y\dxi{(b^0b^x)}+\xi\dxi{(b^0b^y)}-
	\dxi{(b^xb^y)} &=& 0 \ ,  \nonumber\\
&&\\
wW^2(v^x-\xi)\dxi{v^z}-\xi v^z\dxi{p}-
	\xi v^z\dxi{(b^0b^0)}+v^z\dxi{(b^0b^x)}+
	\xi\dxi{(b^0b^z)}-\dxi{(b^xb^z)} &=& 0\ . \nonumber\\
\label{RR8}
\end{eqnarray}

Finally, rewriting the definition of the local sound speed
\begin{equation}
c_s^2 \equiv \frac{1}{h_g}\left.\frac{\partial p_{\rm g}}{\partial
	\rho}\right|_s \ ,
\end{equation}
where $s$ is the specific entropy, in terms of the self-similar variable
\begin{equation}
\dxi{p_{\rm g}}=h_g c_s^2\dxi{\rho} \ ,
\end{equation}
and collecting the different terms in equations (\ref{R1})--(\ref{R8}),
we obtain the following system of seven ODEs in the seven variables
$\rho, p, v^x, v^y, v^z, B^y, B^z$, fully determining the solution across
a rarefaction wave
\begin{eqnarray}
&& 0 = (v^x-\xi)\frac{\mathrm{d}\rho}{\mathrm{d}\xi}+
	\rho\left[(v^x-\xi)W^2v^x+1\right]\frac{\mathrm{d}v^x}
	{\mathrm{d}\xi}+(v^x-\xi)\rho
	W^2v^y\frac{\mathrm{d}v^y}{\mathrm{d}\xi}+(v^x-\xi)\rho
	W^2v^z\frac{\mathrm{d}v^z}{\mathrm{d}\xi} \ , 
\nonumber \\ \label{ode1}\\ 
&& 0 = \dxi{p}-h_g c_s^2\dxi{\rho}+\left(B^2v^x-B^x\eta\right)\dxi{v^x}+
	\left(B^2v^y-B^y\eta\right)\dxi{v^y}+
	\left(B^2v^z-B^z\eta\right)\dxi{v^z}-
\nonumber\\
&&\hskip 7.5cm \left(\frac{B^y}{W^2}+v^y\eta\right)\dxi{B^y}-
	\left(\frac{B^z}{W^2}+v^z\eta\right)\dxi{B^z} \ ,
\nonumber \\ \label{eqnPraref}\\
&& 0 = (1-v^x\xi)\dxi{p}+\left[{B^2_x}(v^x+\xi)-
	2B^x\eta+W^2(v^x-\xi)(w-\eta^2)\right]\dxi{v^x}+
\nonumber\\ 
	&&
	2B_x^2\left\{\left[v^y+\frac{B^y(\xi-v^x)}{2B^x}\right]\dxi{v^y}
	+\left[v^z+\frac{B^z(\xi-v^x)}{2B^x}\right]\dxi{v^z}-
	\frac{(v^x-\xi)}{2 B^x}\left[v^y\dxi{B^y} + v^z\dxi{B^z}\right]\right\}\ , 
	\nonumber \\ \\ 
&&0 = \xi v^y\dxi{p}-
\nonumber\\
	&& B^xB^y\left\{
	\left[(v^x+\xi)-\frac{\eta}{B^x}\right]\dxi{v^x}+
	\left[2v^y+\frac{B^y(\xi-v^x)}{B^x}+
	\frac{W^2(v^x-\xi)(w-\eta^2)}{B^xB^y}-\frac{\eta}{B^y}\right]\dxi{v^y}\right\}-
\nonumber\\ 
	&& B^y[2B^xv^z+B^z(\xi-v^x)]\dxi{v^z}+
	\left[\frac{B^x+W^2(v^x-\xi)(B^yv^y+\eta)}{W^2}\right]\dxi{B^y}+
	B^yv^z(v^x-\xi)\dxi{B^z} \ , 
\nonumber\\ \\ 
&& 0 = \xi v^z\dxi{p} - 
\nonumber\\ 
&& B^xB^z\left\{\left[(v^x+\xi)-\frac{\eta}{B^x}\right]\dxi{v^x}+
	\left[2v^z+\frac{{B^z}(\xi-v^x)}{B^x}+\frac{W^2(v^x-\xi)(w-\eta^2)}{B^xB^z}-
	\frac{\eta}{B^z}\right]\dxi{v^z}\right\}-	
\nonumber \\
&& B^z[2B^xv^y+B^y(\xi-v^x)]\dxi{v^y}+
	B^zv^y(v^x-\xi)\dxi{B^y}+\left[\frac{B^x+W^2(v^x-\xi)
	(B^zv^z+\eta)}{W^2}\right]\dxi{B^z} \ , 
\nonumber \\ \\
	&& 0 = B^y\dxi{v^x}-B^x\dxi{v^y}+(v^x-\xi)\dxi{B^y}  , 
\\ \nonumber \\
\label{ode7}
	&& 0 = B^z\dxi{v^x}-B^x\dxi{v^z}+(v^x-\xi)\dxi{B^z} \ .
\end{eqnarray}

	The system of equations (\ref{ode1})--(\ref{ode7}) can be recast
into a simple matrix form and non-trivial similarity solutions exist only
if the determinant of the matrix of coefficients is zero. This condition
leads to a quartic equation in the self-similar variable $\xi$
\begin{eqnarray}
\tilde{b}_x^2 c_s^2 - \zeta^2 v_x^2 W^2 - (\zeta^2-1) v_x^4 W^4 +\left[2
	\zeta^2 v_x W^2 -2\tilde{b}^0 \tilde{b}_x c_s^2 + 4 (\zeta^2 -1) v_x^3
	W^4\right]  \xi & + &
\nonumber \\
\left[(\tilde{b}^0 - \tilde{b}_x) (\tilde{b}^0+\tilde{b}_x) c_s^2 +
	\zeta^2 (v_x^2-1) W^2 - 6 (\zeta^2-1) v_x^2 W^4\right]  \xi^2 & + &
\nonumber \\
\left[2 \tilde{b}^0 \tilde{b}_x c_s^2 - 2 \zeta^2 v_x W^2 + 4 (\zeta^2-1)
	v_x W^4\right]  \xi^3  + 
%
\left[W^4 + W^2 \zeta^2 \left(1 - W^2 \right)-(\tilde{b}^0)^2 c_s^2\right]  
	\xi^4 & = &0 \ , 
\nonumber \\
\label{quartic}
\end{eqnarray}
where
\begin{equation}
\tilde{b} \equiv \frac{b}{\sqrt{w}} \ ,
\hskip 2.0cm {\rm and} \hskip 2.0cm 
\zeta^2   \equiv c_s^2+\tilde{b}^2 (1 - c_s^2) \ ,
\end{equation}
and whose roots coincide with the eigenvalues of the original system of
equations (\ref{barno})--(\ref{maxwell}). When $B^x=0$, equation
(\ref{quartic}) reduces to a second-order equation whose roots provide
the velocities of the left and right-going fast-waves. In the more
general case when $B^x\neq 0$, however, the quartic cannot be recast as
the product of two quadratic equations (as it is the case in Newtonian
hydrodynamics) and the solution must be found numerically. The
corresponding roots provide the velocities of the left and right-going
slow and fast magnetosonic rarefaction waves, respectively.

	Using the appropriate root for $\xi$, the system of ODEs
(\ref{ode1})--(\ref{ode7}) can be rewritten in terms of the total
pressure to obtain a new system of six ODEs to be integrated from the
value of pressure ahead the rarefaction to the one behind it\footnote{The
number of equations to be solved reduces from seven to six because when
using the total pressure as the self-similar variable one equation
becomes then trivial, {\it i.e.}  $dp/dp=1$.}. The explicit expressions
of these equations are rather lengthy and do not provide any important
information; for this reason we report them in Appendix~\ref{appendixb}.

\subsection{Solution across an Alfv\`en discontinuity}
\label{saad}

The solution across Alfv\`en discontinuities is found by imposing the
continuity of $\rho$ and $p$ and then solving the system of equations
(\ref{eqs3})-(\ref{eqs5}) and (\ref{eqs7})-(\ref{eqs8}), using $V_s=V_A$,
where $V_A \equiv v_x+B_x/[W^2(\eta\mp\sqrt{w})]$ is the Alfv\`en
velocity for left ($-$) and right ($+$) going waves, respectively. Since
$\rho$ and $p$ are continuous across the Alfv\`en discontinuity, a
solution needs to be found only for the three components of $\vec{v}$ and
for the tangential components of the magnetic field $B_y$ and $B_z$. In
general, and because no analytic solution was found, we solve the
corresponding system of equations (\ref{eqs3})-(\ref{eqs5}),
(\ref{eqs7})-(\ref{eqs8}) numerically with a Newton-Raphson scheme. No
major difficulties have been found in determining an accurate solution
provided that the waves are all well separated and that a sufficiently
accurate initial guess is provided ({\it cf.} solution in
Figure~\ref{BG_AL1}). For the latter we have used an approximate Riemann
solver based on the Harten-Lax-van Leer-Einfeldt (HLLE)
algorithm~(\cite{harten83,einfeldt88}) and a moderate truncation error
({\it i.e.}  using about 800 gridpoints for the tests reported here).
However, considerable difficulties have been encountered if the
waves are very close to each other. This is the case, for instance, of
test number 5 of Balsara (2001), in which the left-going Alfv\`en
discontinuity and the left-going slow rarefaction wave have very similar
propagation velocities ({\it cf.} solution in Figure~\ref{BA5}). The
exact solution found in this case has a truncation error which is small,
but larger that those reached in the other tests ({\it cf.} data in
Table~\ref{tab:balsara5}).

\section{Tangential Magnetic Field Approach: ``${\boldsymbol B_t}$-method''}
\label{btmethod}

	As done in Sect.~\ref{pmethod}, in what follows we describe in
detail the approach referred to as the $B_t$-method, in which we
calculate all of the variables in the Riemann fan using as unknowns the
values of the tangential components of the magnetic field, {\it i.e.}
$B^y$ and $B^z$. As mentioned in the Introduction, much of the
inspiration in the development and use of this method comes from the
corresponding approach developed by Ryu and Jones (1995) in
nonrelativistic MHD. However, important differences are also present.

	In particular, in Newtonian MHD the problem can be solved
using the norm of the tangential component of the magnetic field $B_t
\equiv \sqrt{B^2_y+B^2_z}$ and the rotation angle $\psi \equiv
\arctan(B^z/B^y)$ across Alfv\`en discontinuities. This is because
$B_t$ is conserved across Alfv\`en discontinuities and $\psi$ is
constant across fast and slow-waves (see \cite{jeffrey66}). As a
result, the relevant system of equations is solved using as unknowns
the values of $B_t$ in regions R2-R3, R4-R5, R6-R7 of the Riemann fan
in Figure~\ref{mhd_zones} and the angle $\psi$ in regions R3-R6. At
the contact discontinuity it is then necessary to solve a system of
four equations, given by the continuity of $\vec{v}$ and of $p$, in
the same four unknowns. This can be solved using root-finding
techniques such as the Newton-Raphson method. Finally, when $B_x=0$,
the presence of only two fast waves and a tangential discontinuity
makes the solution of the problem even simpler (see Ryu and Jones 1995
for details).

	In relativistic MHD, on the other hand, the value of $B_t$ can
be discontinuous across Alfv\`en waves and the angle $\psi$ can vary
across fast and slow-waves; it is then not possible to solve the
system using the same method. Note also that the equations reported
below both for shock and rarefactions waves are strictly valid only if
$B^x \ne 0$ and indeed should be used only in regions R4 and R5 of the
Riemann fan shown in Figure~\ref{mhd_zones}. In these regions, only
slow-waves are present and these do not appear when $B^x=0$.

\subsection{Solution across a shock front}
\label{sasf_btm}

	To calculate the solution across a shock front within the
$B_t$-method we start by considering the same system of equations in
Section~\ref{sasf_pm}, but we solve equations (\ref{S1})--(\ref{S3})
considering $B^y$ and $B^z$ as the unknown quantities. From equations
(\ref{eqs7}) and (\ref{eqs8}) we express the post-shock values of $v^y$
and $v^z$:
\begin{eqnarray}
v^y_b &=& \frac{1}{B^x}\left[B^y_a\frac{J}{W_s D_a}-B^y_b \left(
	\frac{J}{W_s D_a}+v^x_a-v^x_b\right)+B^x v^y_a\right] \ , 
\\ \nonumber \\ 
v^z_b &=& \frac{1}{B^x}\left[B^z_a\frac{J}{W_s D_a}-B^z_b 
	\left( \frac{J}{W_s D_a}+v^x_a-v^x_b\right)+B^x v^z_a\right] \ .
\end{eqnarray}
Using now equation (\ref{eqs1}) to obtain the post-shock value of $D$
\begin{equation}
D_b = \frac{D_a J}{J+D_a W_s \left(v^x_a-v^x_b\right)} \ ,
\end{equation}
and calculating the post-shock value of the total pressure using the
invariance of $h_g B_n$, {\it i.e.} $[[h_g B_n]]=0$ (see Anile 1989),
we can express all of the quantities as a function of the post-shock
values of $v^x$, $B^y$, $B^z$, and of the shock-velocity $V_s$. An
analytic solution for the post-shock value of $v^x$ in terms of
the other post-shock quantities was sought but not found, forcing to
the numerical solution of one of the equations
(\ref{eqs3})--(\ref{eqs5}). Furthermore, in analogy with what done in
the $p$-method, we calculate the value of the shock velocity by
solving numerically equation (\ref{vs1}). 

	Finally, it may be useful to point out that the numerical
solution of equation (\ref{vs1}) is at times complicated by the existence
of two acceptable roots in the interval of velocities in which the value
of the slow shock velocity has to be found ({\it i.e.}  between the value
of $v^x$ and the value of the Alfv\`en velocity). Because only one of
these two roots will lead to a convergent exact solution, a careful
selection needs to be made. {The existence of these two roots could
be related to a known problem in Newtonian MHD where the use of the
tangential components of the magnetic field as the post-shock independent
variables can lead to the presence of more than one solution ({\it cf.},
for instance, \cite{jeffrey64}). This problem seems to be present also in
relativistic MHD (\cite{KO03}), but it has not represented a serious
drawback for the approach followed here. More work is needed to determine
whether the use of the tangential components of the magnetic field as the
post-shock independent variables is really optimal or whether different
choices are preferable.}

\subsection{Solution across a rarefaction wave}
\label{sarw}

	To calculate the solution across a rarefaction wave within the
$B_t$-method we use the same set of ODEs (\ref{R1})--(\ref{R8}) discussed
in Section~\ref{sarw_pm}, with the only but important difference that we
do not use $\xi$ as self-similar variable but, rather, the norm of the
tangential components of the magnetic field $B_t$. More specifically, we
use equations (\ref{R1})--(\ref{R3}) together with equations
(\ref{R7})--(\ref{R8}) and substitute the derivative with respect to
$\xi$ with the one with respect to $B_t$. In addition to these equations,
which provide a solution for variables $\rho, p, v^x, v^y$ and $v^z$, we
express explicitly the relation between the norm and the tangential
components in terms of the angle $\psi$
\begin{eqnarray}
B^y &=& \cos{\psi} B_t \ , \\ \nonumber \\
B^z &=& \sin{\psi} B_t \ ,
\end{eqnarray}
and rewrite them as ODEs having $B_t$ as the self-similar variable
\begin{eqnarray}
\label{constpsi1}
\frac{\mathrm{d}B^y}{\mathrm{d}B_t} &=& \cos{\psi} \ , 
\\ \nonumber \\
\label{constpsi2}
\frac{\mathrm{d}B^z}{\mathrm{d}B_t} &=& \sin{\psi} \ .
\end{eqnarray}
Note that in deriving equations (\ref{constpsi1})--(\ref{constpsi2}),
an implicit assumption has been made: {\it i.e.} that the angle $\psi$
is constant across the rarefaction wave and thus that the tangential
magnetic field does not rotate across the rarefaction wave. With the
use of the supplementary equations
(\ref{constpsi1})--(\ref{constpsi2}), the resulting system of ODEs is
complete and can be solved numerically using standard techniques for
the solution of a system of coupled ODEs. In practice, the integration
is started ahead of the rarefaction and is progressed toward the
contact discontinuity, where $B_t$ is given by the values of $B^y$ and
$B^z$ chosen at the contact discontinuity. {In all of the tests
reported here (with the exception of test number 5 of Balsara 2001;
see Section~\ref{num_bxneq0_tests} for a discussion), the assumption
$\psi={\rm const.}$ is valid. This is probably related to the choice
of the initial conditions used in these tests and in particular to the
fact that $v_A^y = v_A^z$, $B_A^y = B_A^z$, or $v_A^z=B_A^z = 0$,
where $A=(${\em left}, {\em right}$)$, so that the initial states are
essentially invariant after the exchange of $y$ with $z$ or the $z$
components of $v$ and $B$ remain equal to zero in all the regions.}

	It should be noted that also in relativistic hydrodynamics the
velocity components tangential to a nonlinear wave can change their norm
across the wave, in contrast with what happens in Newtonian
hydrodynamics. Considering for simplicity the case for a shock wave in
the limit of zero magnetic field, equations (\ref{eqs4})--(\ref{eqs5})
reduce to $[[S^y/D]] = 0 = [[S^z/D]]$, indicating that the ratio
$v^y/v^z$ remains unchanged through shocks so that the tangential
velocity 3-vector can change its norm but does not rotate. This property,
which applies also across rarefaction waves, is not present across
Newtonian nonlinear waves, in which the tangential 3-velocity vector does
not rotate, nor changes its norm: $[[v^y]]= 0 =[[v^z]]$.

	Although the condition $\psi={\rm const.}$ is exact in
nonrelativistic MHD, it may not be valid in relativistic regimes, where
the tangential magnetic field is instead free to rotate across the slow
rarefaction. In this case, a new strategy needs to be implemented and the
simplest one consists of using the angle $\psi$ as the self-similar
variable so that the system of equations (\ref{R1})--(\ref{R8}) can be
expressed in terms of this new integration variable. In addition, the
supplementary differential equation for one of the components of the
tangential magnetic field can be obtained through the algebraic relation
\begin{equation}
B^y=\frac{\cos\psi}{\sin\psi} B^z \ ,
\end{equation}
and its derivative with respect to $\psi$
\begin{equation}
\frac{\mathrm{d}B^y}{\mathrm{d}\psi} =
	\frac{\cos\psi}{\sin\psi} \frac{\mathrm{d}B^z}{\mathrm{d}\psi} - 
	\frac{1}{\sin^2\psi}B^z \ .
\label{eq:psi}
\end{equation}
	The integration of the system of ODEs is done starting from the
value of $\psi$ given by the ratio of the tangential components of the
magnetic field ahead of the rarefaction wave, up to the value given by
the amplitudes of $B^y$ and $B^z$ at the contact
discontinuity. Furthermore, as for the $p$-method, also within the
$B_t$-method the values of the variable $\xi$ are obtained from the
quartic equation (\ref{quartic}).

	A representative example of this effect is shown in
 Figure~\ref{BG_AL1_normb}, where we plot the exact solution of the
 generic Alfv\`en test at time $t=1.5$ ({\it cf.} Table~\ref{tab:tests}
 for the initial conditions of this test). In particular, the left panel
 of Figure~\ref{BG_AL1_normb} shows the norm of the tangential magnetic
 field $B_t$, while the right panel the angle $\psi \equiv
 \arctan{(B^z/B^y)}$. Note how both quantities vary across all the fast,
 slow and Alfv\`en waves.

\section{Numerical Implementation and Representative Results}
\label{numerical}

	Since the properties of the magnetic field components in the
initial states lead to considerably different Riemann problems ({\it cf.}
the two Riemann fans in Figures~\ref{mhd_zones_nullBx}
and~\ref{mhd_zones}), we will discuss separately the numerical solution
in the cases in which $B^x=0$ and $B^x \ne 0$, emphasizing the properties
of some of the most representative tests.

\subsection{Tangential Initial Magnetic Field: $B^x=0$}
\label{num_bxeq0}

	As discussed in Section~\ref{strategy}, when $B^x=0$ the Riemann
problem consists of only two fast-waves and of a tangential discontinuity
across which only $v^x$ and $p$ are continuous ({\it cf.}
Figure~\ref{mhd_zones_nullBx}). It should be noted that the condition of
continuity of the total pressure across the tangential discontinuity does
not necessarily extend also to the gas pressure and, indeed, the latter
is in general discontinuous ({\it cf.}  Figures~\ref{KOST2}
and~\ref{BG1}). In essence, the numerical solution of the Riemann problem
when $B^x=0$ proceeds as follows: given the initial left and right states
({\it i.e.} regions R1 and R4 of Figure~\ref{mhd_zones_nullBx}), we
follow the procedure used in relativistic hydrodynamics and determine two
unknown states as function of the common total pressure in regions R2 and
R3 ($p$-method). The jump in the normal component of the velocity at the
tangential discontinuity is then checked and a new guess for the total
pressure found. This procedure is iterated until the solution is found
with the desired accuracy. The numerical approach used is a combination
of Newton-Raphson and bisection methods, starting from a value for the
total pressure which is the average of the initial left and right
states. Furthermore, to decide whether the wave considered is a shock or
a rarefaction, we compare the values of the total pressure ahead of and
behind the wave, solving the set of equations across a shock if the
guessed value is larger than the total pressure ahead of the wave and
thus in the initial state. We note that this procedure could be improved
if an approach similar to the one discussed by Rezzolla et al. (2001,
2003) is implemented, which would exploit the values of the initial
relative velocity to predict the wave-pattern produced.

	It is also worth noting that even though the numerical strategy
discussed so far is very similar to the one used in relativistic
hydrodynamics, the equations to be solved in MHD are much more complex
and, more importantly, their computational cost markedly larger. This is
essentially because an analytic expression for the shock velocity was not
found, so that the latter must be determined numerically.

\begin{table}
\begin{center}
\begin{tabular}{|r|c|c|c|c|c|c|c|c|}
\cline{1-9} & & & & & & & & \\
{\bf Test type}~~~~~~~ & $\rho$ & $p_{\rm g}$ & $v^x$ & $v^y$ & $v^z$ &
$B^x$ & $B^y$ & $B^z$ \\
& & & & & & & & \\ \cline{1-9} & & & & & & & & \\
{\bf Komissarov: Shock-Tube 2}~~~($\Gamma=4/3$)  & & & & & & & & \\
{\it left state} ~~~& 1.0 & 30.0 & 0.0 & 0.0 & 0.0 & 0.0 & 20.0 & 0.0 \\
{\it right state} ~~~& 0.1 & 1.0  & 0.0 & 0.0 & 0.0 & 0.0 & 0.0  & 0.0 \\
\cline{1-9} & & & & & & & & \\
{\bf Generic Shock-Tube}~~~($\Gamma=5/3$) & & & & & & & & \\ 
{\it left state}  ~~~& 1.0  & 0.01 & 0.1 & 0.3 & 0.4 & 0.0 & 6.0  & 2.0 \\
{\it right state} ~~~& 0.01 & 5000 & 0.5 & 0.4 & 0.3 & 0.0 & 5.0  & 20.0 \\
\cline{1-9} 
\end{tabular}
\end{center}
\smallskip
\caption{Initial conditions for the tests of the exact Riemann solver when
the magnetic field has zero normal component, {\it i.e.} $B^x=0$.}
\label{tab:testsfornullBx} 
\end{table}
%
\begin{table}
\begin{center}
\begin{tabular}{|c|c|c|c|c|c|c|c|}
\cline{1-8} & & & & & & & \\
 & $\rho$ & $p$ & $v^x$ & $v^y$ & $v^z$ & $B^y$ & $B^z$ \\
& & & & & & & \\ \cline{1-8} & & & & & & & \\ 
R1 & 0.1000E+01 &    0.2300E+03 &    0.0000E+00 &    0.0000E+00 &    0.0000E+00 &    0.2000E+02 &    0.0000E+00 \\
R2 & 0.2410E+00 &    0.1611E+02 &    0.8497E+00 &    0.0000E+00 &    0.0000E+00 &    0.9141E+01 &    0.0000E+00 \\
R3 & 0.6426E+00 &    0.1611E+02 &    0.8497E+00 &    0.0000E+00 &    0.0000E+00 &    0.0000E+00 &    0.0000E+00 \\
R4 & 0.1000E+00 &    0.1000E+01 &    0.0000E+00 &    0.0000E+00 &    0.0000E+00 &    0.0000E+00 &    0.0000E+00 \\
\cline{1-8}
\end{tabular}
\end{center}
\smallskip
\caption{First significant digits for the exact solution of the test
shock-tube 2 of Komissarov (1999) computed with an accuracy of
$10^{-12}$. The left column indicates the regions in which the solution
is computed ({\it cf.} Fig.~\ref{mhd_zones_nullBx}).}
\label{tab:kost2} 
\end{table}
%
\begin{table}
\begin{center}
\begin{tabular}{|c|c|c|c|c|c|c|c|}
\cline{1-8} & & & & & & & \\
& $\rho$ & $p$ & $v^x$ & $v^y$ & $v^z$ & $B^y$ & $B^z$ \\
& & & & & & & \\ \cline{1-8} & & & & & & & \\ 
R1 & 0.1000E+01 &    0.1819E+02 &    0.1000E+00 &    0.3000E+00 &    0.4000E+00 &    0.6000E+01 &    0.2000E+01 \\
R2 & 0.1581E+01 &    0.4459E+02 &   -0.3073E+00 &    0.3082E+00 &    0.2927E+00 &    0.9582E+01 &    0.3194E+01 \\
R3 & 0.5489E-03 &    0.4459E+02 &   -0.3073E+00 &    0.7488E+00 &    0.5556E+00 &    0.1023E+01 &    0.4092E+01 \\
R4 & 0.1000E-01 &    0.5138E+04 &    0.5000E+00 &    0.4000E+00 &    0.3000E+00 &    0.5000E+01 &    0.2000E+02 \\
\cline{1-8} 
\end{tabular}
\end{center}
\smallskip
\caption{The same as Table~\ref{tab:kost2} but for the generic shock-tube test computed
  with an accuracy of $10^{-11}$.}
\label{tab:bgt1}
\end{table}

\subsubsection{Representative Tests for $B^x=0$}
\label{num_bxeq0_tests}

	Because initial states with a zero normal magnetic field lead to
a Riemann problem that is comparatively much simpler to solve, an
independent numerical code has been built for this case and it has been
tested to reproduce known results in relativistic hydrodynamics, as well
as a test proposed by Komissarov (1999) (this is referred to as the
``shock-tube'' test 2). We have also considered an additional, more
generic shock-tube test in which all of the quantities in the initial
states are nonzero and in which $\vec{v} \cdot \vec{B}\neq0$ (this is
referred to as the ``generic shock-tube'' test)\footnote{We note that a
Riemann problem with $B^x=0$, but with $\vec{v} \cdot \vec{B}\neq0$
cannot be solved with the exact solution recently proposed by Romero et
al. (2005).}.

\begin{figure}
\begin{center}
      \includegraphics[width=0.45\textwidth]{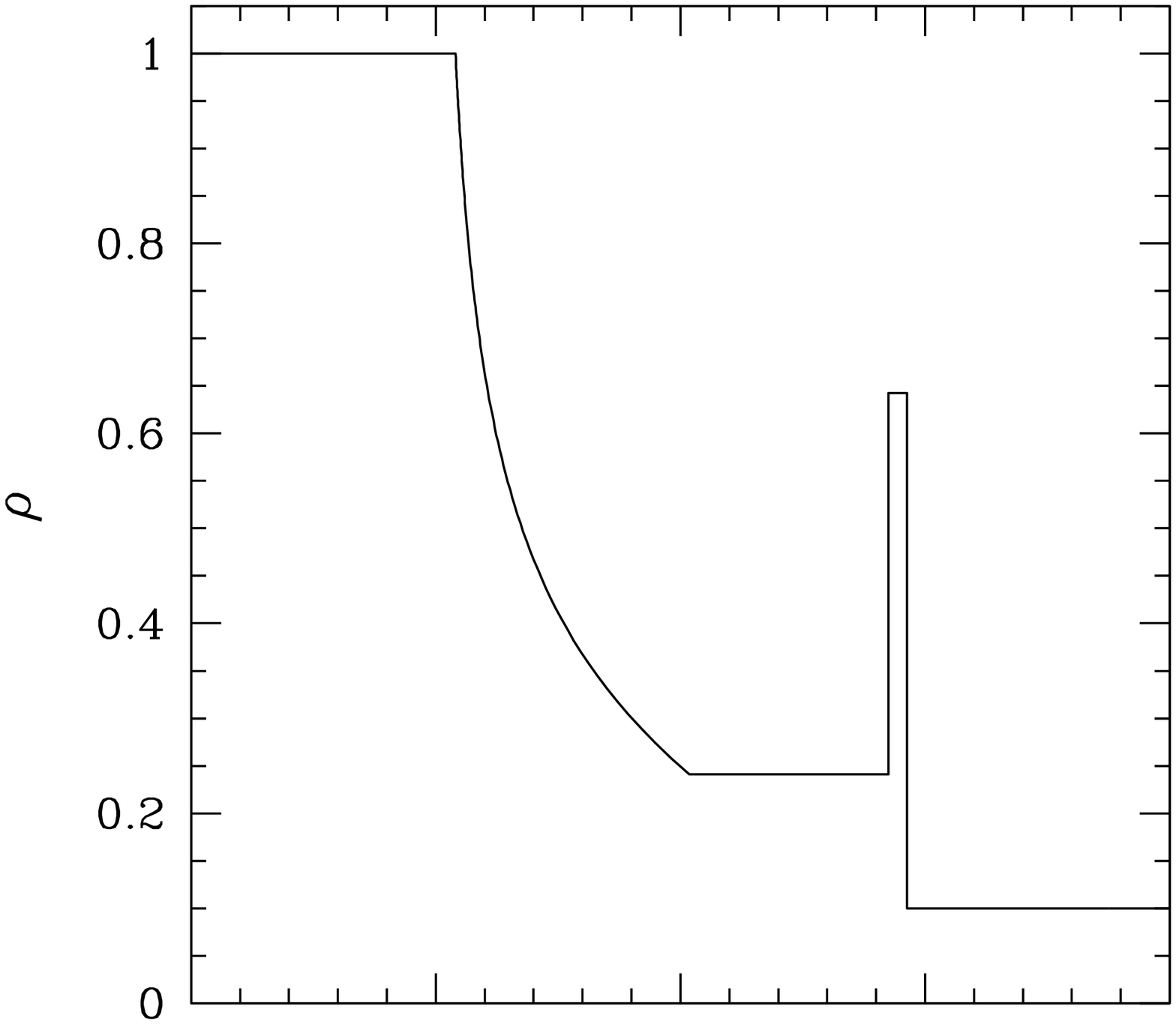}
      \hskip 0.75cm 
      \includegraphics[width=0.45\textwidth]{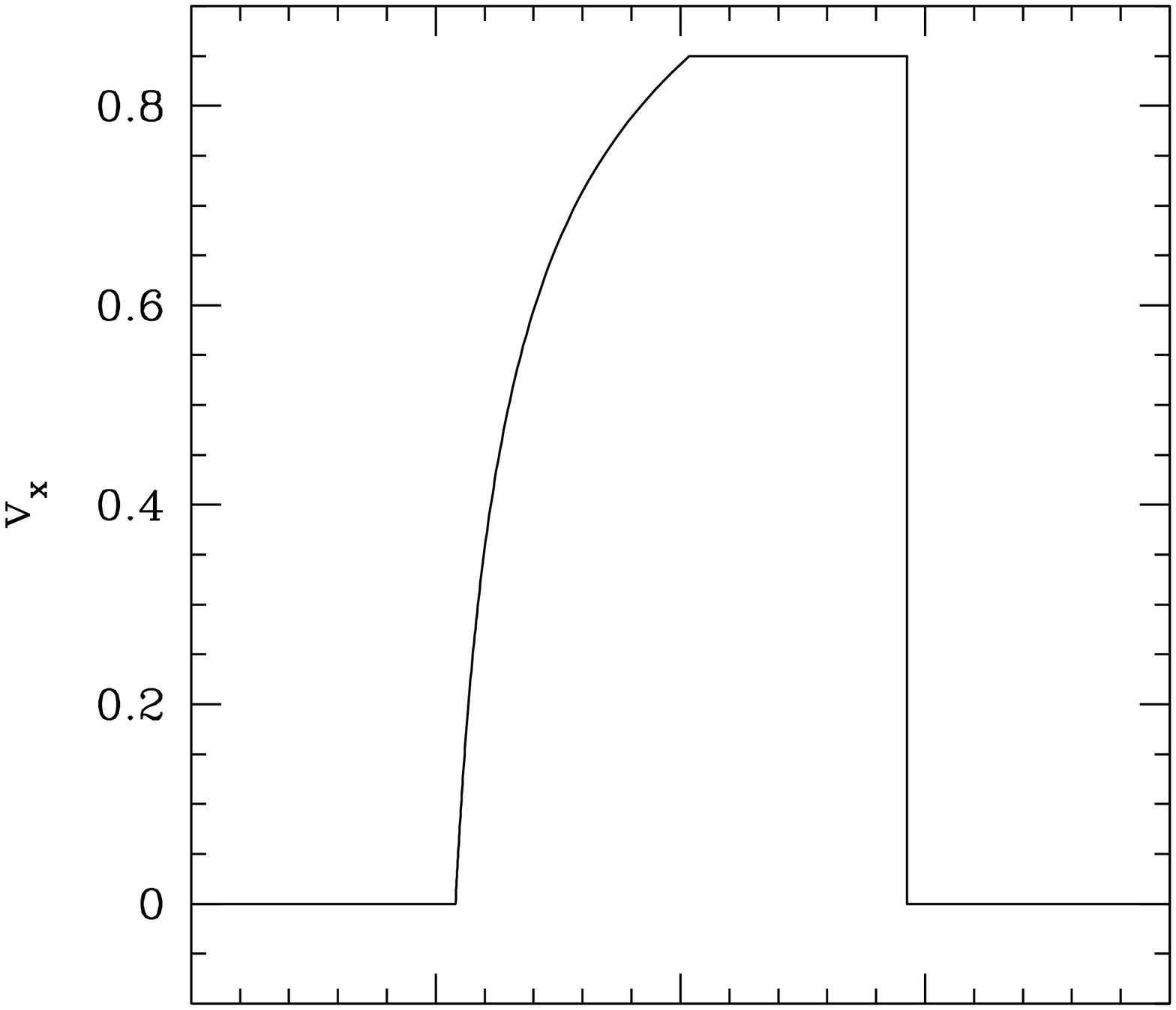}
      \vskip -1.0cm 
      \includegraphics[width=0.45\textwidth]{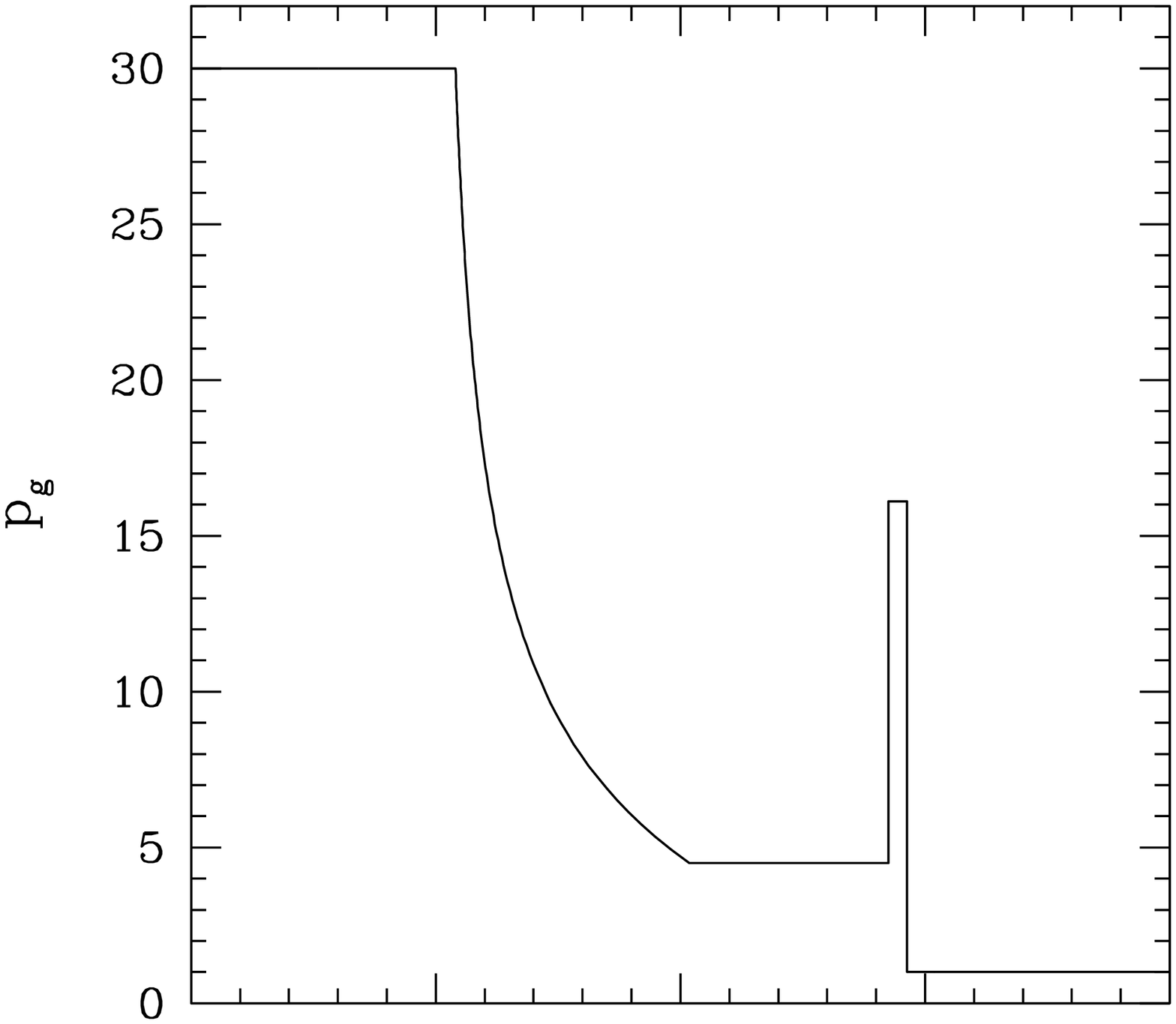}
      \hskip 0.75cm 
      \includegraphics[width=0.45\textwidth]{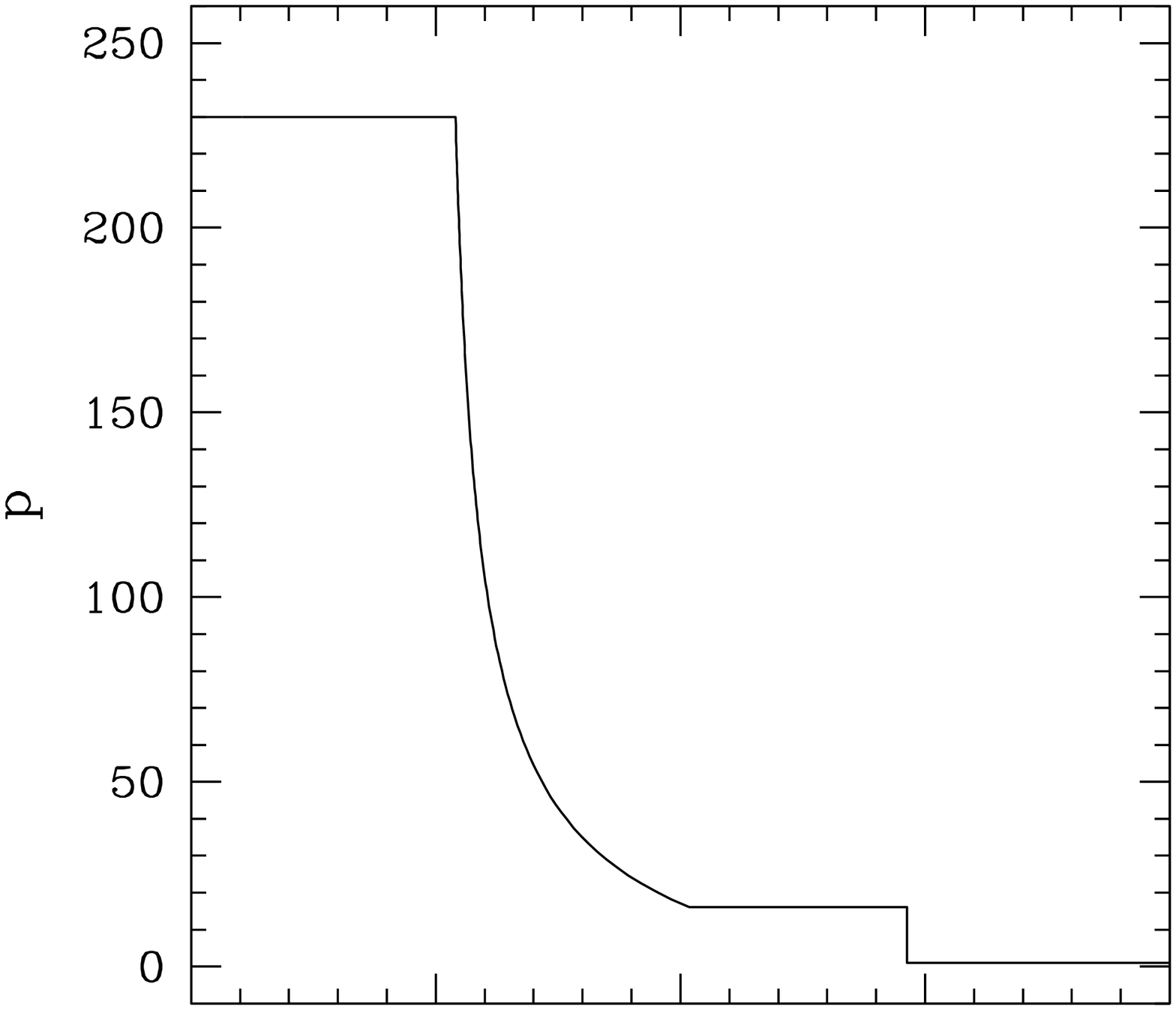}
      \vskip -1.0cm 
      \includegraphics[width=0.45\textwidth]{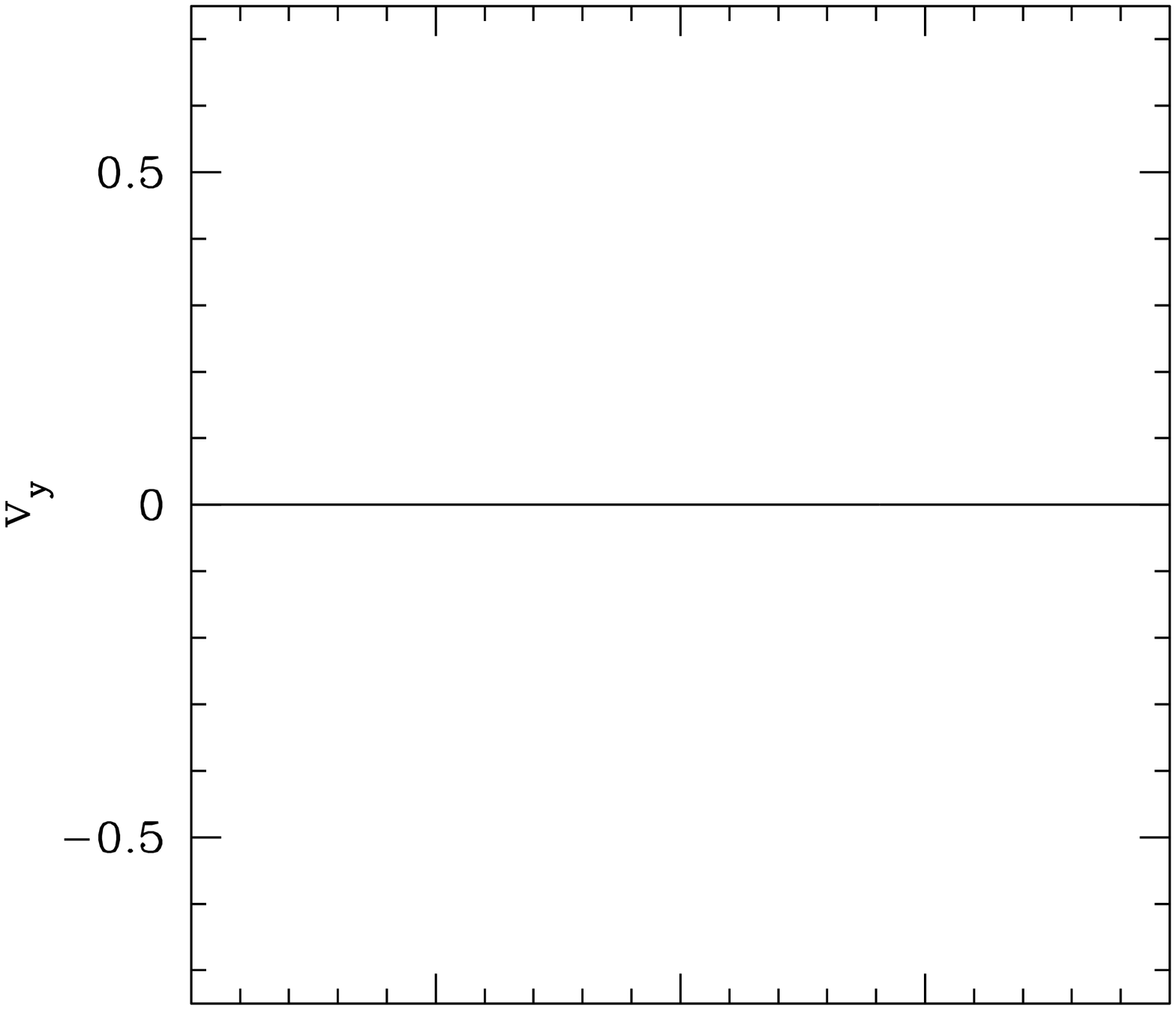}
      \hskip 0.75cm 
      \includegraphics[width=0.45\textwidth]{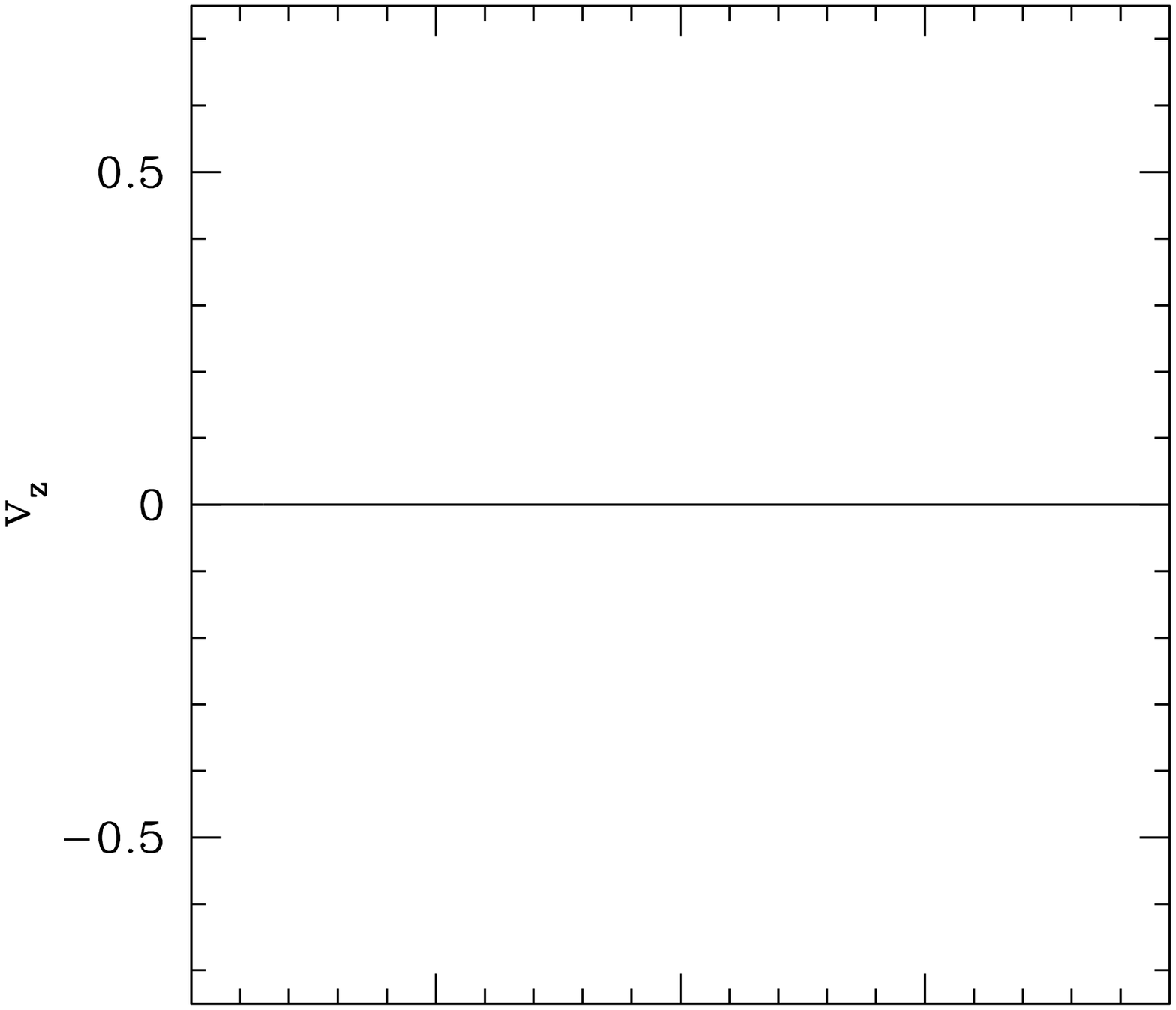}
      \vskip -1.0cm 
      \includegraphics[width=0.45\textwidth]{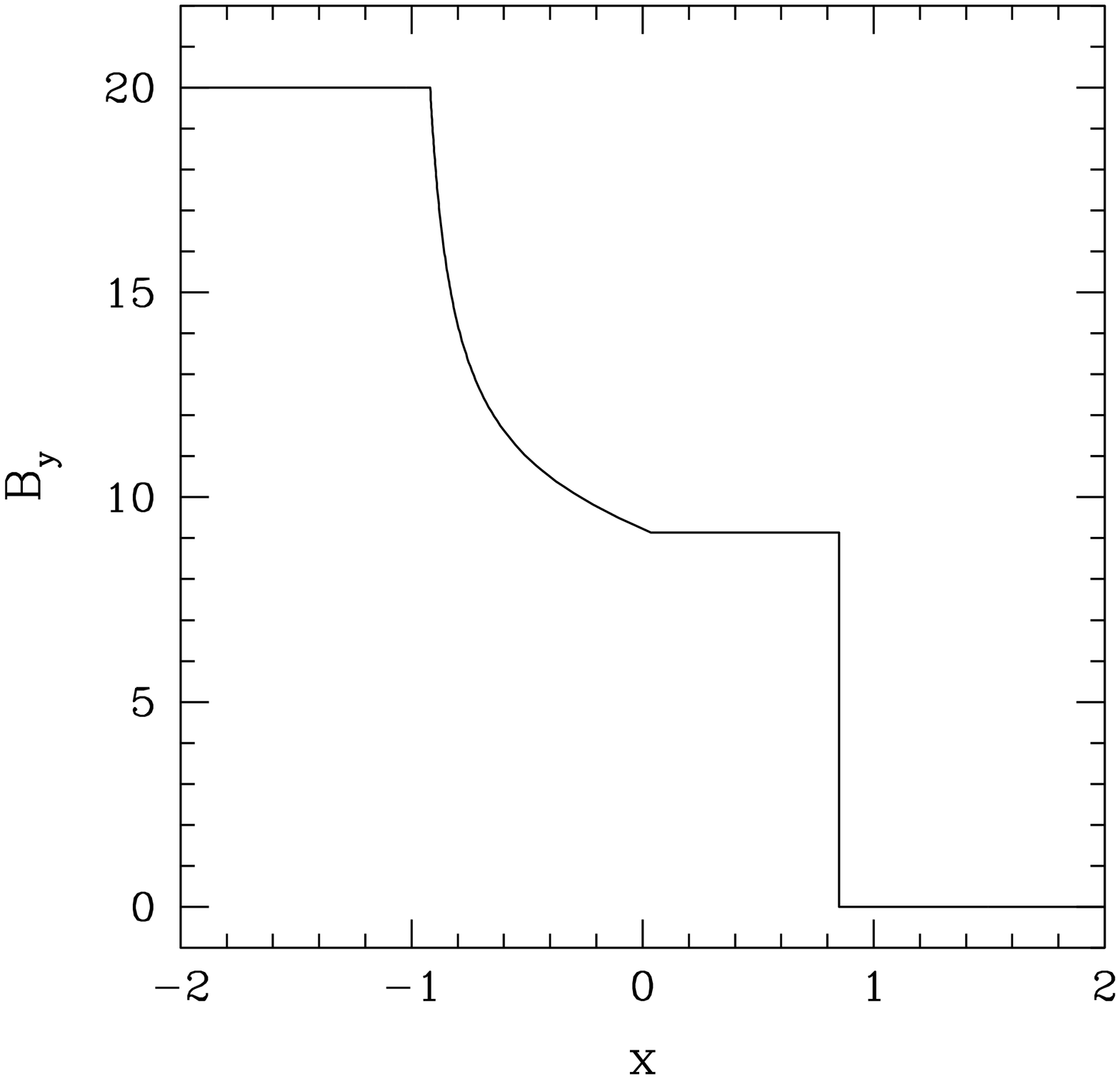}
      \hskip 0.75cm 
      \includegraphics[width=0.45\textwidth]{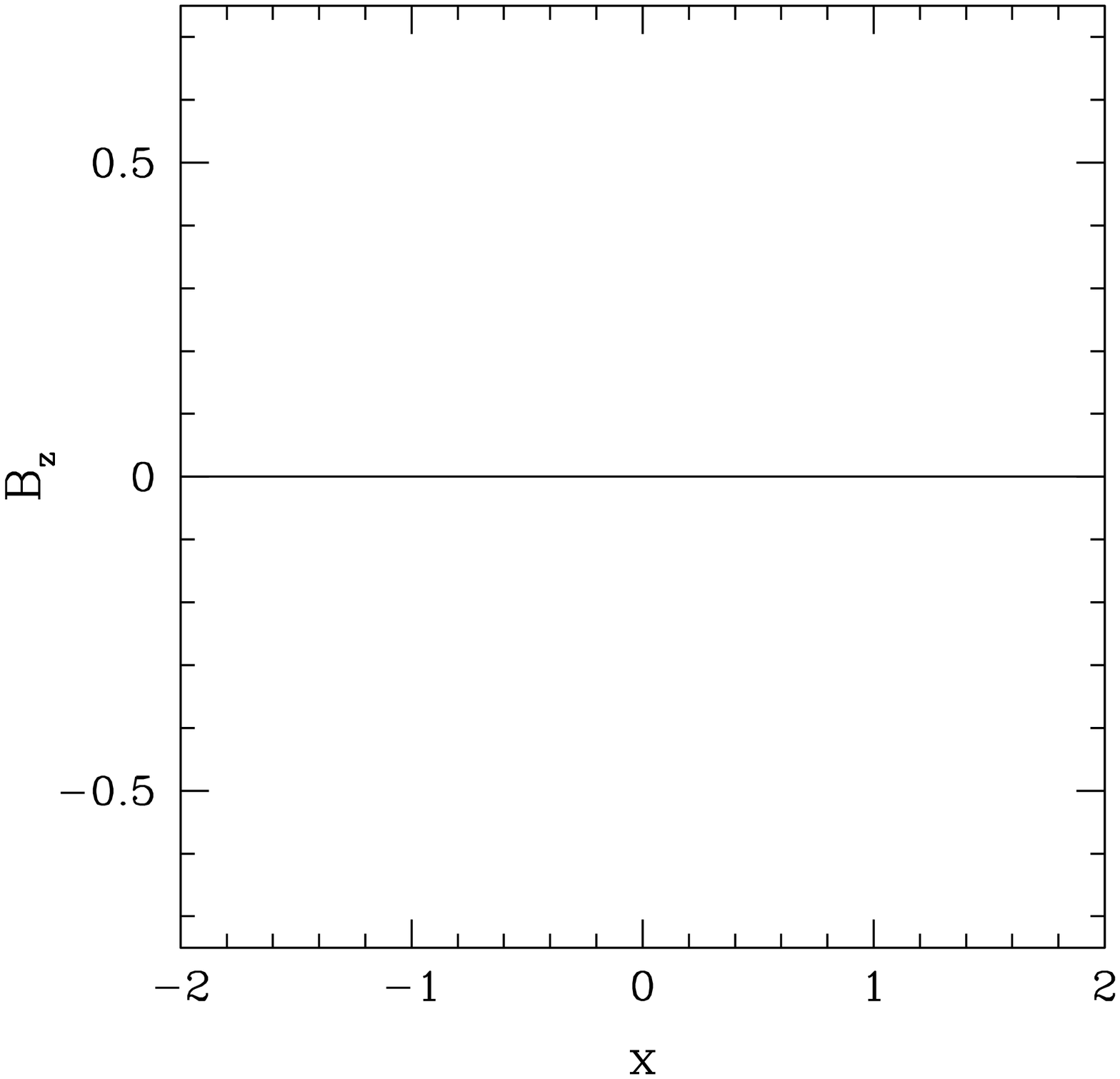}
  \caption{\label{KOST2}Exact solution of the test shock-tube 2 of
    Komissarov (1999) at time $t=1.0$. The solution is composed of a
    left-going rarefaction wave, a tangential discontinuity and a
    right-going shock.}
  \end{center}
\end{figure}

\begin{figure}
\begin{center}
      \includegraphics[width=0.45\textwidth]{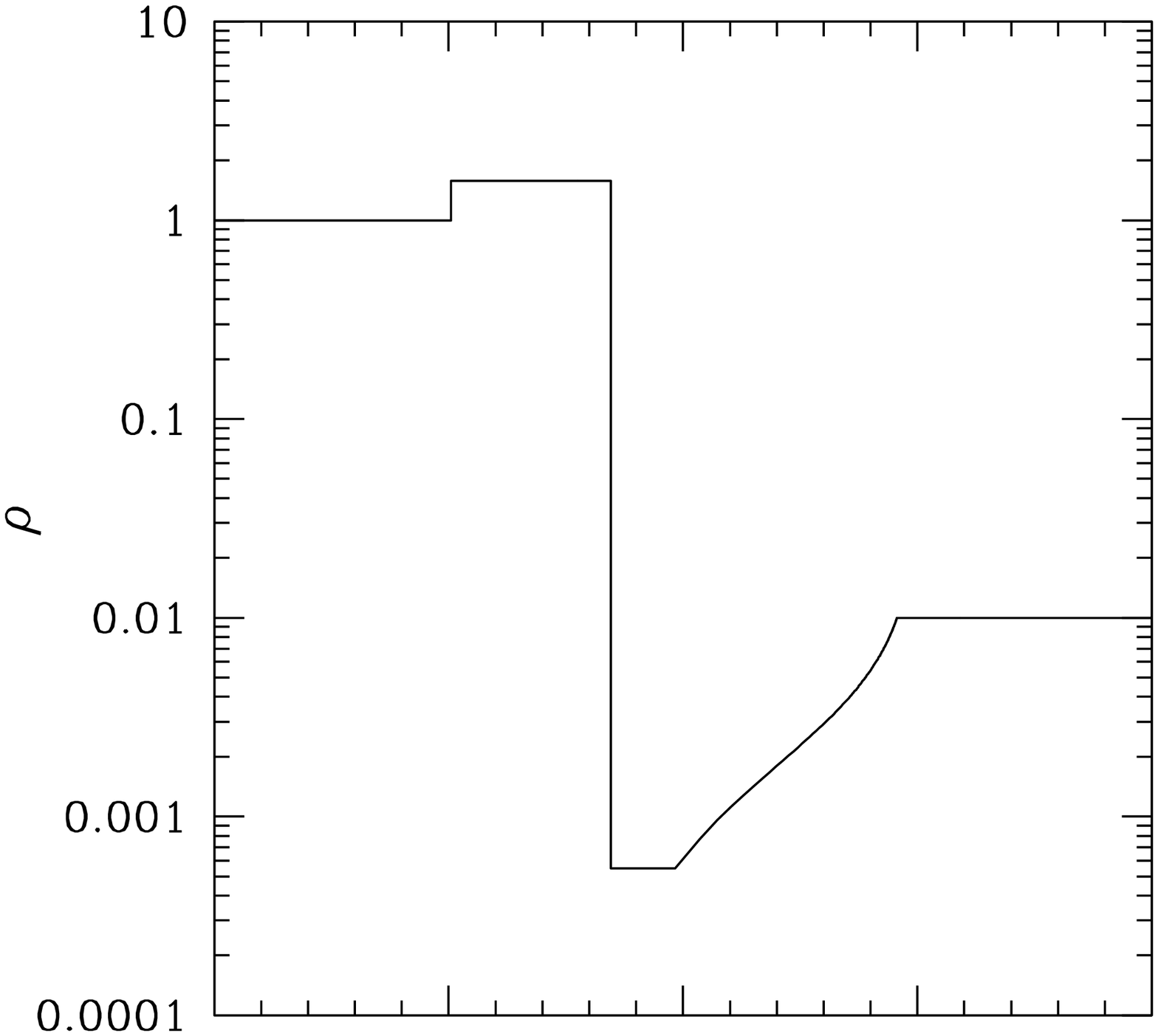}
      \hskip 0.75cm 
      \includegraphics[width=0.45\textwidth]{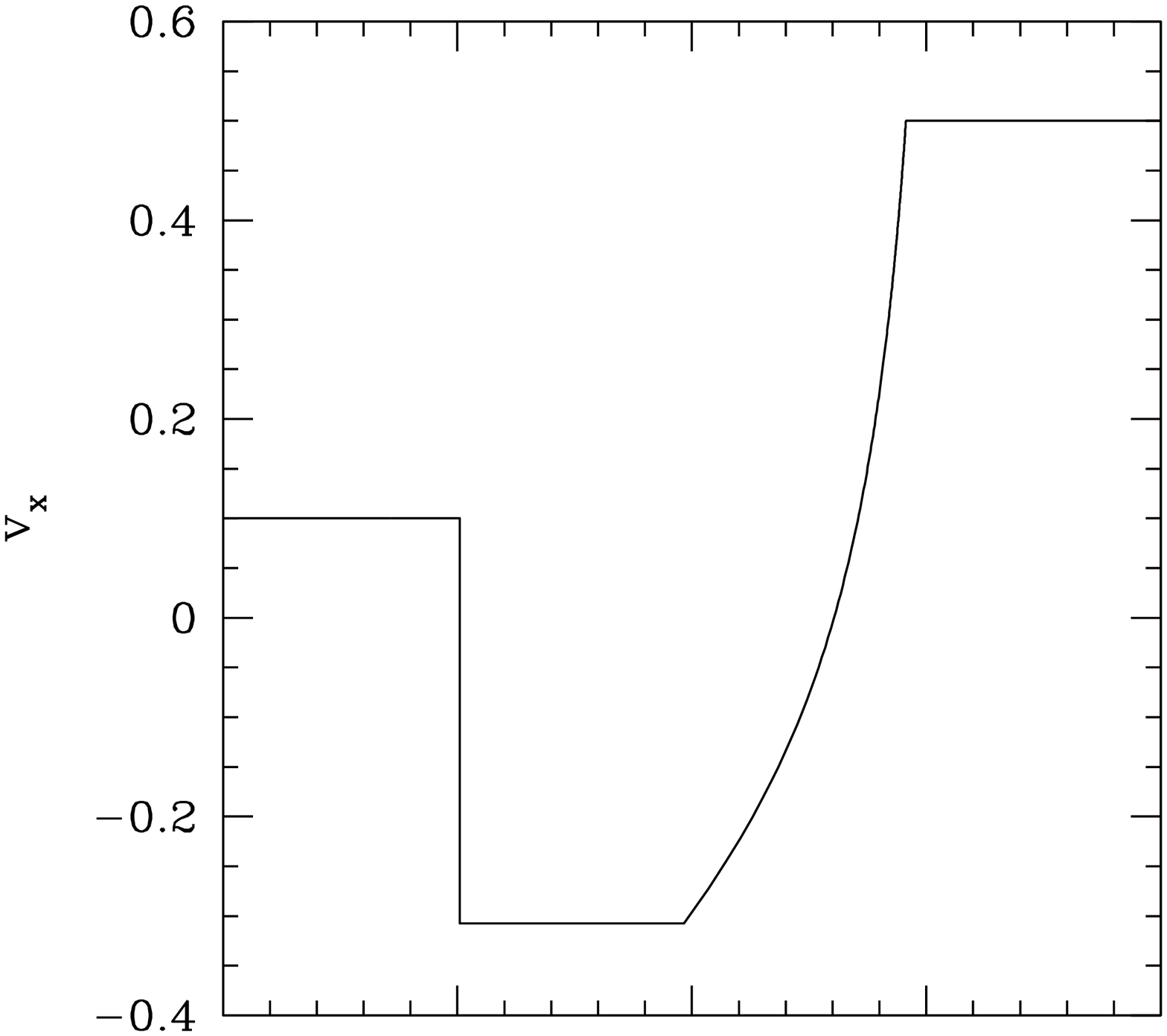}
      \vskip -1.0cm 
      \includegraphics[width=0.45\textwidth]{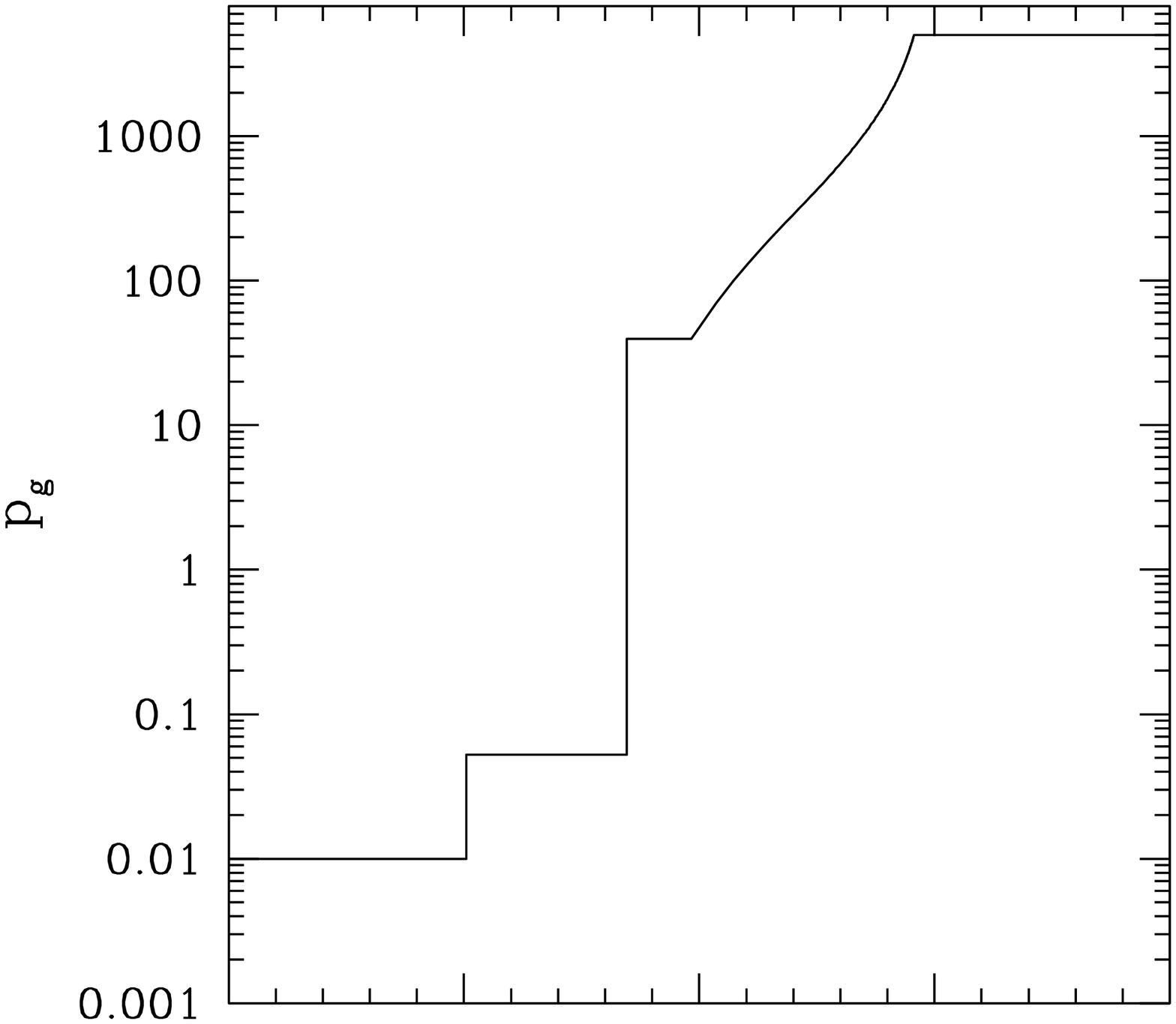}
      \hskip 0.75cm 
      \includegraphics[width=0.45\textwidth]{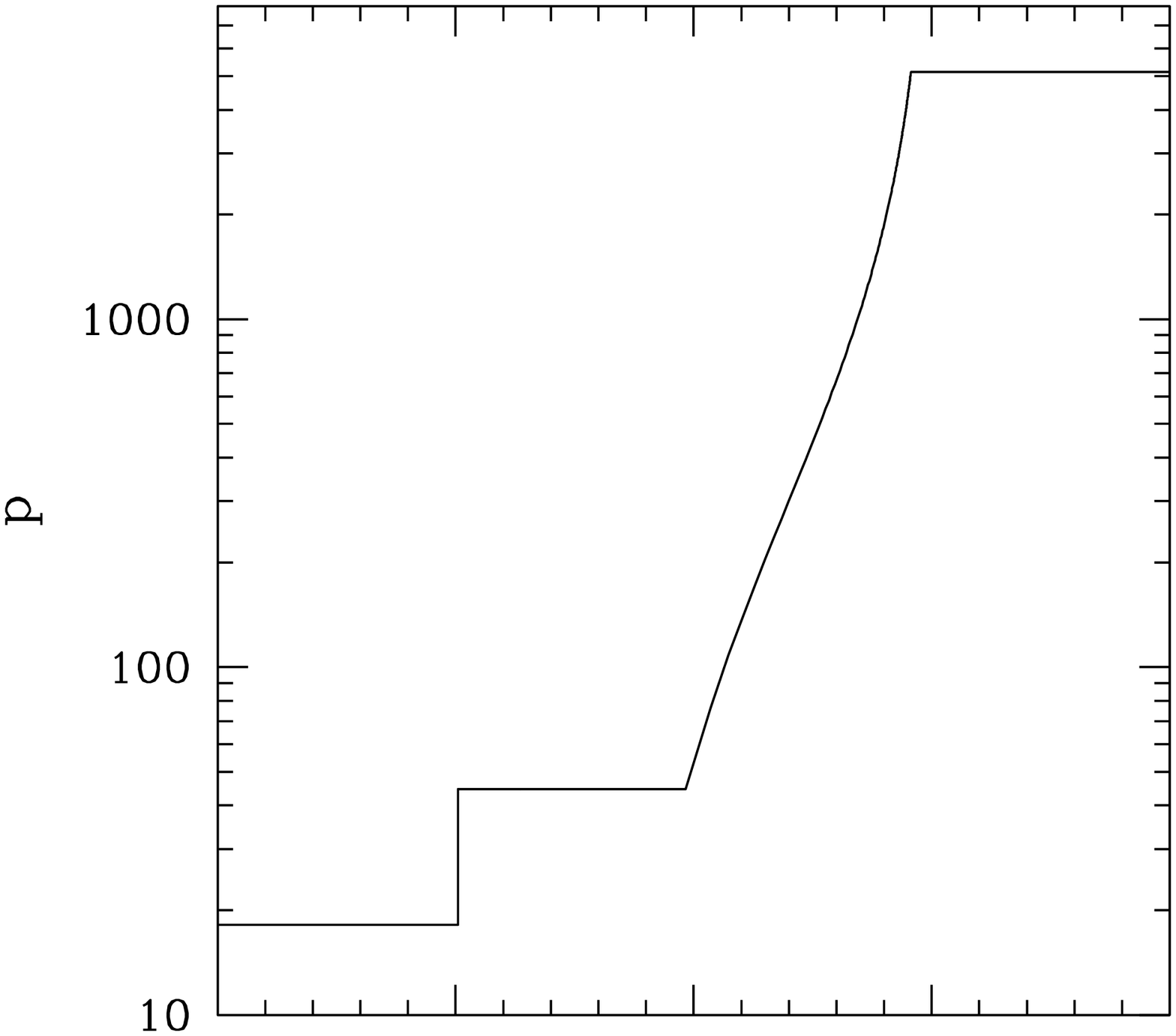}
      \vskip -1.0cm 
      \includegraphics[width=0.45\textwidth]{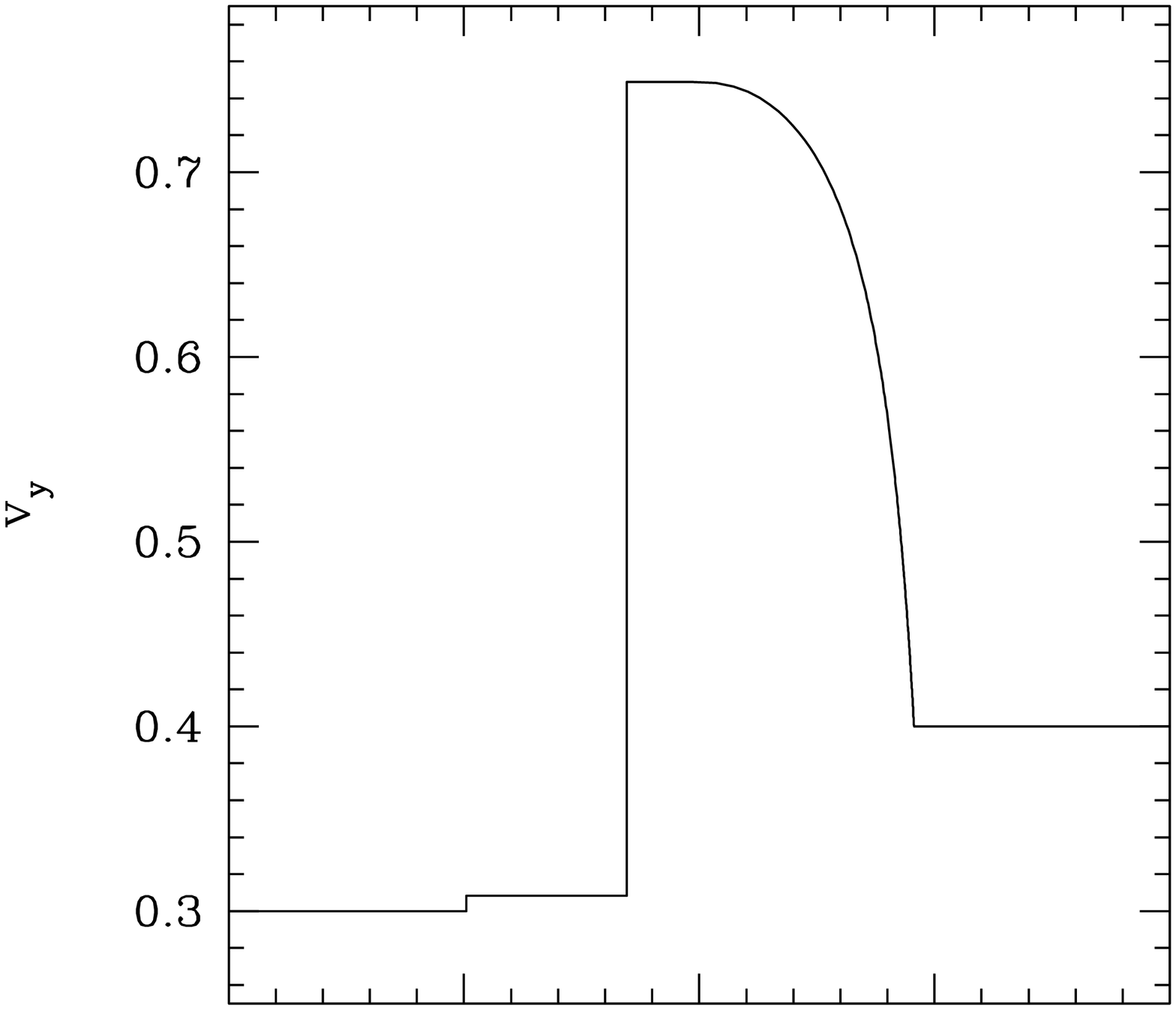}
      \hskip 0.75cm 
      \includegraphics[width=0.45\textwidth]{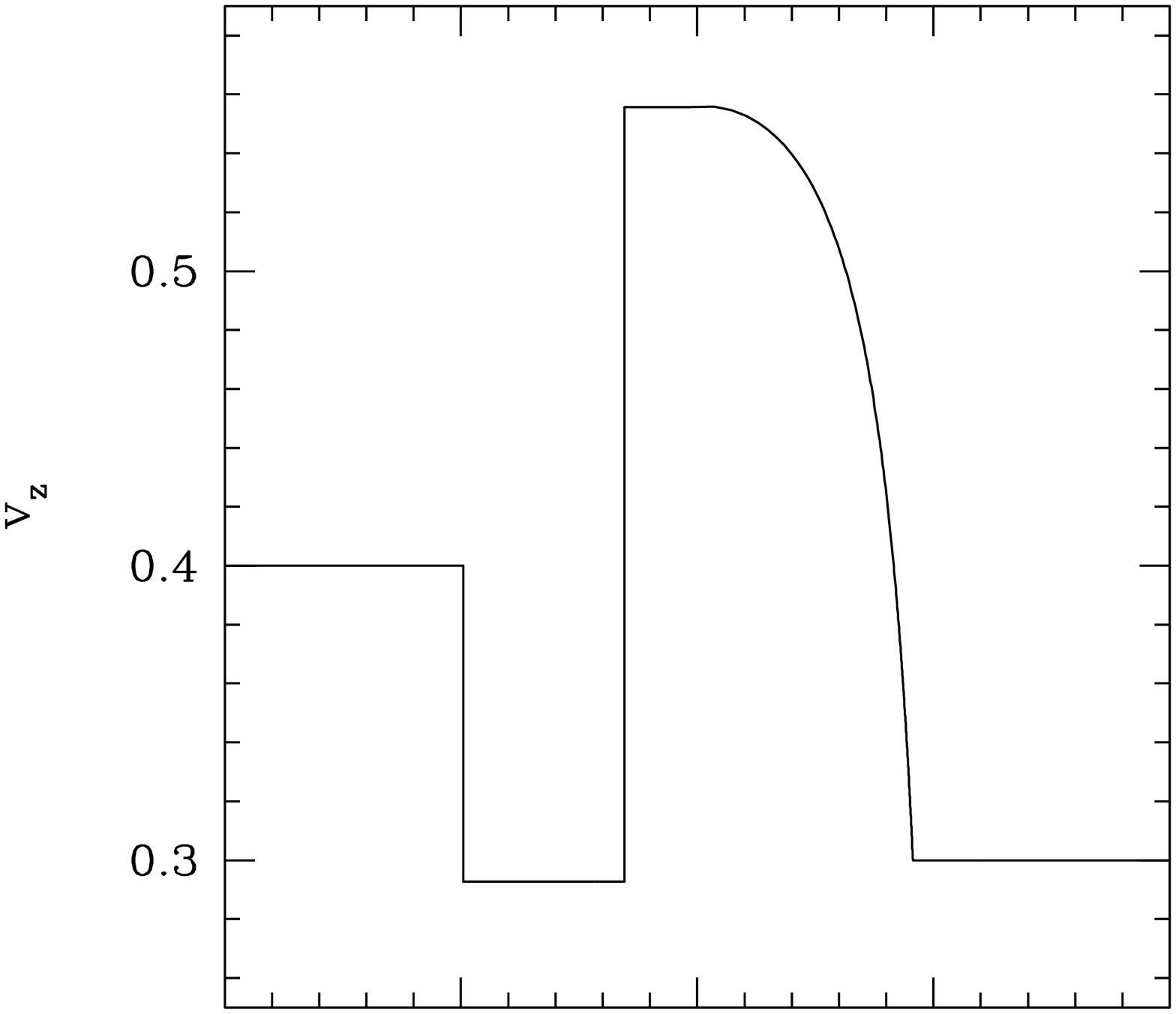}
      \vskip -1.0cm 
      \includegraphics[width=0.45\textwidth]{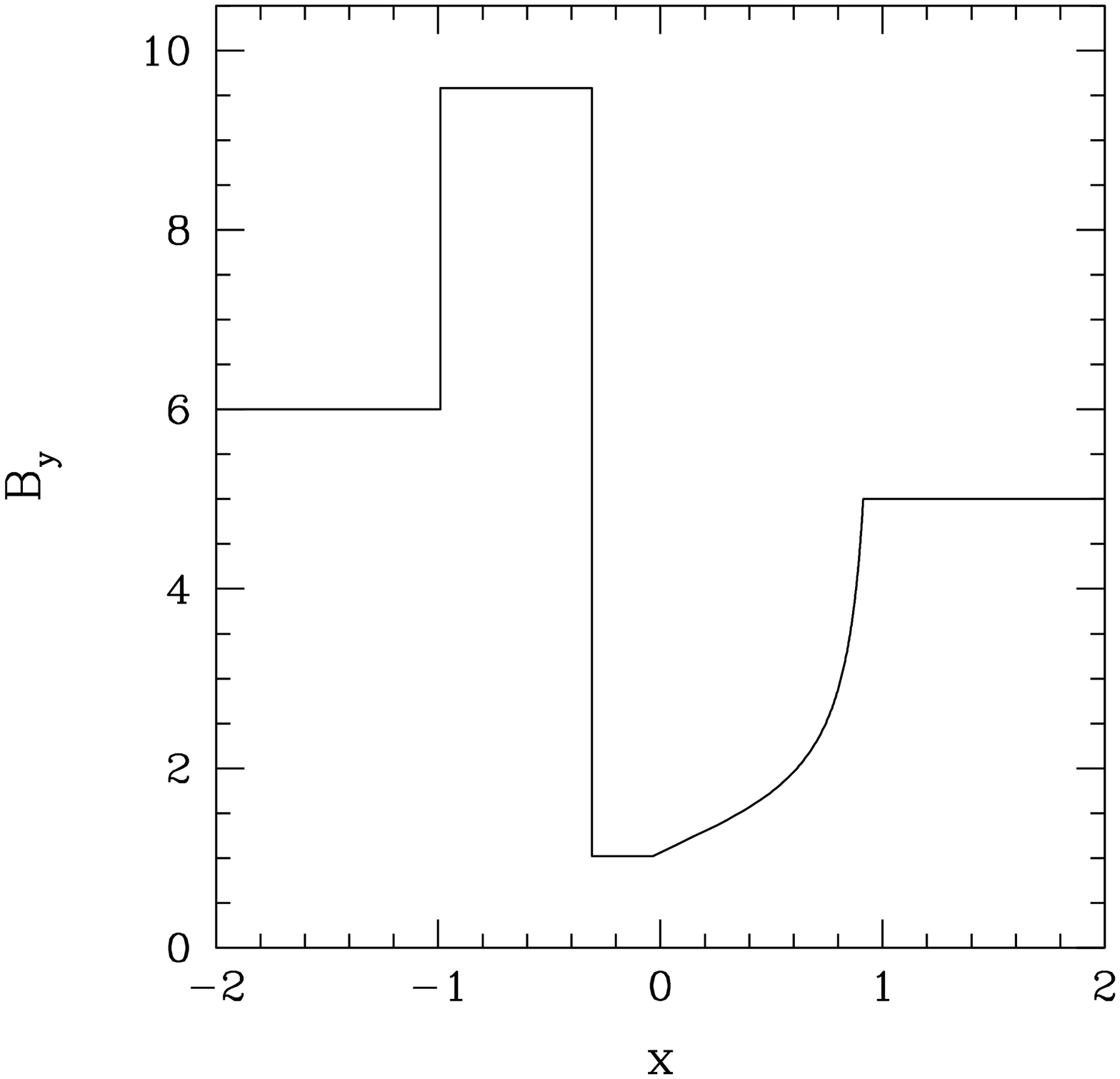}
      \hskip 0.75cm 
      \includegraphics[width=0.45\textwidth]{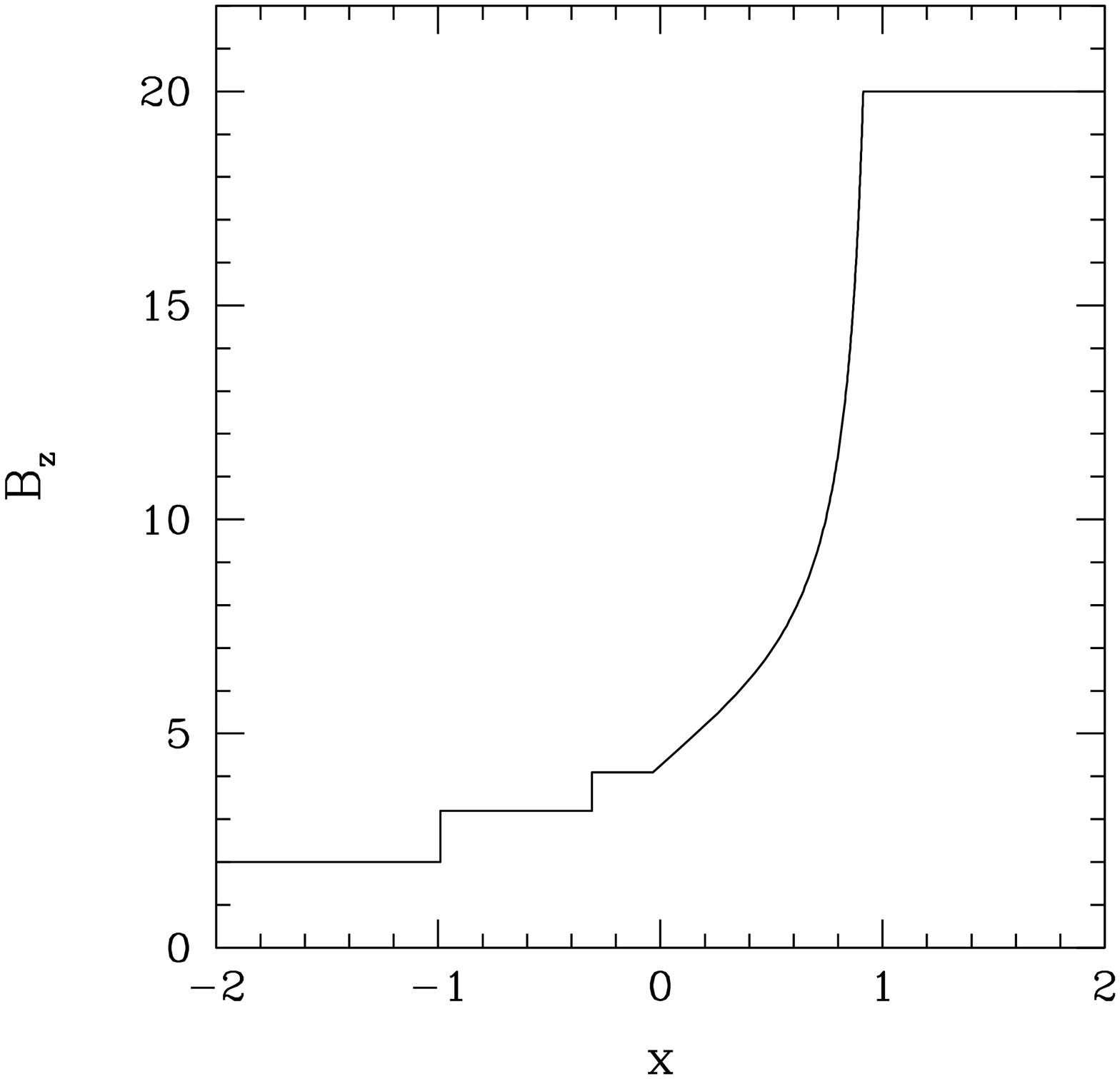}
  \caption{\label{BG1}Exact solution of the generic shock-tube test at
  time $t=1.0$. The solution is composed of a left-going shock, a
  tangential discontinuity and a right-going rarefaction wave}
  \end{center}
\end{figure}

	Because the procedure for calculating the solution in this case
is particularly simple and well tested from relativistic hydrodynamics,
the algorithm employed has shown to be very robust and no failures were
encountered in the calculation of any quantity. We list in Table
\ref{tab:testsfornullBx} the set of initial conditions used in the tests
solved, while we report in Tables~\ref{tab:kost2} and \ref{tab:bgt1} the
first significant digits for the exact solution of the same tests,
reporting in all cases the accuracy obtained (which usually is $\lesssim
10^{-11}$).  Finally, the full solutions in space of the various Riemann
problems listed in Table \ref{tab:testsfornullBx} and for the quantities
$\rho$, $v^x$, $p_{\rm g}$, $p$, $v^y$, $v^z$, $B^y$, and $B^z$ are shown
in Figures~\ref{KOST2} and \ref{BG1} at the indicated representative
times.

%

\subsection{Generic Initial Magnetic field: $B^x\neq 0$}
\label{num_bxneq0}

	As discussed in Section~\ref{strategy}, when $B^x \ne 0$ the
Riemann problem consists of seven different waves: two fast-waves, two
slow-waves, two Alfv\`en discontinuities and a central contact
discontinuity across which only the density can be discontinuous ({\it
cf.} Figure~\ref{mhd_zones}). In essence, the numerical solution of the
Riemann problem when $B^x \ne 0$ proceeds as follows: starting from the
initial left and right states ({\it i.e.} regions R1 and R8 of
Figure~\ref{mhd_zones}), we compute the states after the fast-waves
(regions R2 and R7), then we determine the jumps across the Alfv\`en
discontinuities (regions R3 and R6) and finally we solve the equations
for the slow-waves (regions R4 and R5). As a result of this sequence, the
jump conditions in all the physical variables in the two states across
the contact discontinuity are computed and if the solution obtained in
this way does not reach the desired accuracy, the procedure is iterated.

	We also recall that when $B^x \ne 0$, the numerical solution
is found using a hybrid method which adopts different sets of
equations according to the region in which the Riemann problem has to
be solved. In particular, to compute the states after the fast-waves
and across the Alfv\`en discontinuities we use as unknown the total
pressure ($p$-method; Section~\ref{pmethod}) and to discriminate
between shocks and rarefaction waves we evaluate the jump in the total
pressure in a way similar to the case when $B^x=0$. To compute the
states after the slow-waves, on the other hand, we use the tangential
components of the magnetic field ($B_t$-method;
Section~\ref{btmethod}) and to decide whether a wave is a shock or a
rarefaction we evaluate the jump in the norm of the magnetic field
bearing in mind that it must decrease across slow shocks and increase
otherwise. Then at the contact discontinuity we compute the jumps in
the total pressure and in the components of three-velocity and if they
are above a certain accuracy we iterate by changing the values of the
total pressure, used in regions R2-R3 and R6-R7, and of the tangential
components of the magnetic field, used in regions R4-R5.

	It is worth underlining that the solution of the Riemann problem
with generic initial states is considerably more demanding than when
$B^x=0$ and not only because of the more numerous waves present. Indeed,
the most severe difficulty is due to the fact that the set of equations
to be solved becomes particularly stiff near the solution. A careful
investigation of the several cases considered has in fact revealed that,
in general, the functional behavior of the quantities whose roots are
sought, changes very rapidly near the roots, stretching the ability of
standards root-finding algorithms. As a result, it is not uncommon that
the solution cannot be found if the iteration for the search of the root
starts from a guess which is not sufficiently close to the exact
solution. To avoid such failures and to provide a first guess which is
reasonably accurate, we have used as a guide the solution provided by the
HLLE approximate Riemann solver\footnote{Note that this is not necessary
when $B^x=0$ since in this case the solution can also be quite far from
the exact one and yet the iterative scheme does not show problems in
converging to it.}. In practice, the approximate solution should be
accurate to a few percent in the regions away from the waves, where the
states are almost constant (very close to the waves the errors are of
course much larger). Using this guess has proven to be sufficient to
obtain a solution in all of the cases considered, but of course there is
no guarantee that a solution will be straightforwardly found for all of
the possible initial states. Our experience when the solution could not
be immediately obtained is that an increase in the accuracy of the
approximate Riemann solver is in general sufficient to yield a convergent
and accurate solution.

\begin{table}
\begin{center}
\begin{tabular}{|r|c|c|c|c|c|c|c|c|}
\cline{1-9} & & & & & & & & \\
{\bf Test type}~~~~~~~&  $\rho$ & $p_{\rm g}$ & $v^x$  & $v^y$  & $v^z$ 
&  $B^x$ & $B^y$  & $B^z$ \\
& & & & & & & & \\ \cline{1-9} & & & & & & & & \\
{\bf Komissarov: Shock-Tube Test 1}~~~($\Gamma=4/3$) & & & & & & & & \\
{\it left state}  ~~~& 1.0 & 1000.0 & 0.0 & 0.0 & 0.0 & 1.0 & 0.0  & 0.0 \\
{\it right state} ~~~& 0.1 & 1.0    & 0.0 & 0.0 & 0.0 & 1.0 & 0.0  & 0.0 \\
\cline{1-9} & & & & & & & & \\
{\bf Komissarov: Collision Test}~~~($\Gamma=4/3$) & & & & & & & & \\
{\it left state}  ~~~& 1.0 & 1.0  & $5/\sqrt{26}$  & 0.0 & 0.0 & 10.0 & 10.0   & 0.0 \\
{\it right state} ~~~& 1.0 & 1.0  & $-5/\sqrt{26}$ & 0.0 & 0.0 & 10.0 & -10.0  & 0.0 \\
\cline{1-9} & & & & & & & & \\
{\bf Balsara Test 1 (Brio \& Wu)}~~~($\Gamma=2$) & & & & & & & & \\
{\it left state}  ~~~& 1.000 & 1.0  & 0.0 & 0.0 & 0.0 & 0.5 & 1.0   & 0.0 \\
{\it right state} ~~~& 0.125 & 0.1  & 0.0 & 0.0 & 0.0 & 0.5 & -1.0  & 0.0 \\
\cline{1-9} & & & & & & & & \\
{\bf Balsara Test 2}~~~($\Gamma=5/3$) & & & & & & & & \\
{\it left state}  ~~~& 1.0 & 30.0 & 0.0 & 0.0 & 0.0 & 5.0 & 6.0  & 6.0 \\
{\it right state} ~~~& 1.0 & 1.0  & 0.0 & 0.0 & 0.0 & 5.0 & 0.7  & 0.7 \\
\cline{1-9} & & & & & & & & \\
{\bf Balsara Test 3} ~~~($\Gamma=5/3$) & & & & & & & & \\
{\it left state}  ~~~& 1.0 & 1000.0 & 0.0 & 0.0 & 0.0 & 10.0 & 7.0  & 7.0 \\
{\it right state} ~~~& 1.0 & 0.1    & 0.0 & 0.0 & 0.0 & 10.0 & 0.7  & 0.7 \\
\cline{1-9} & & & & & & & & \\
{\bf Balsara Test 4} ~~~($\Gamma=5/3$) & & & & & & & & \\
{\it left state}  ~~~& 1.0 & 0.1  & 0.999  & 0.0 & 0.0 & 10.0 & 7.0   & 7.0 \\
{\it right state} ~~~& 1.0 & 0.1  & -0.999 & 0.0 & 0.0 & 10.0 & -7.0  & -7.0 \\
\cline{1-9} & & & & & & & & \\
{\bf Balsara Test 5} ~~~($\Gamma=5/3$) & & & & & & & & \\
{\it left state}  ~~~& 1.08 & 0.95 & 0.40  & 0.3  & 0.2 & 2.0 & 0.3   & 0.3 \\
{\it right state} ~~~& 1.00 & 1.0  & -0.45 & -0.2 & 0.2 & 2.0 & -0.7  & 0.5 \\
\cline{1-9} & & & & & & & & \\
{\bf Generic Alfv\`en Test} ~~~($\Gamma=5/3$) & & & & & & & & \\
{\it left state}  ~~~& 1.0  & 5.0  & 0.0 & 0.3  & 0.4 & 1.0 & 6.0 & 2.0 \\
{\it right state} ~~~& 0.9  & 5.3  & 0.0 & 0.0  & 0.0 & 1.0 & 5.0 & 2.0 \\
\cline{1-9} 
\end{tabular}
\end{center}
\smallskip
\caption{Initial conditions for the tests of the exact Riemann solver when
the magnetic field has nonzero normal component, {\it i.e.} $B^x \neq 0$.}
\label{tab:tests}
\end{table}

\subsubsection{Representative Tests for $B^x\neq 0$}
\label{num_bxneq0_tests}

\begin{figure}
\begin{center}
      \includegraphics[width=0.45\textwidth]{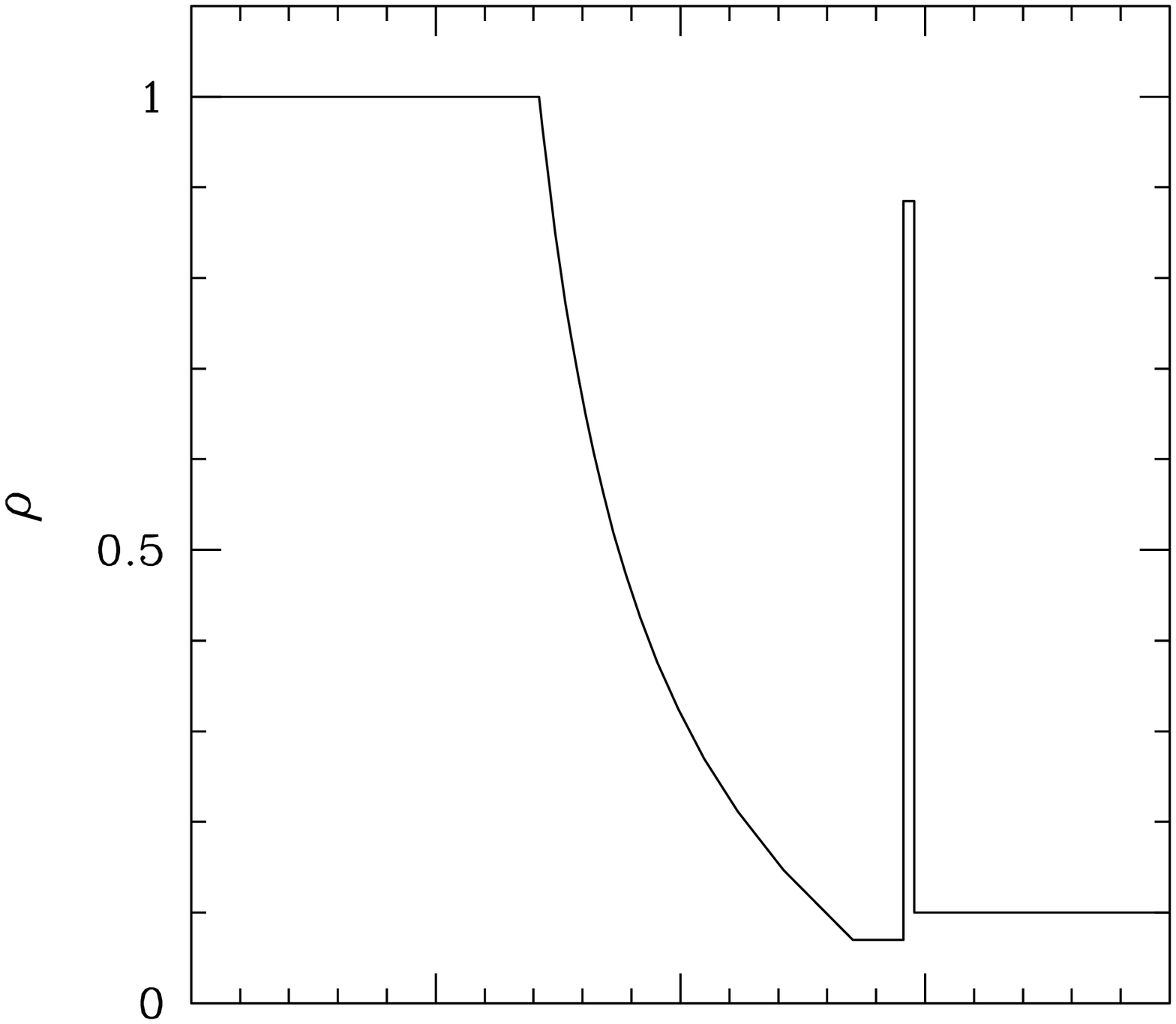}
      \hskip 0.75cm 
      \includegraphics[width=0.45\textwidth]{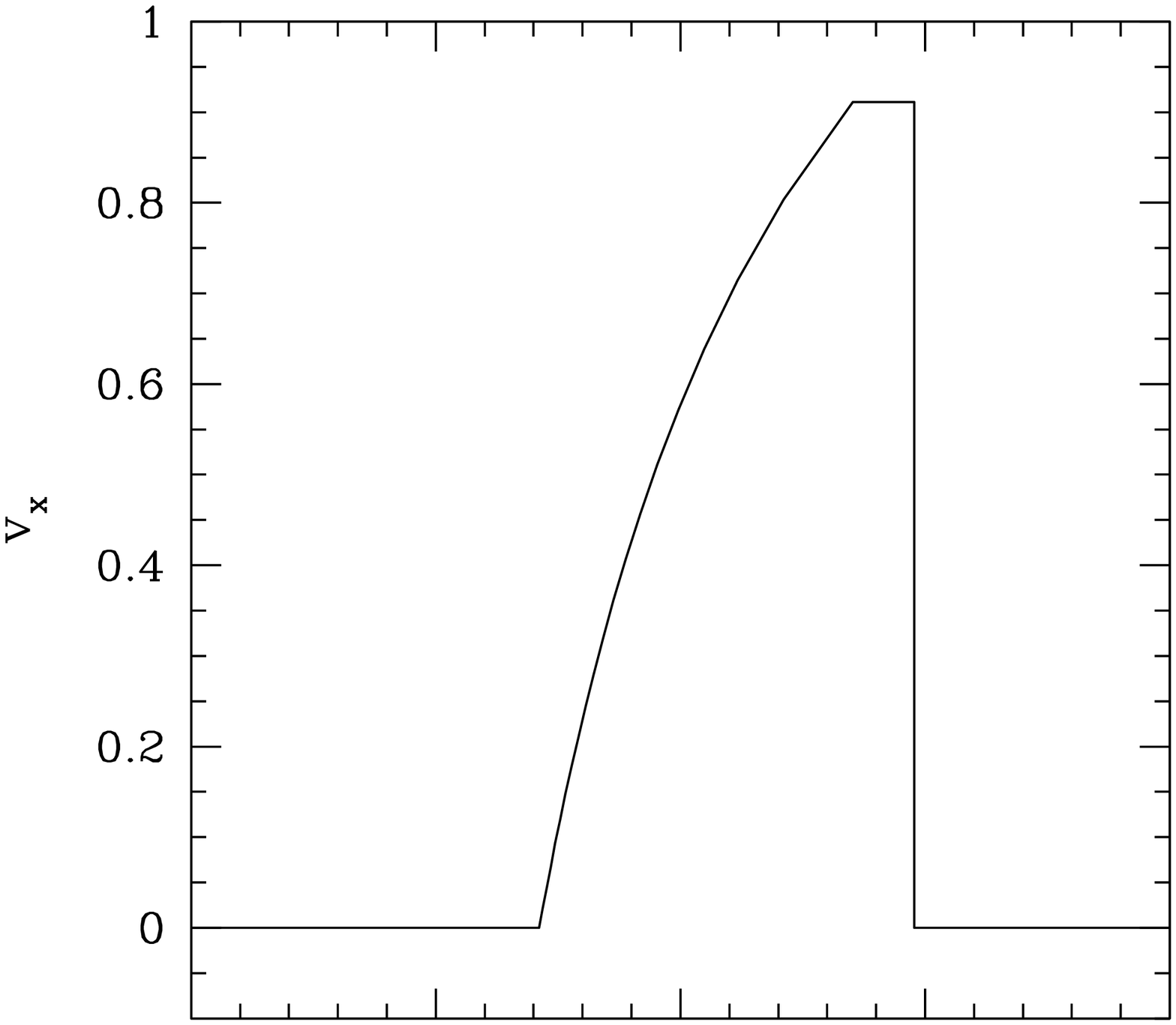}
      \vskip -1.cm 
      \includegraphics[width=0.45\textwidth]{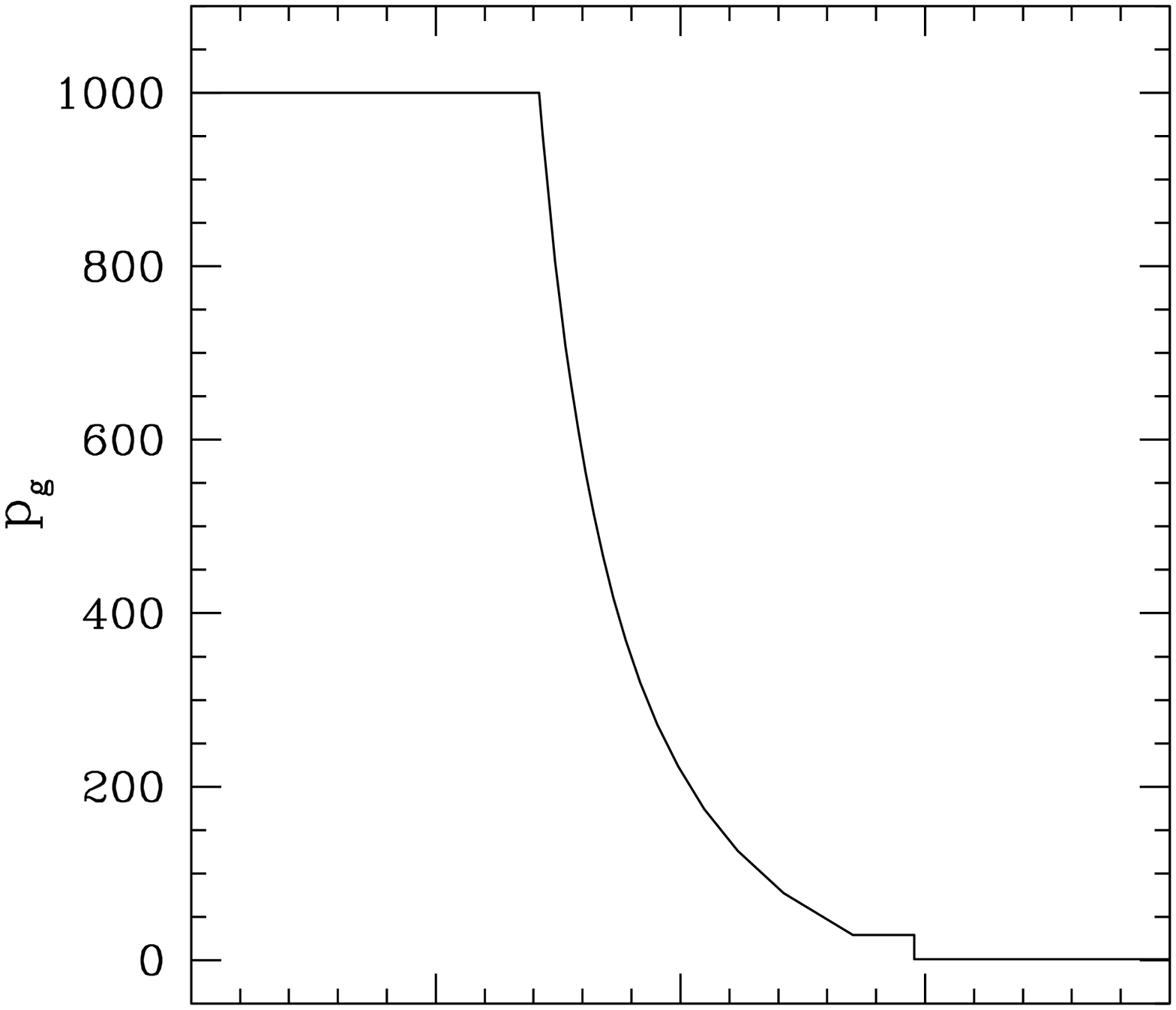}
      \hskip 0.75cm 
      \includegraphics[width=0.45\textwidth]{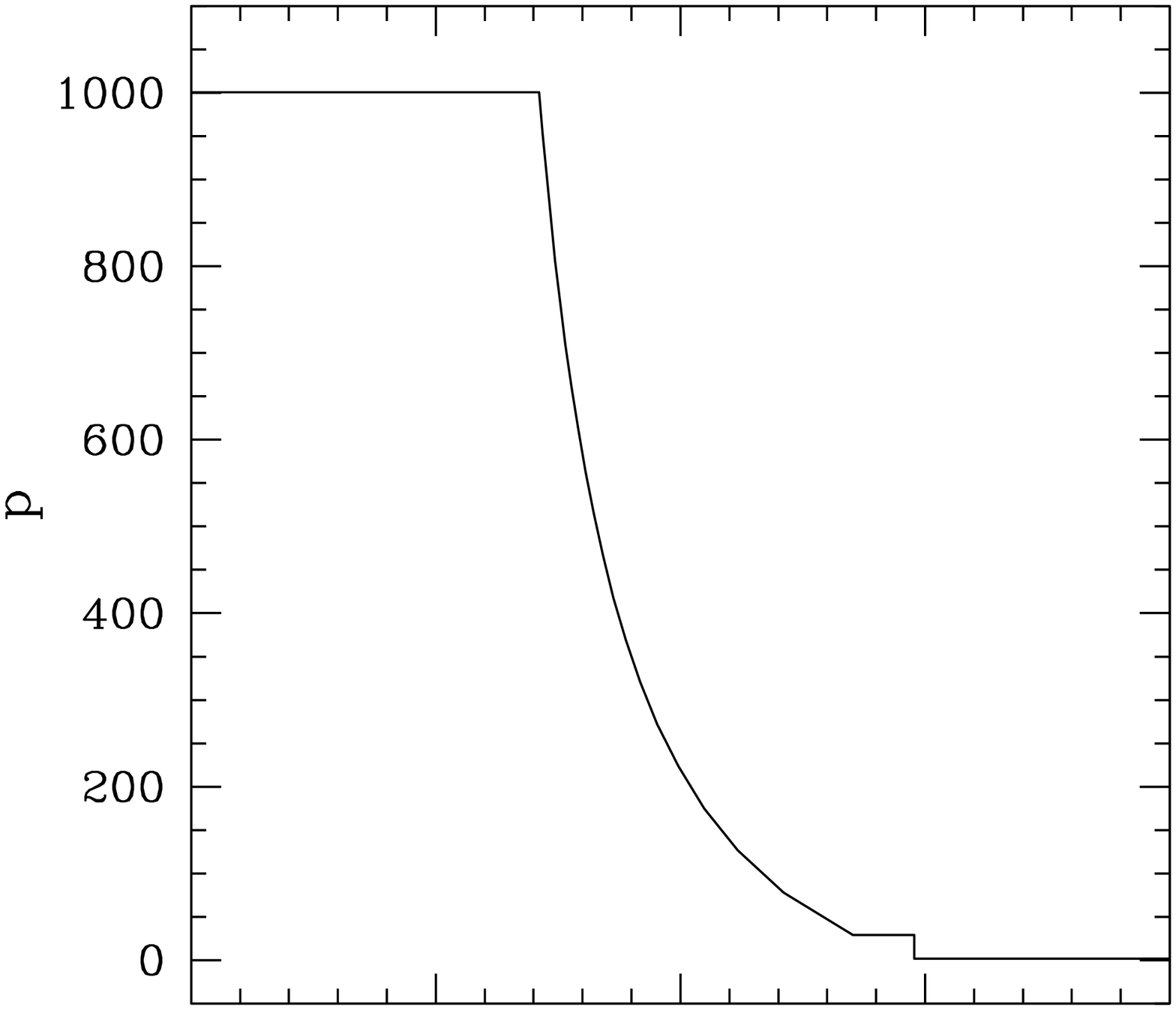}
      \vskip -1.cm 
      \includegraphics[width=0.45\textwidth]{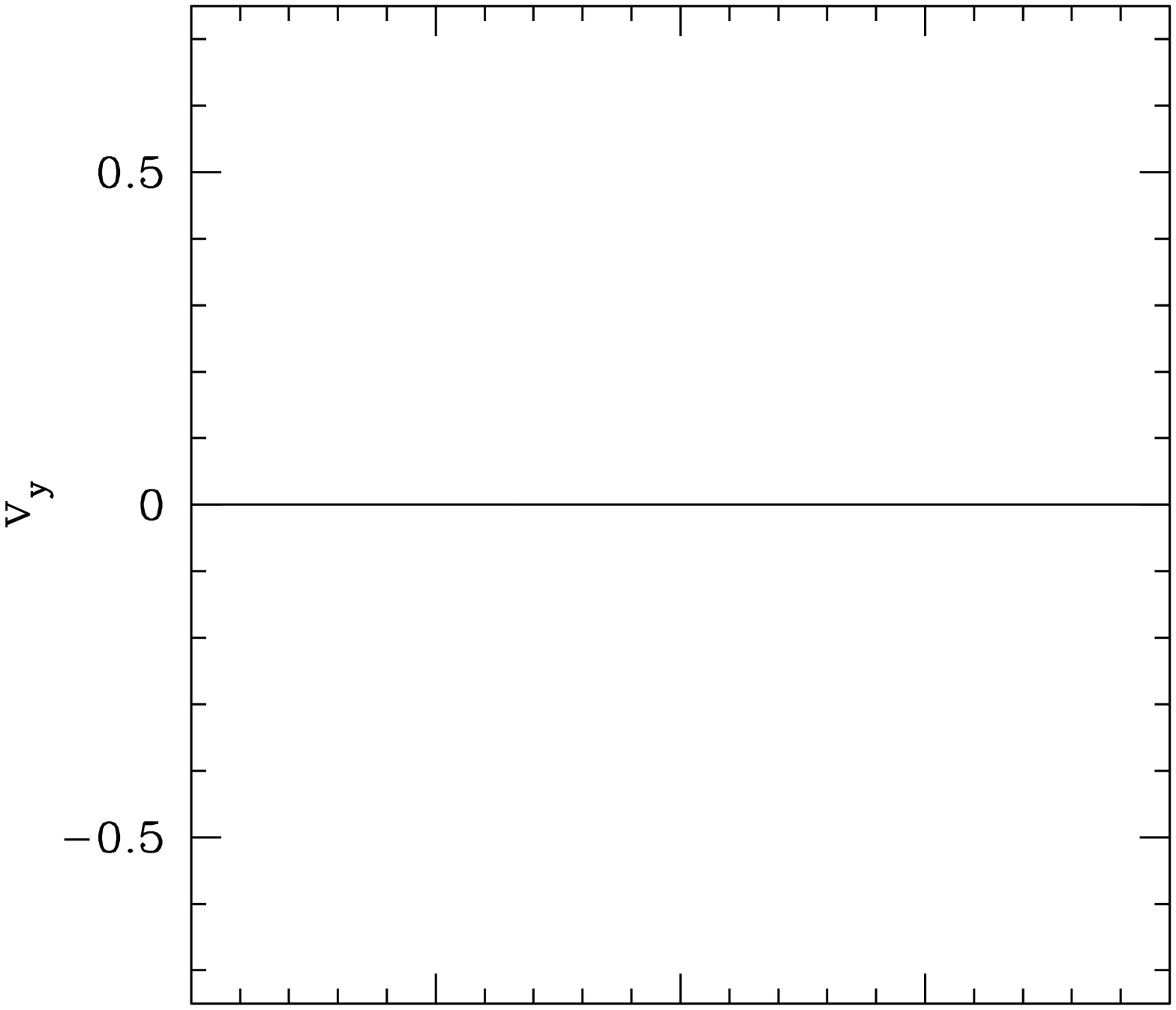}
      \hskip 0.75cm 
      \includegraphics[width=0.45\textwidth]{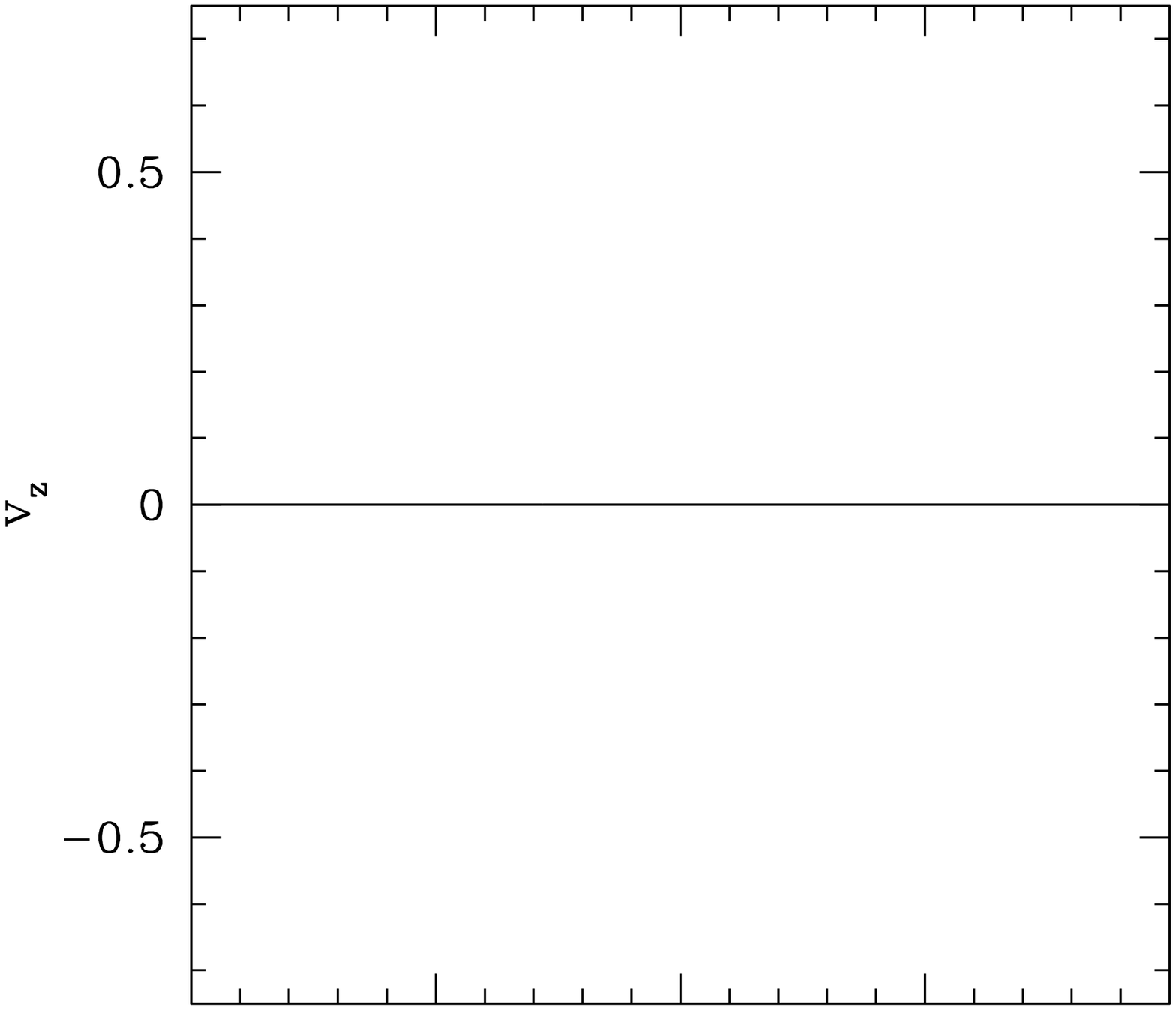}
      \vskip -1.cm 
      \includegraphics[width=0.45\textwidth]{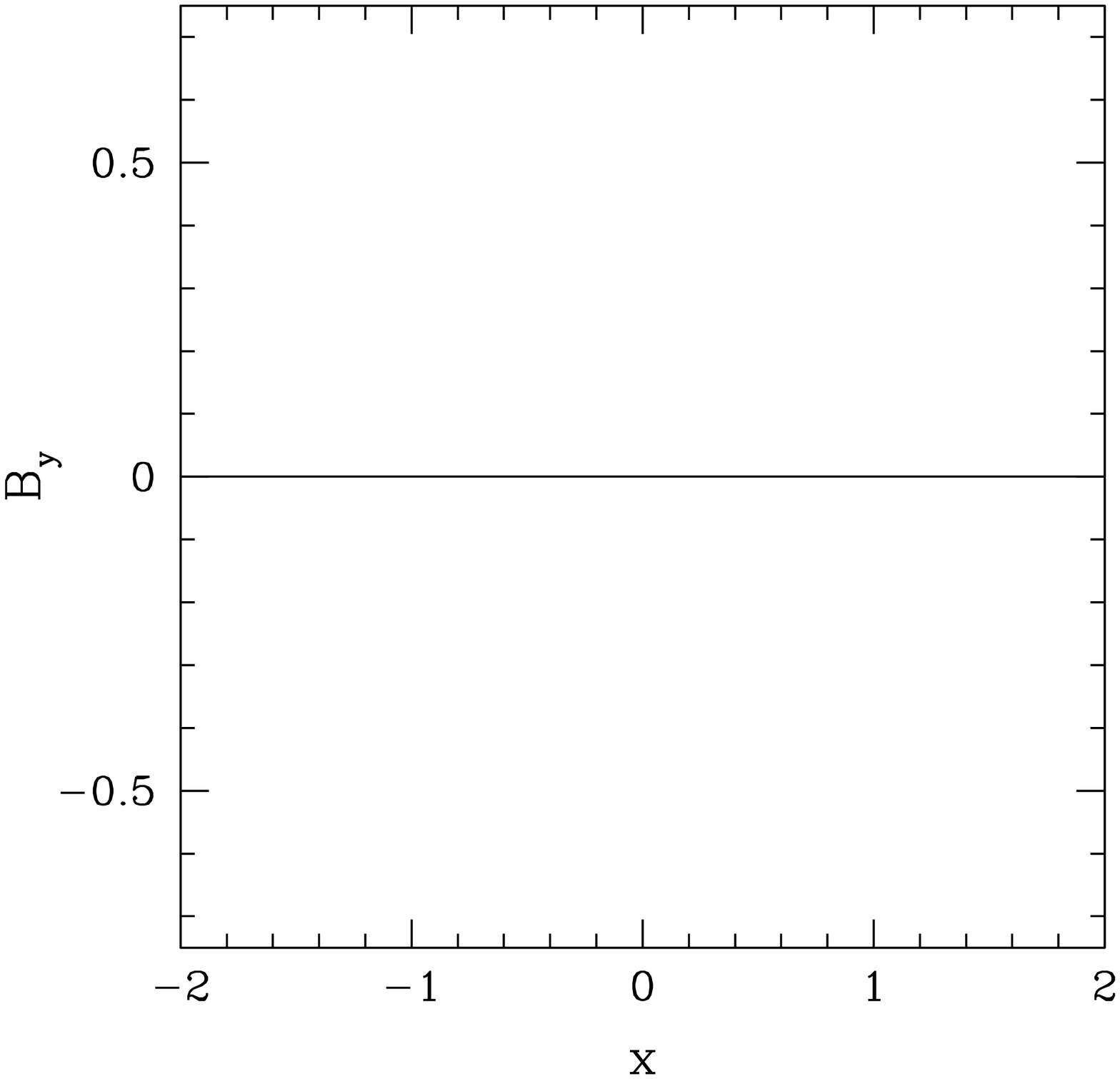}
      \hskip 0.75cm 
      \includegraphics[width=0.45\textwidth]{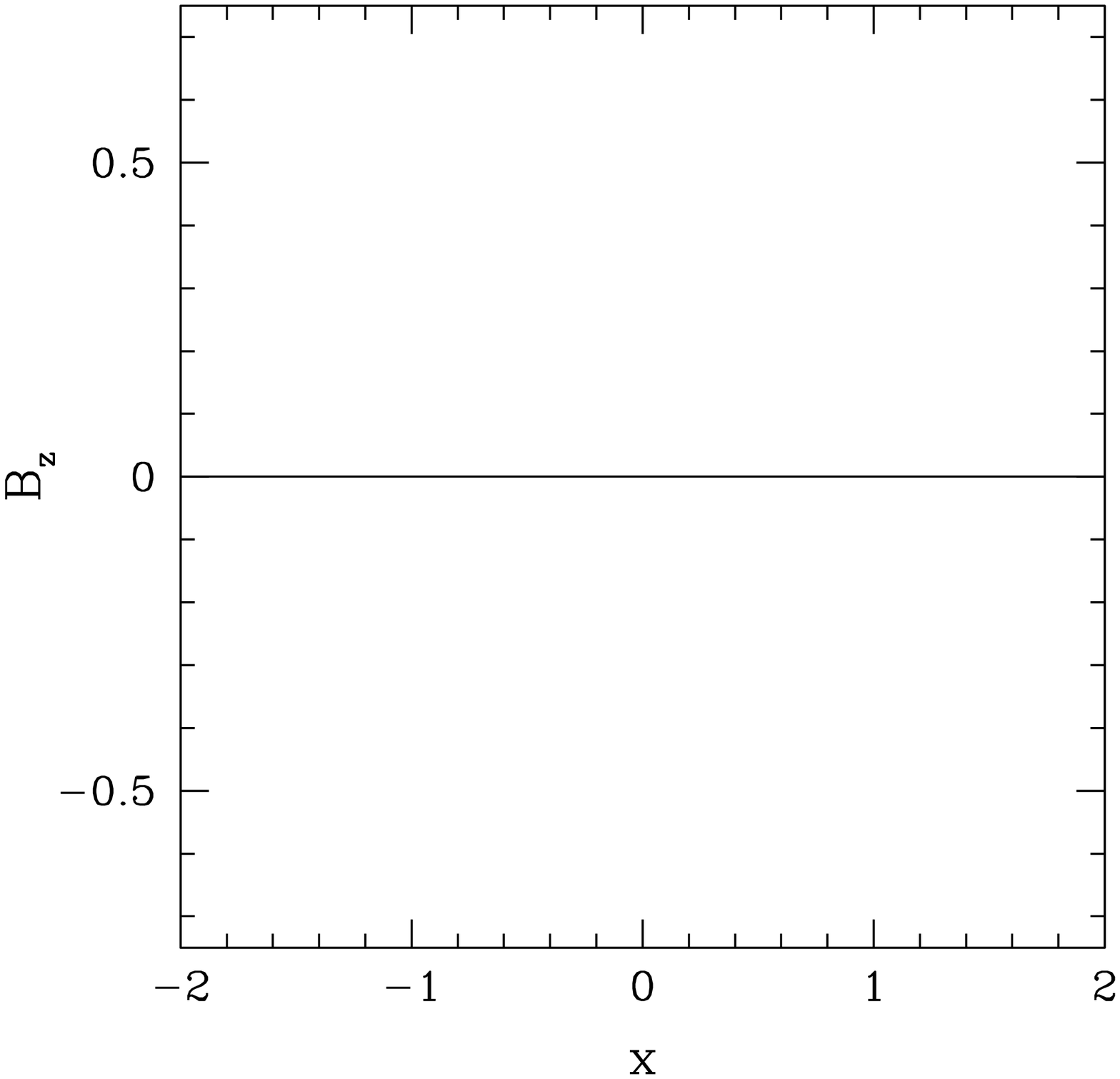}
  \caption{\label{ST1} Exact solution of the test shock-tube 1 of
    Komissarov (1999) at time $t=1.0$. The solution is composed of a
    left-going fast rarefaction, of a contact discontinuity and of a
    right-going fast shock.}
  \end{center}
\end{figure}

\begin{figure}
\begin{center}
      \includegraphics[width=0.45\textwidth]{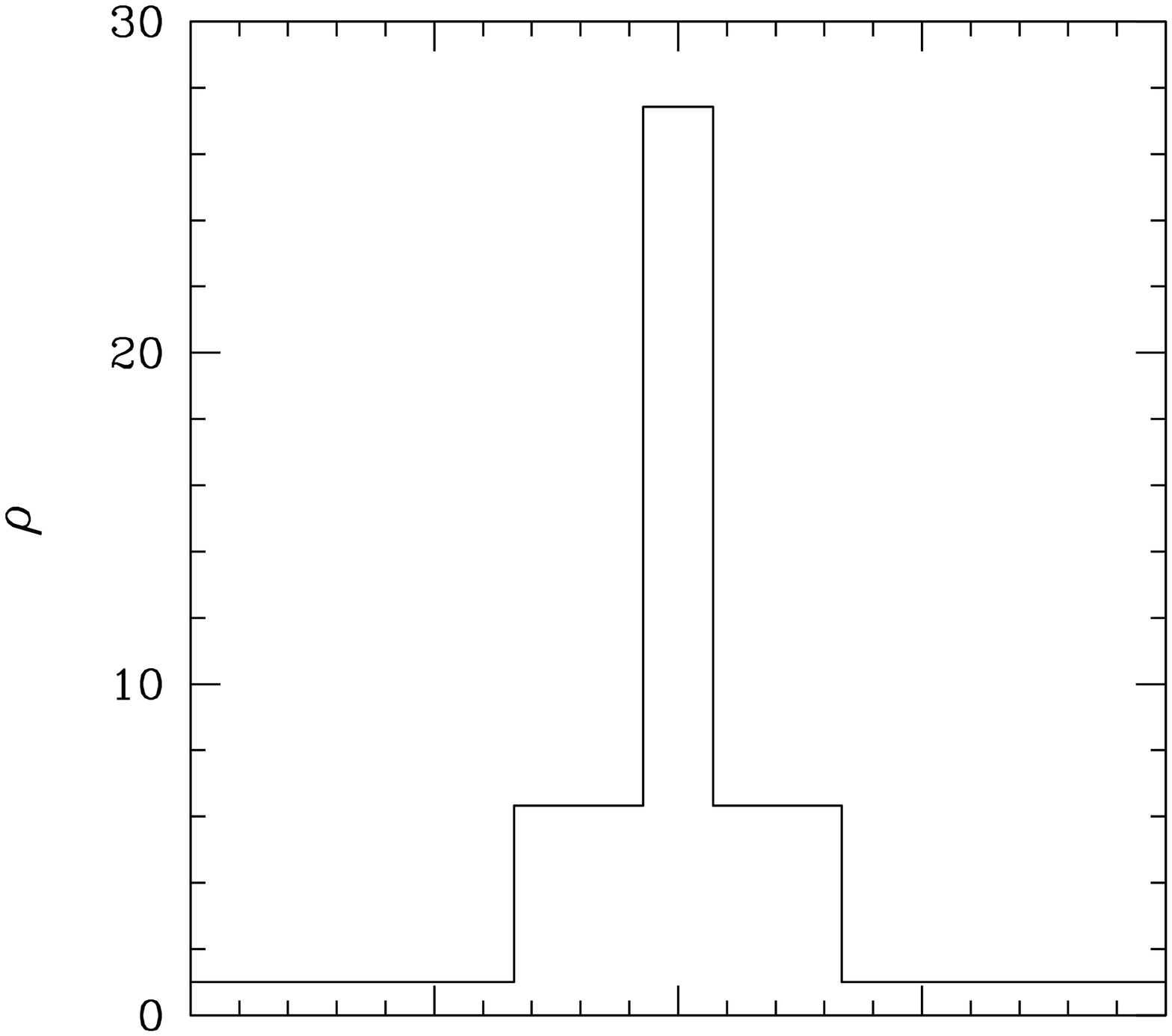}
      \hskip 0.75cm 
      \includegraphics[width=0.45\textwidth]{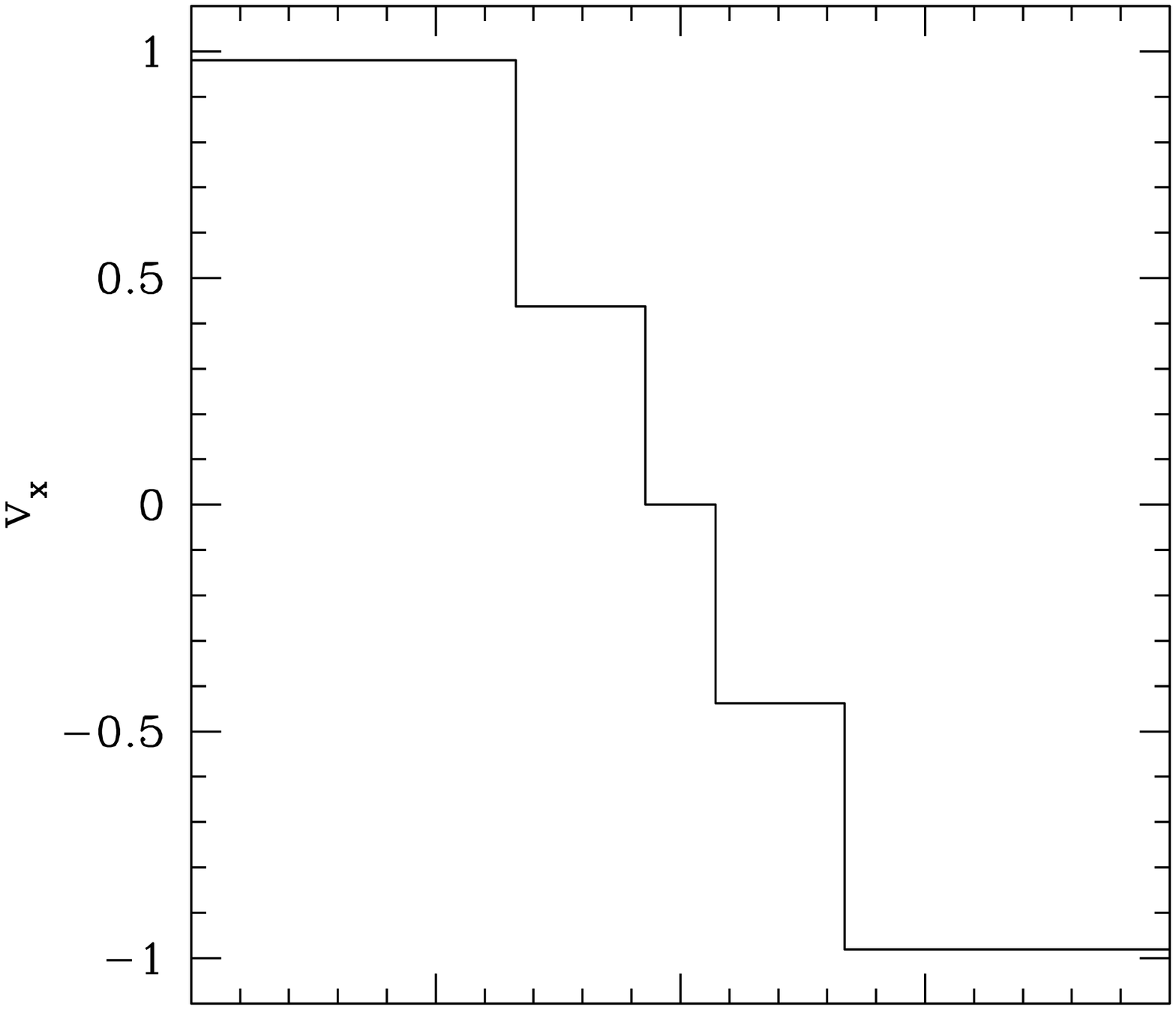}
      \vskip -1.cm 
      \includegraphics[width=0.45\textwidth]{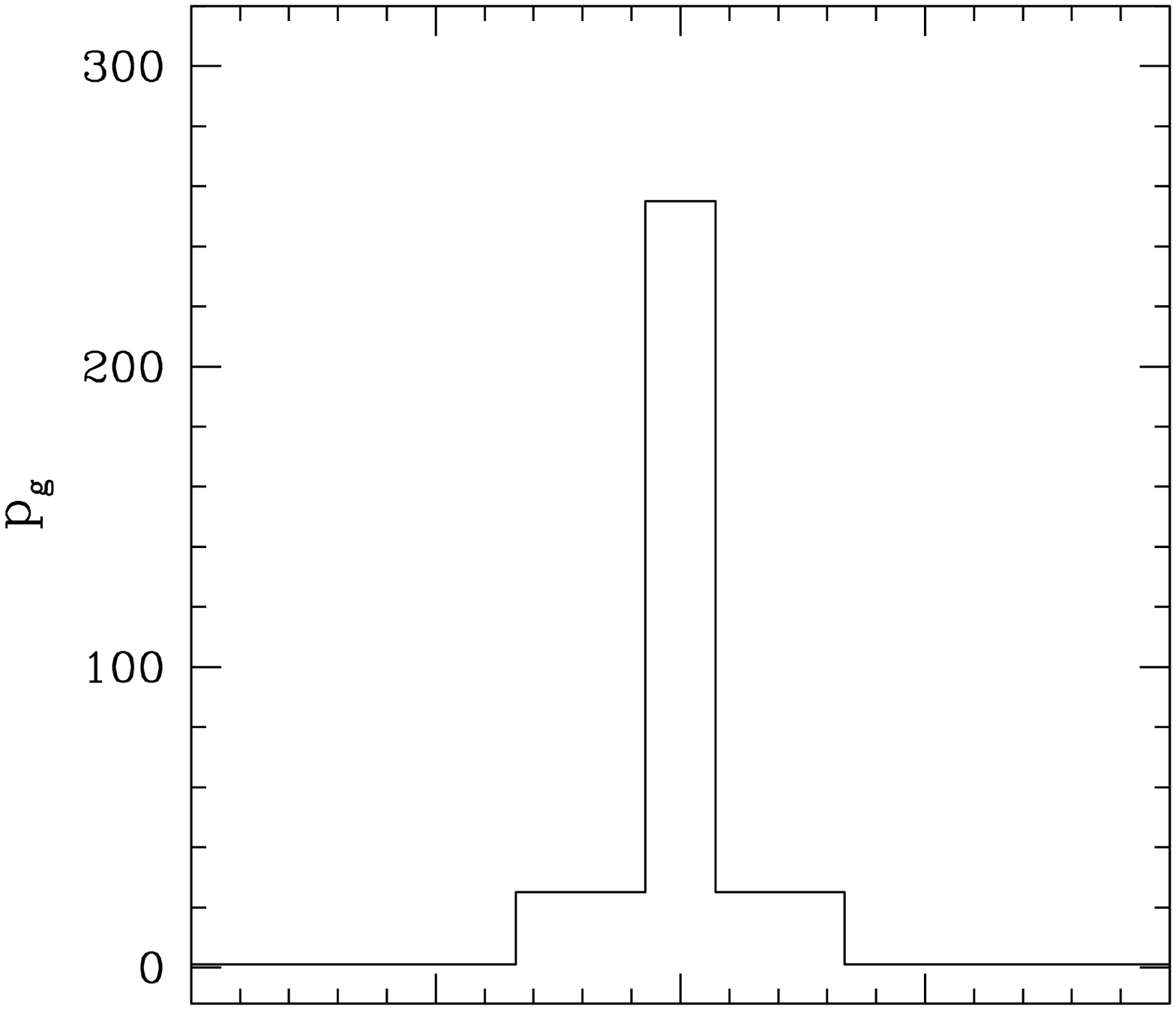}
      \hskip 0.75cm 
      \includegraphics[width=0.45\textwidth]{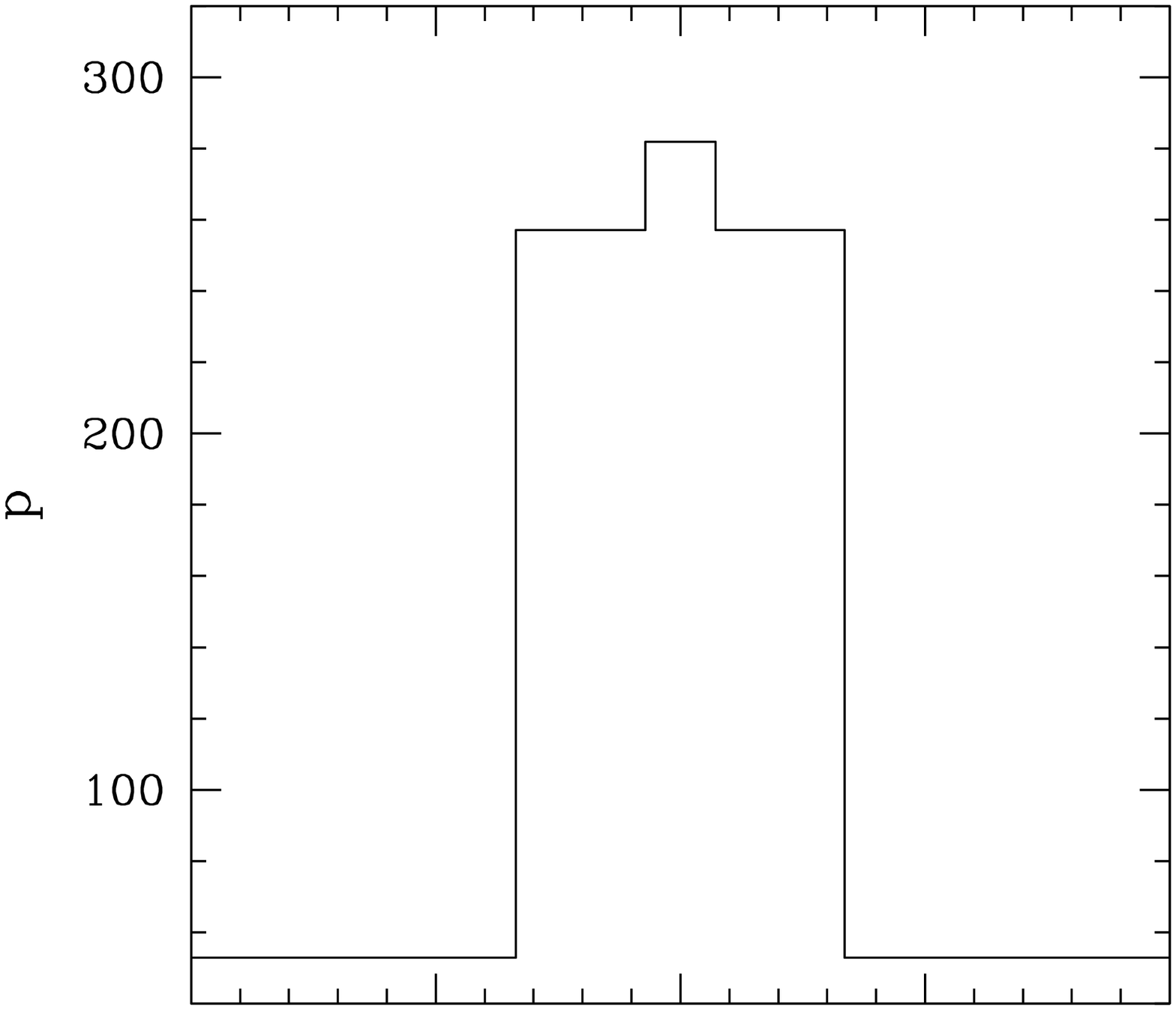}
      \vskip -1.cm 
      \includegraphics[width=0.45\textwidth]{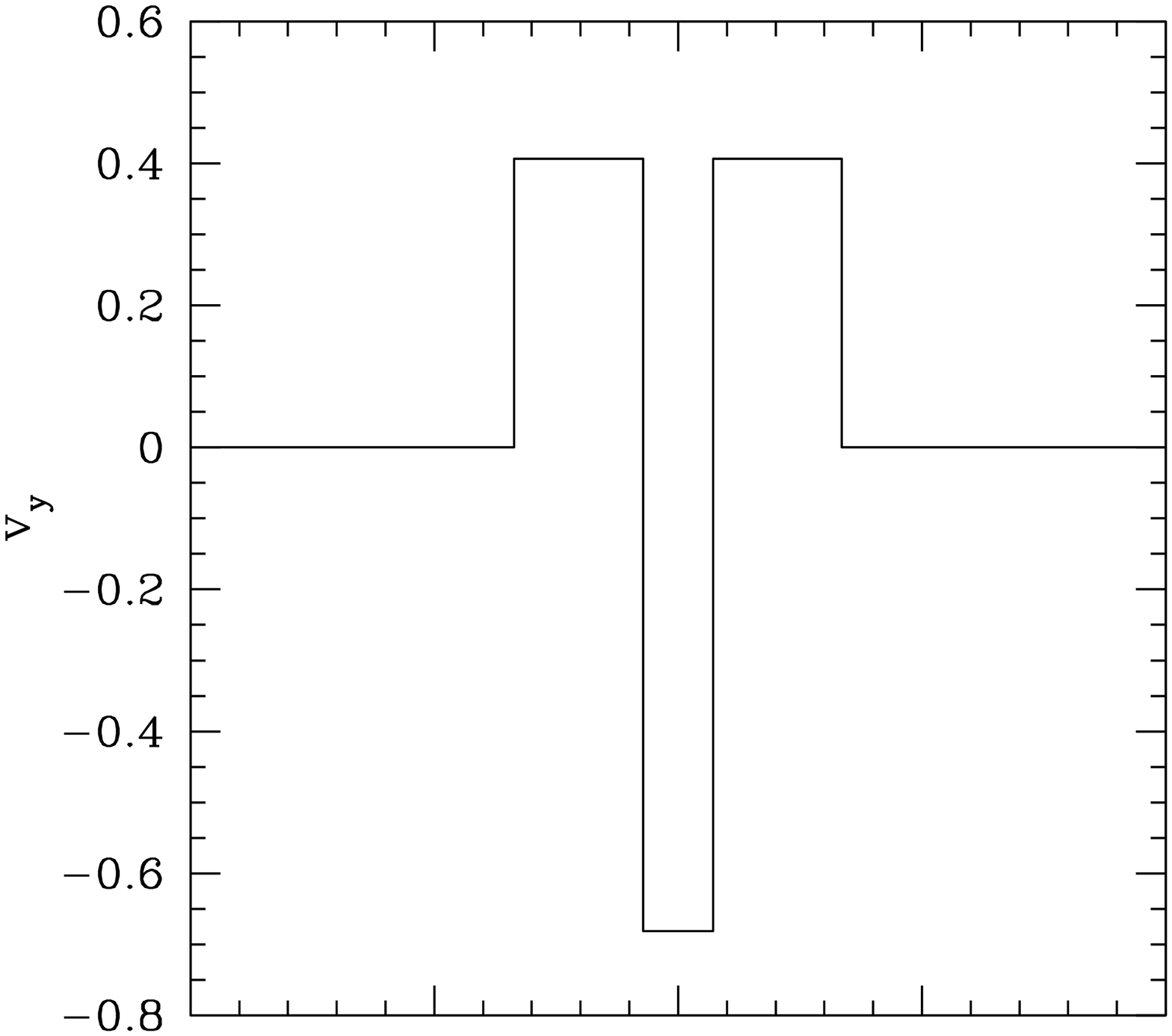}
      \hskip 0.75cm 
      \includegraphics[width=0.45\textwidth]{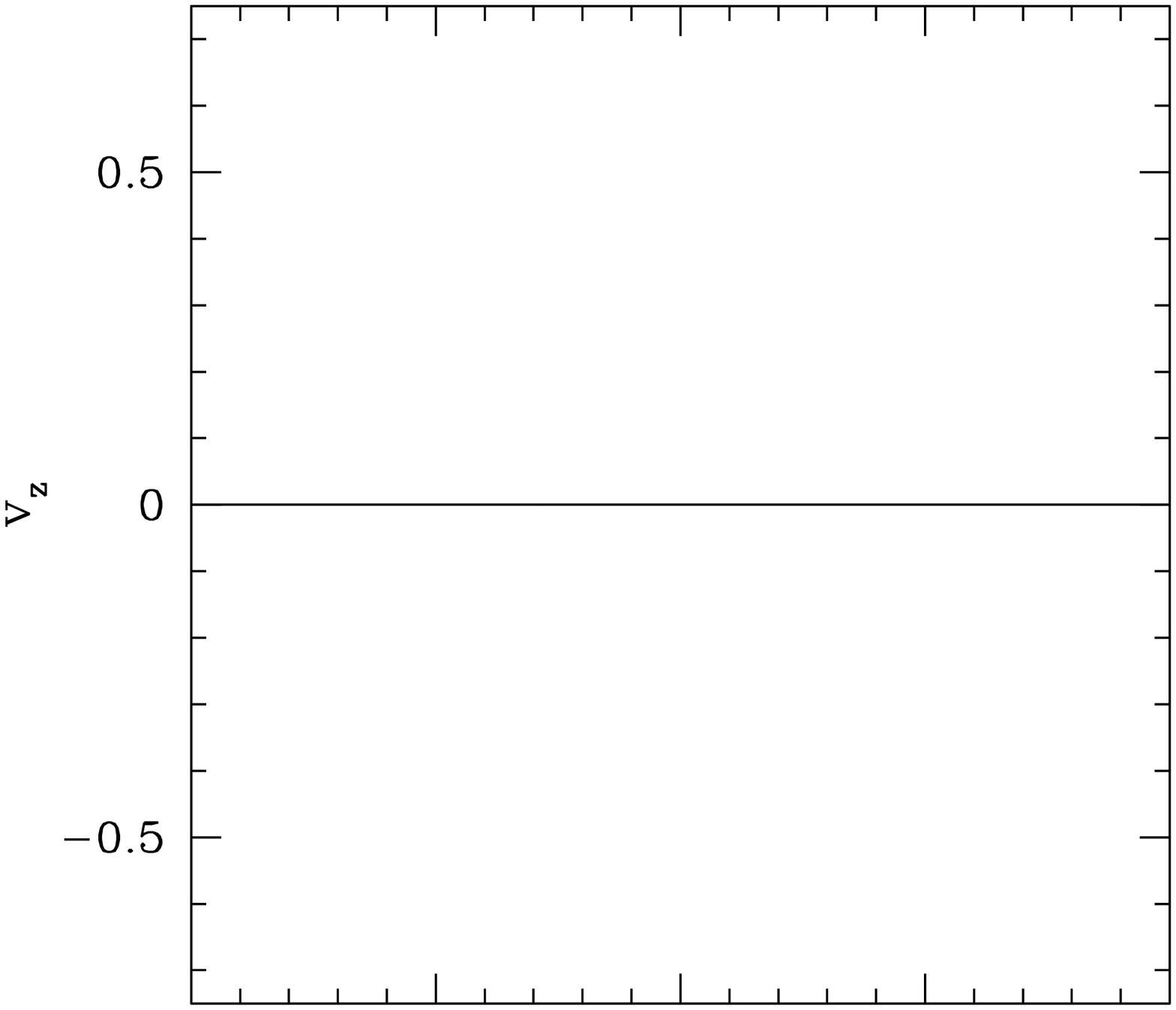}
      \vskip -1.cm 
      \includegraphics[width=0.45\textwidth]{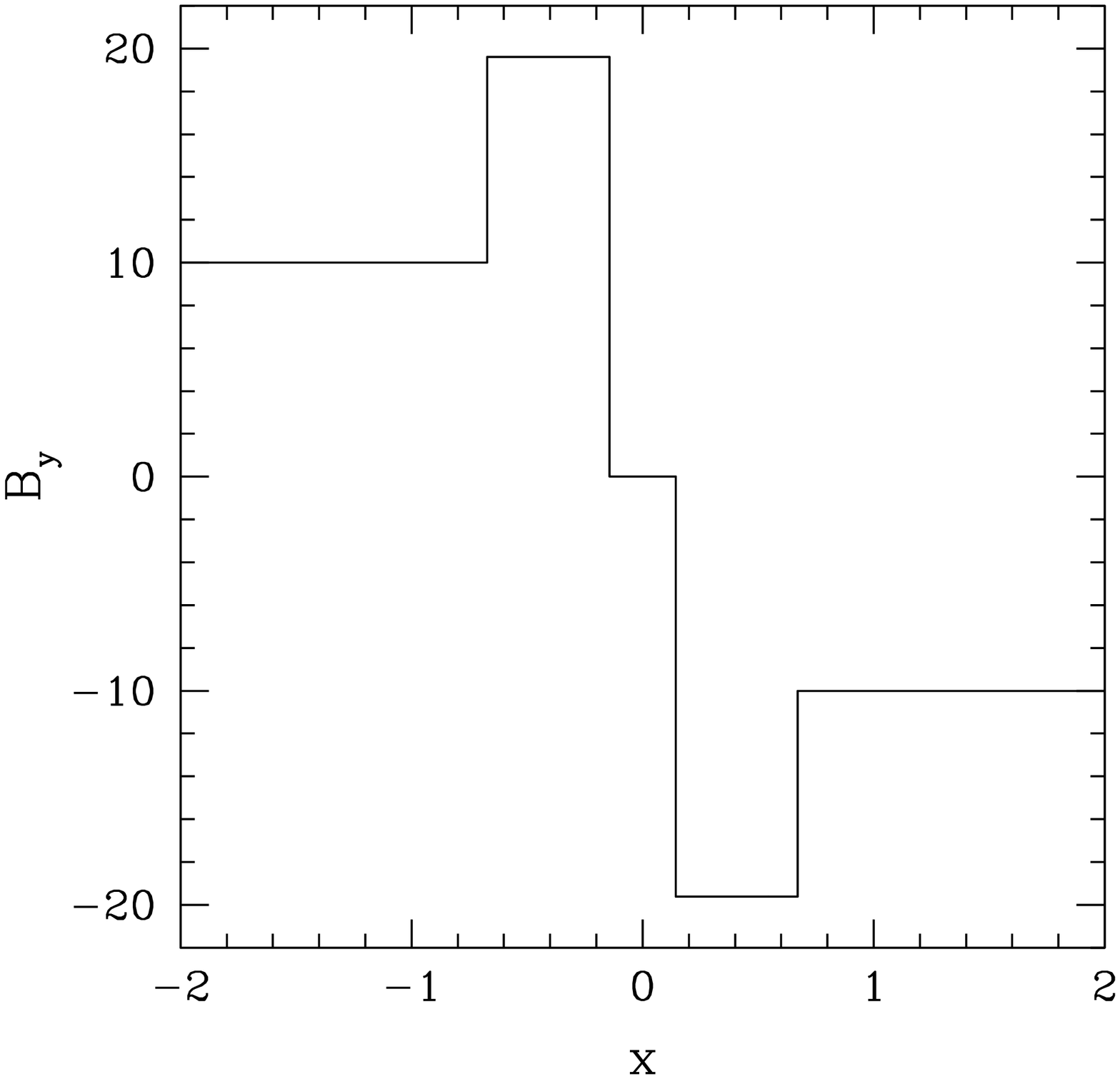}
      \hskip 0.75cm 
      \includegraphics[width=0.45\textwidth]{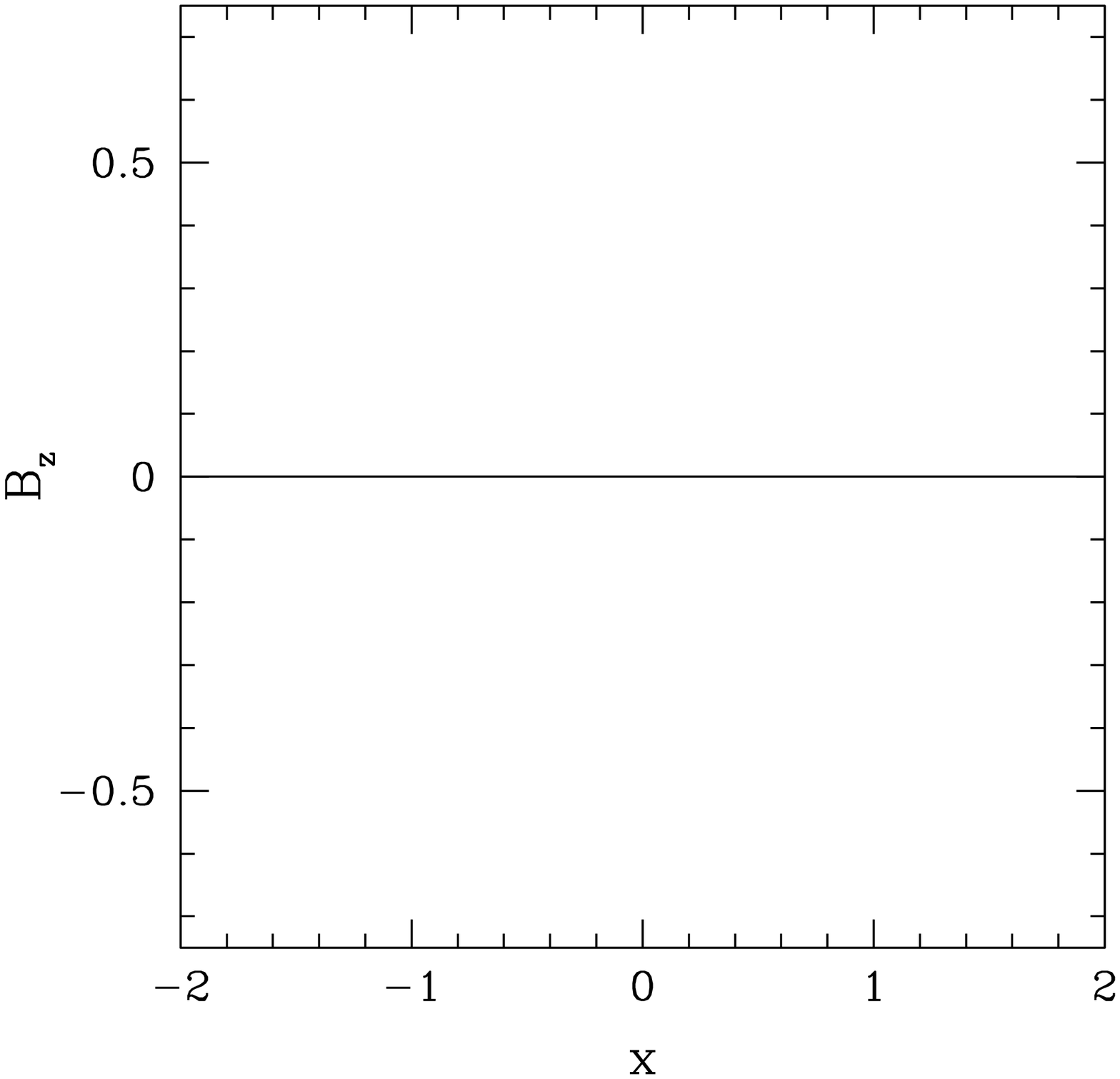}
  \caption{\label{Coll} Exact solution of the collision test of
    Komissarov (1999) at time $t=1.22$. The solution is composed of a
    left-going fast shock, of a left-going slow shock, a right-going slow
    shock and of a right-going fast shock.}
  \end{center}
\end{figure}

\begin{figure}
\begin{center}
      \includegraphics[width=0.45\textwidth]{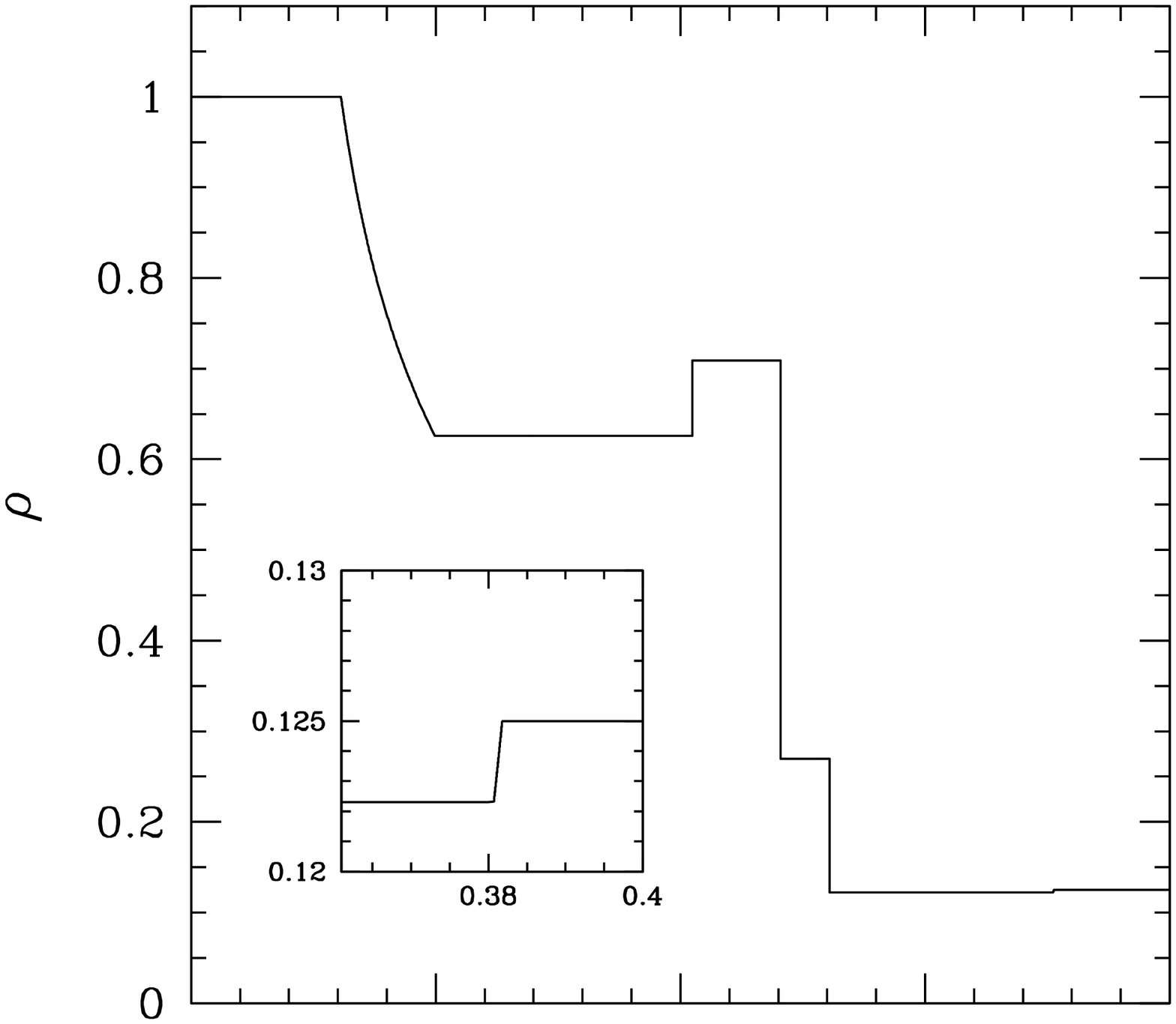}
      \hskip 0.75cm 
      \includegraphics[width=0.45\textwidth]{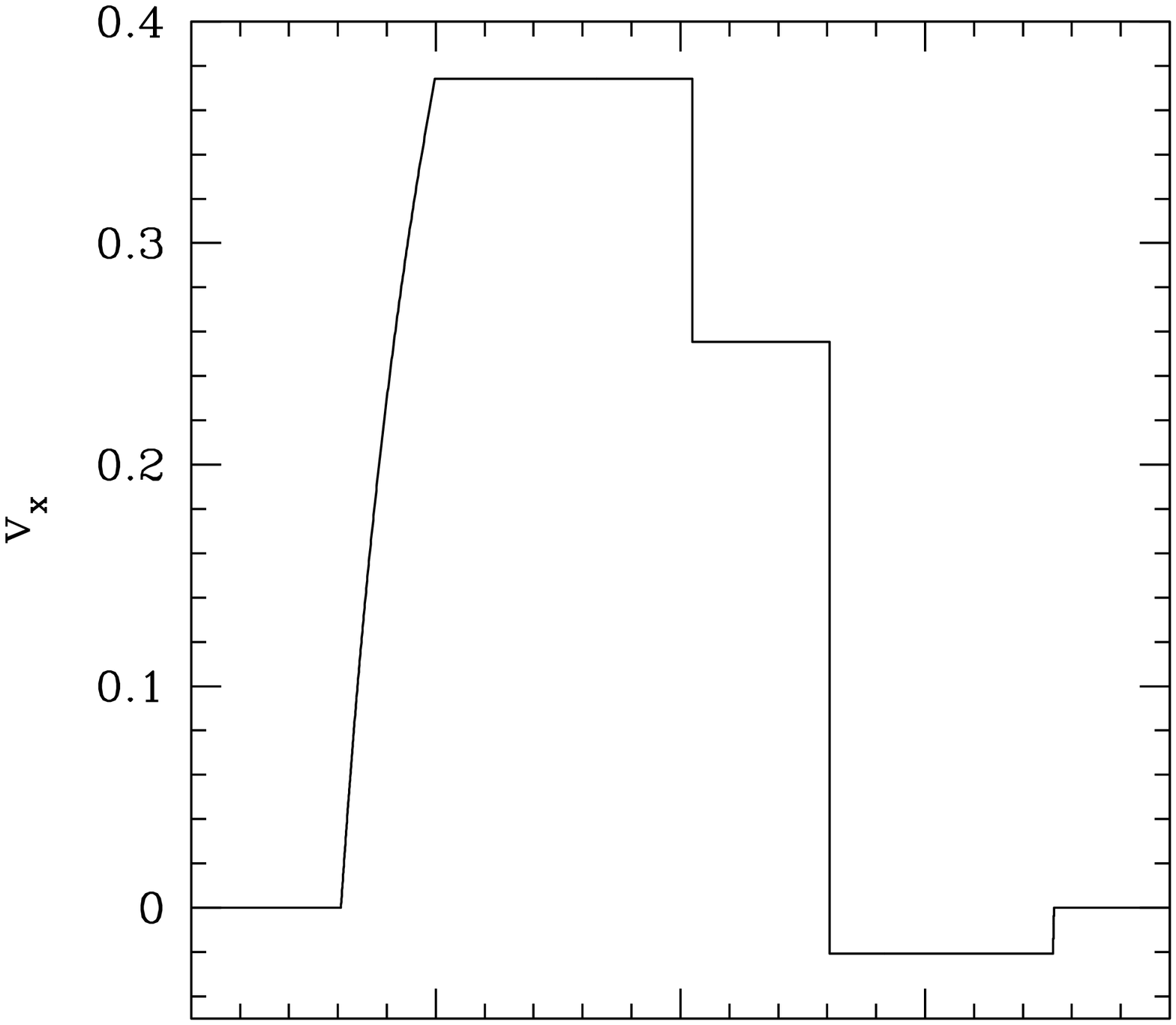}
      \vskip -1.cm 
      \includegraphics[width=0.45\textwidth]{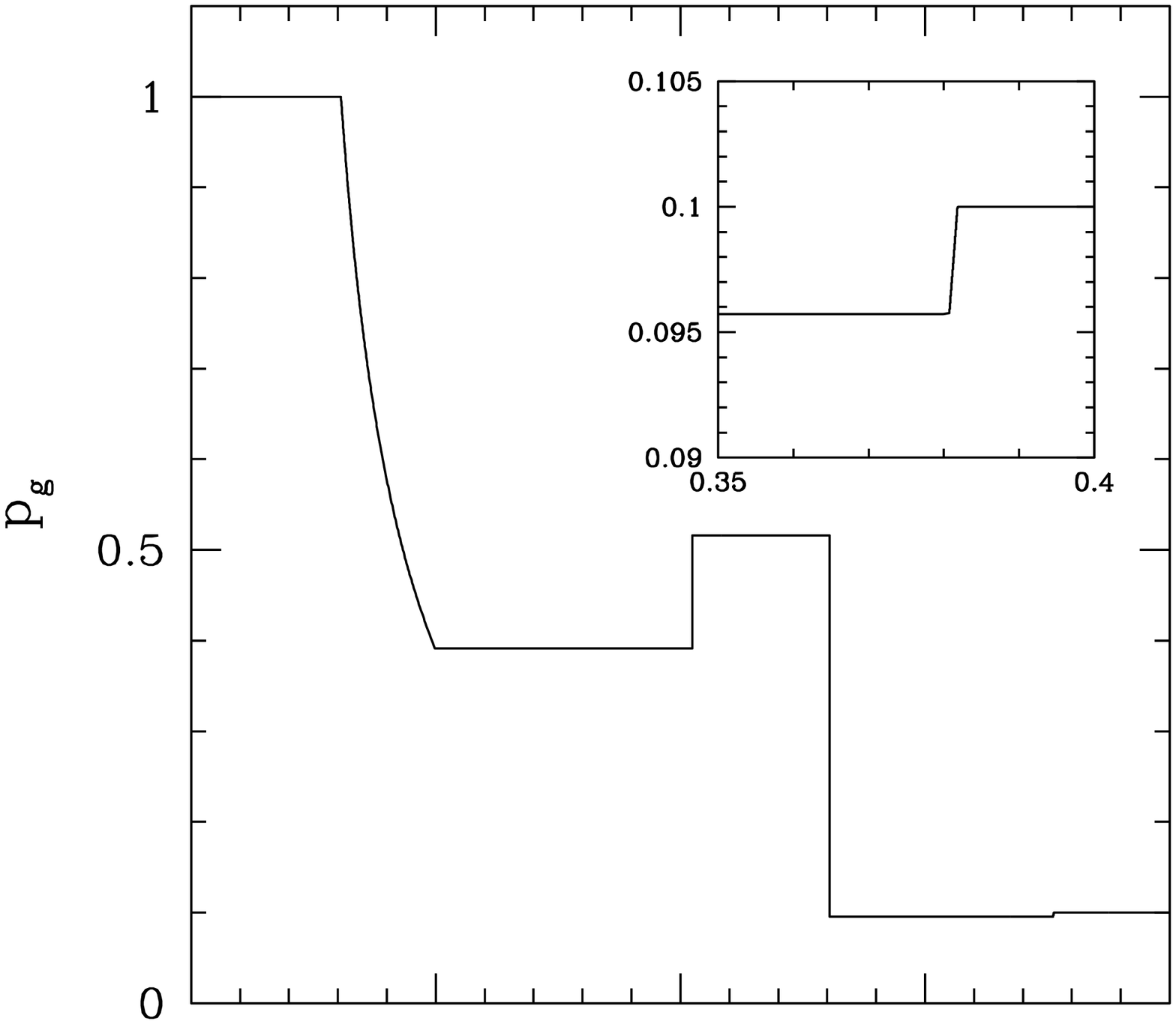}
      \hskip 0.75cm 
      \includegraphics[width=0.45\textwidth]{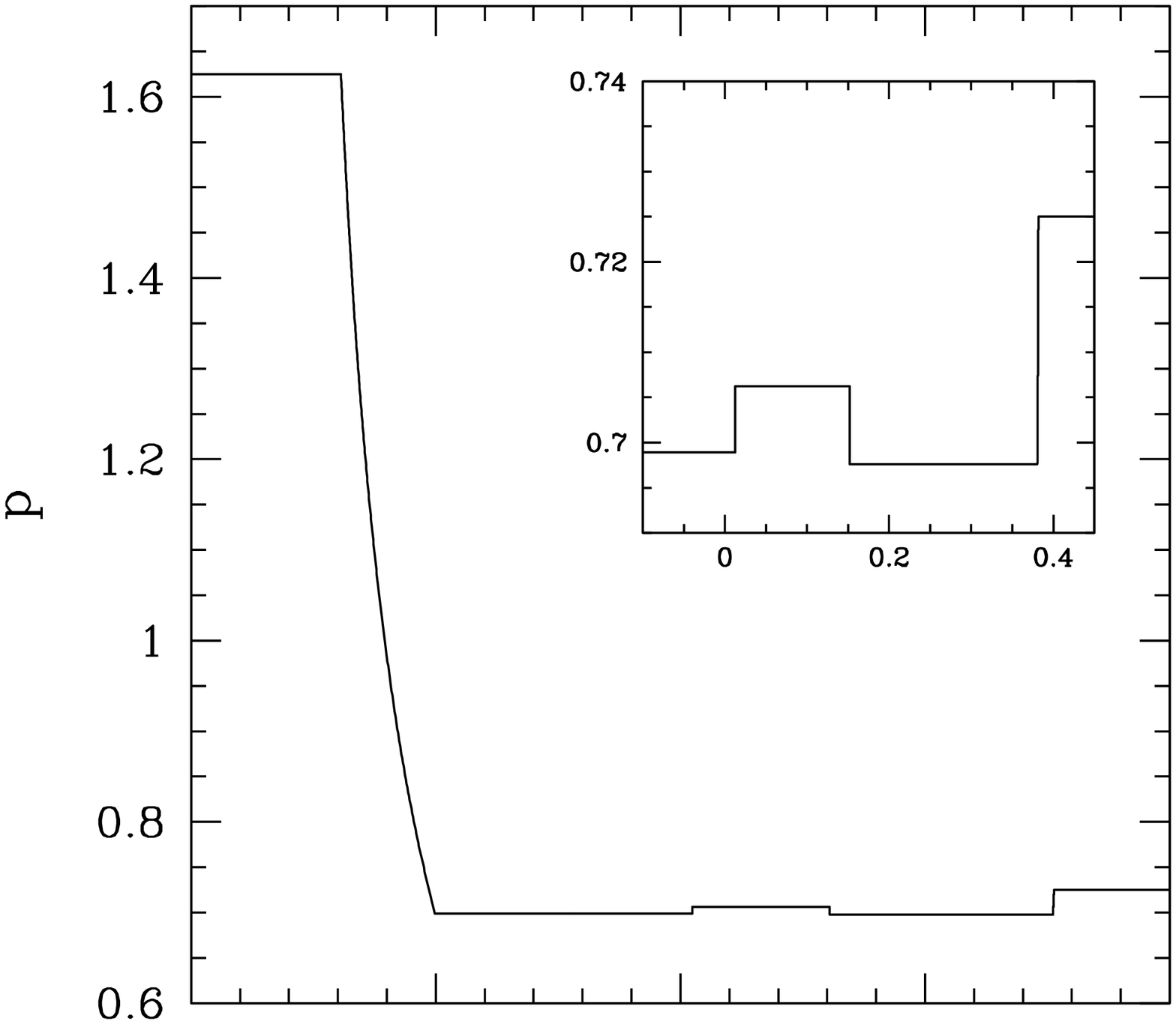}
      \vskip -1.cm 
      \includegraphics[width=0.45\textwidth]{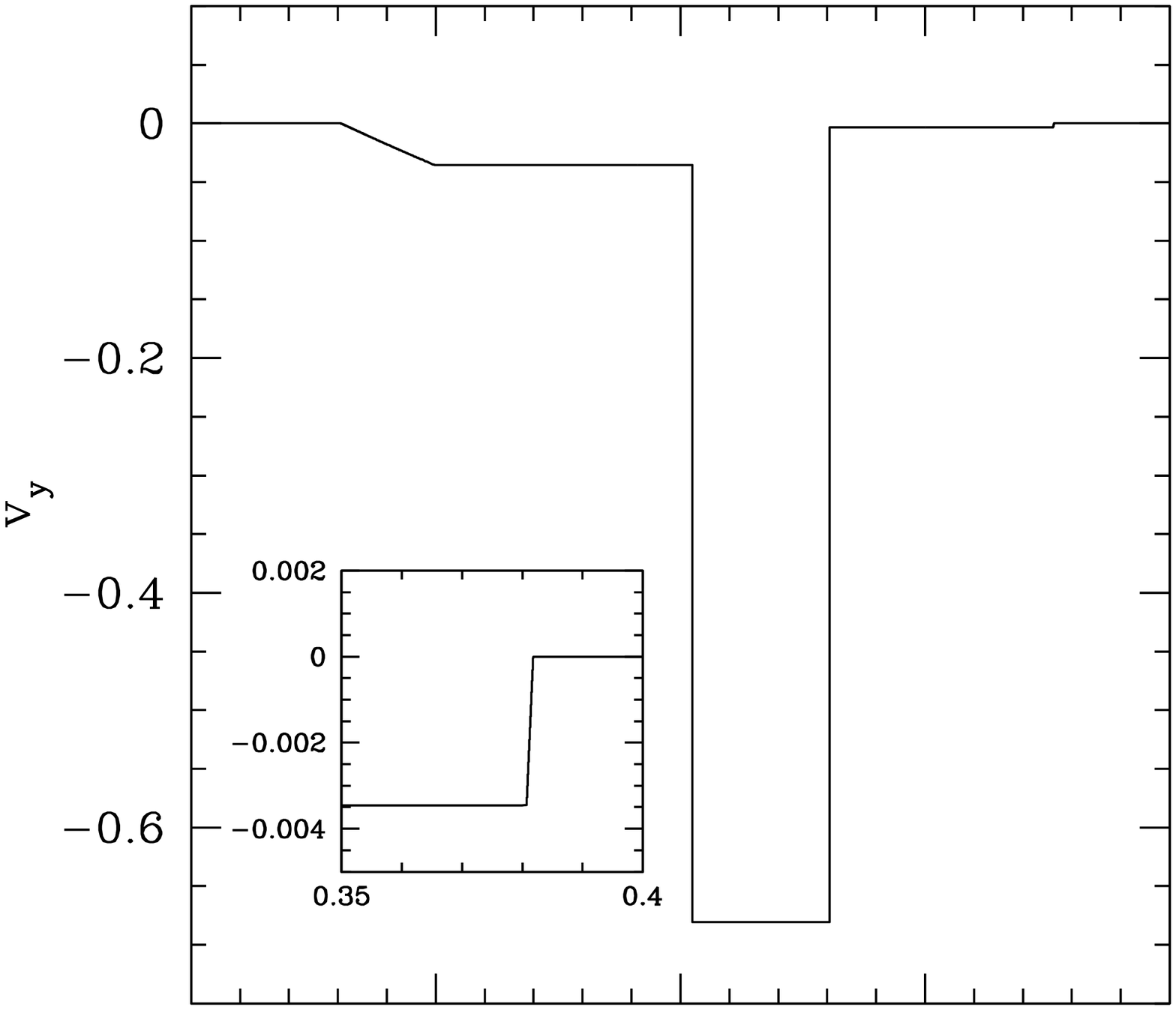}
      \hskip 0.75cm 
      \includegraphics[width=0.45\textwidth]{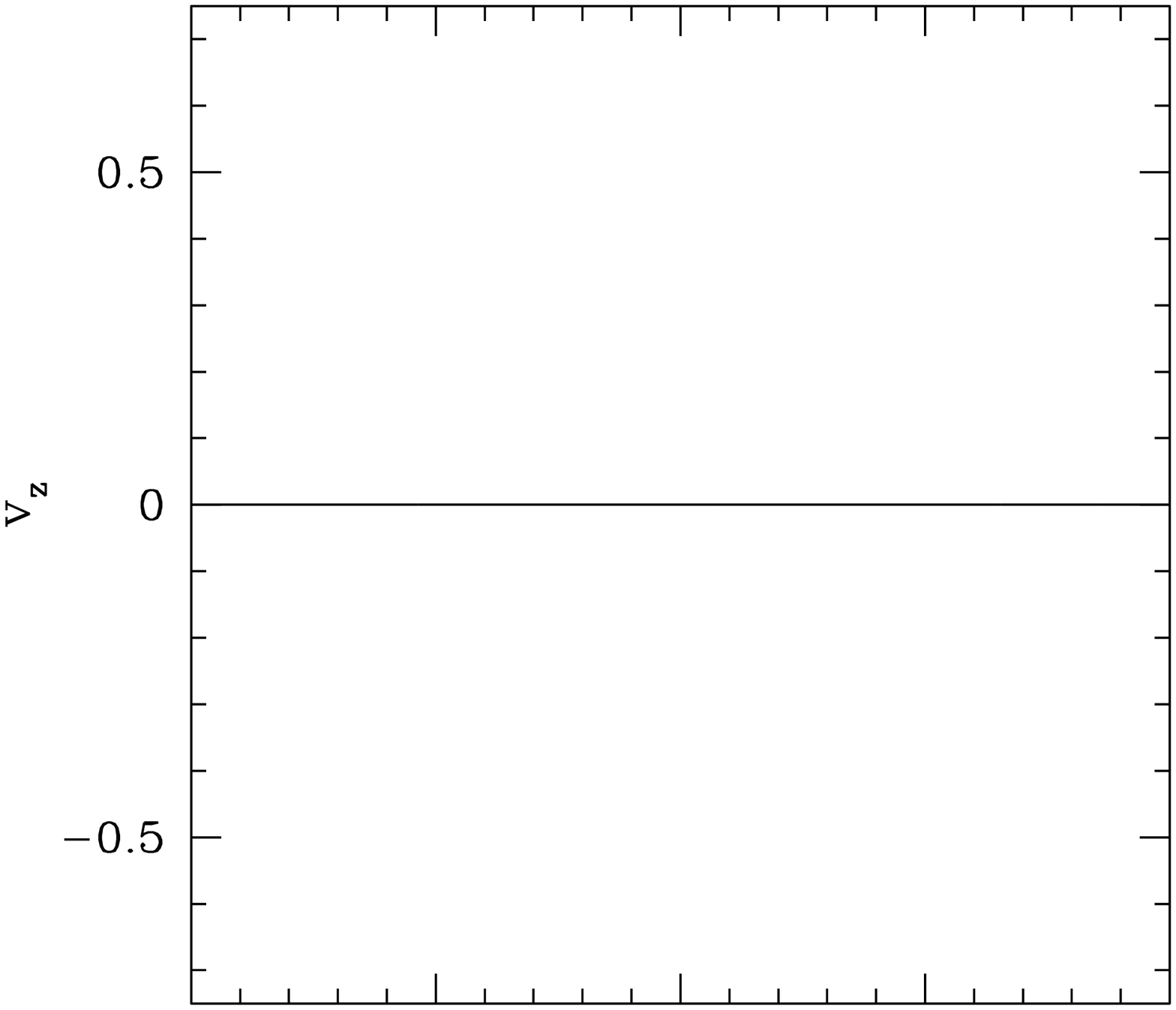}
      \vskip -1.cm 
      \includegraphics[width=0.45\textwidth]{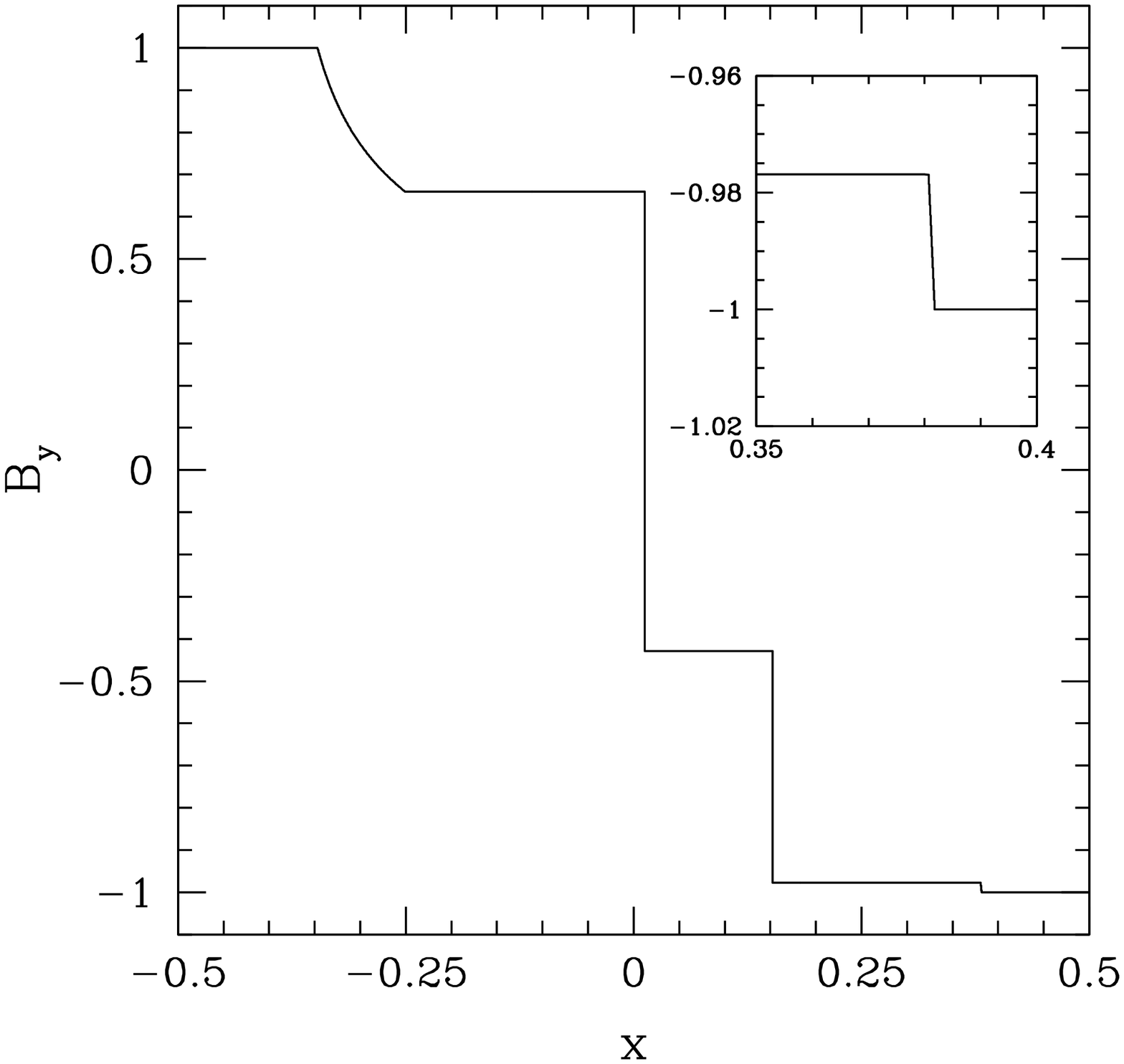}
      \hskip 0.75cm 
      \includegraphics[width=0.45\textwidth]{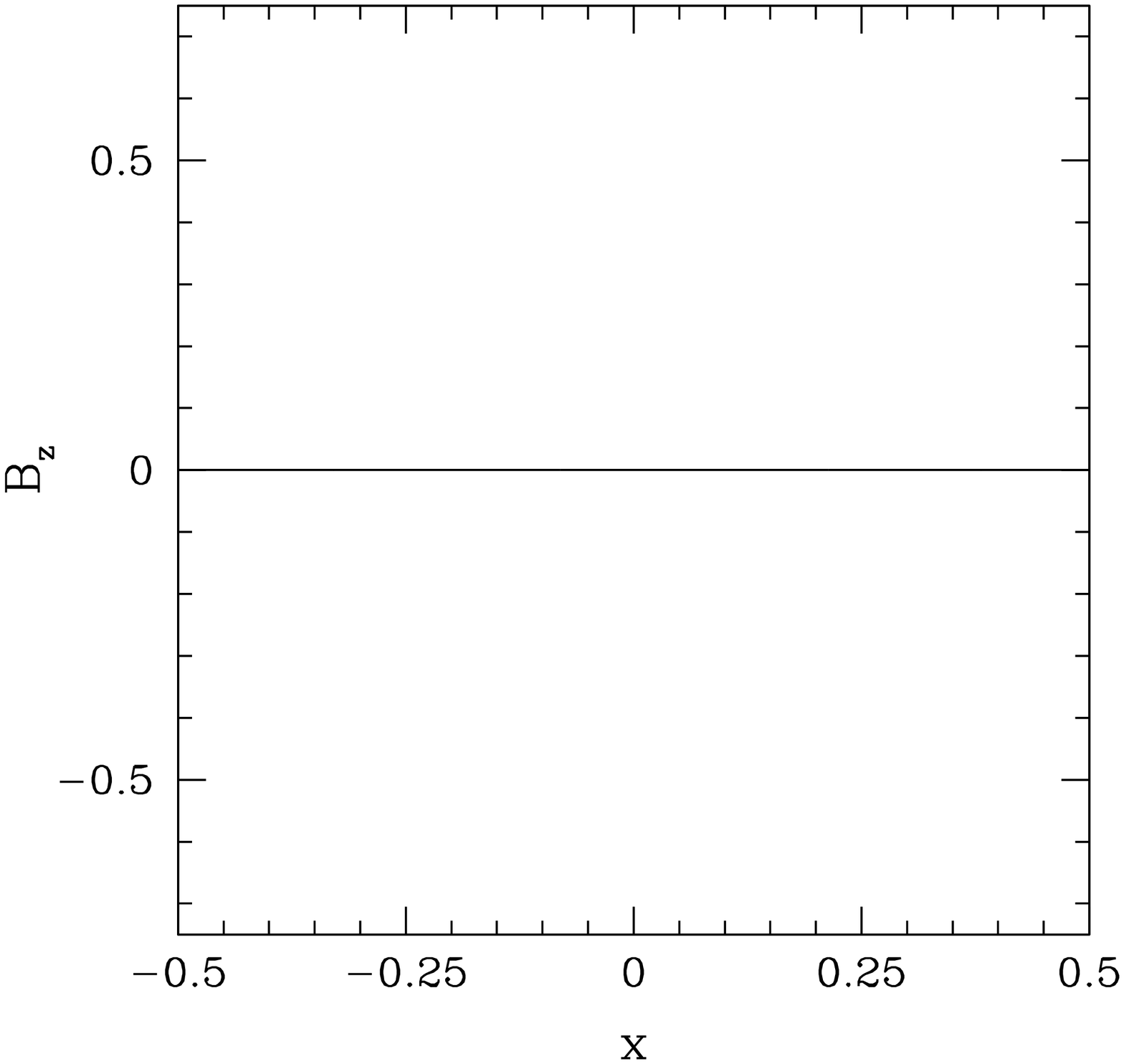}
  \caption{\label{BA1} Exact solution of the test number 1 of Balsara
  (2001) at time $t=0.4$ and which represents the relativistic version of
  the Brio-Wu test (1988). The solution is composed of a left-going fast
  rarefaction, of a left-going slow shock, of a contact discontinuity, of
  a right-going slow shock and of a right-going fast rarefaction. Note
  the absence of a slow compound-wave which cannot be found by
  construction in our exact solver, but that appears in the solution of
  the HLLE approximate Riemann solver (not shown).}
  \end{center}
\end{figure}

\begin{figure}
\begin{center}
      \includegraphics[width=0.45\textwidth]{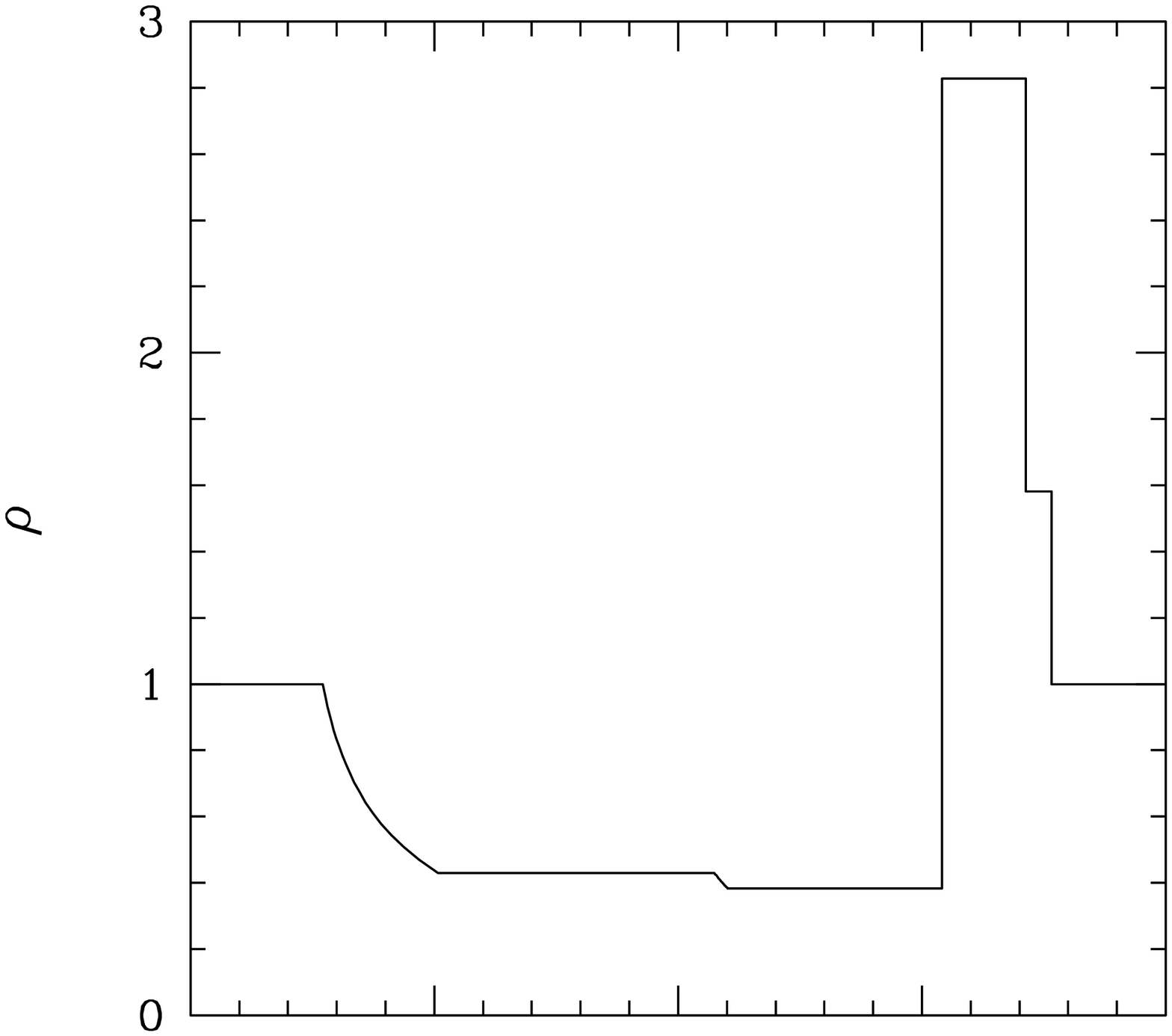}
      \hskip 0.75cm 
      \includegraphics[width=0.45\textwidth]{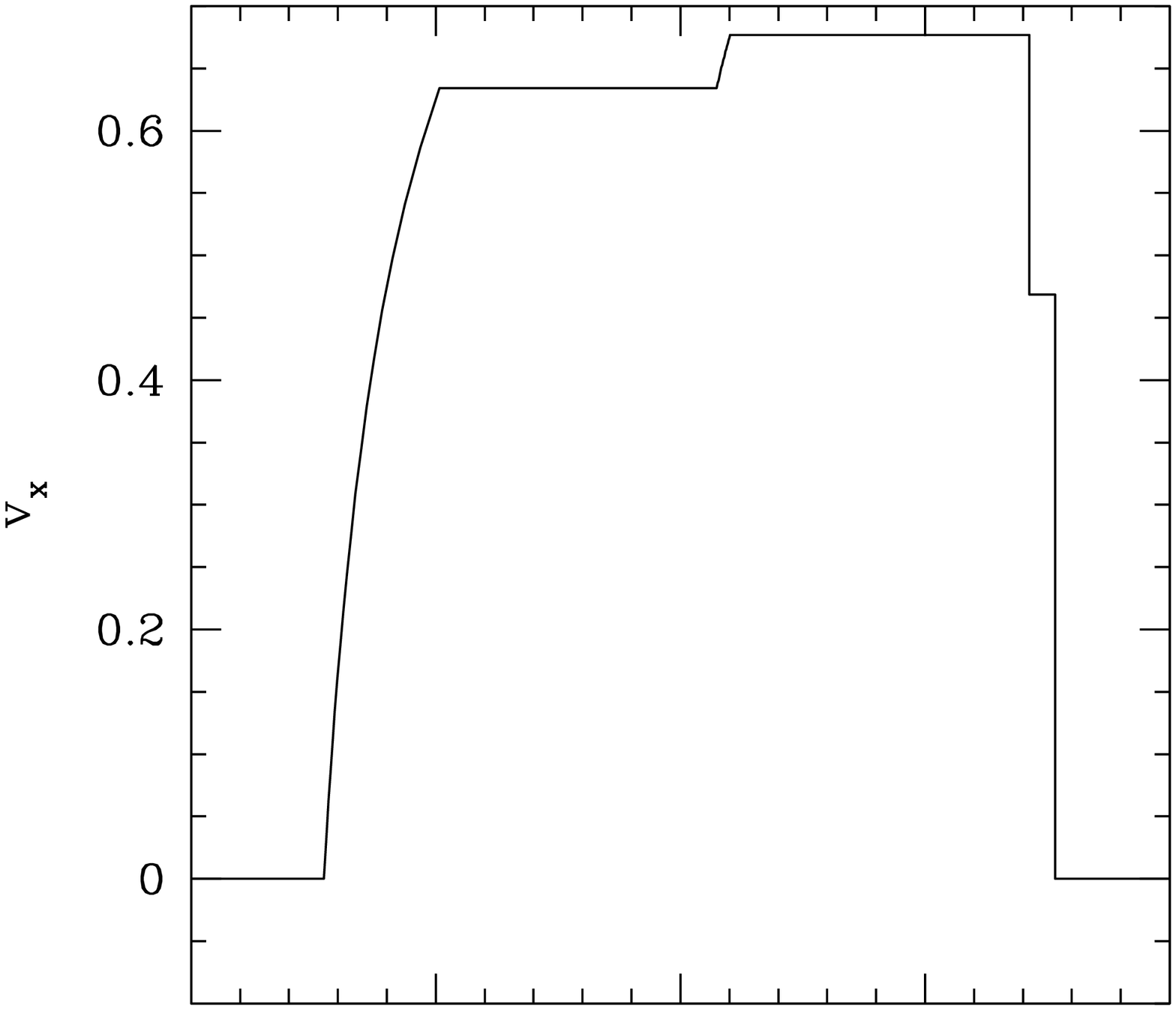}
      \vskip -1.cm 
      \includegraphics[width=0.45\textwidth]{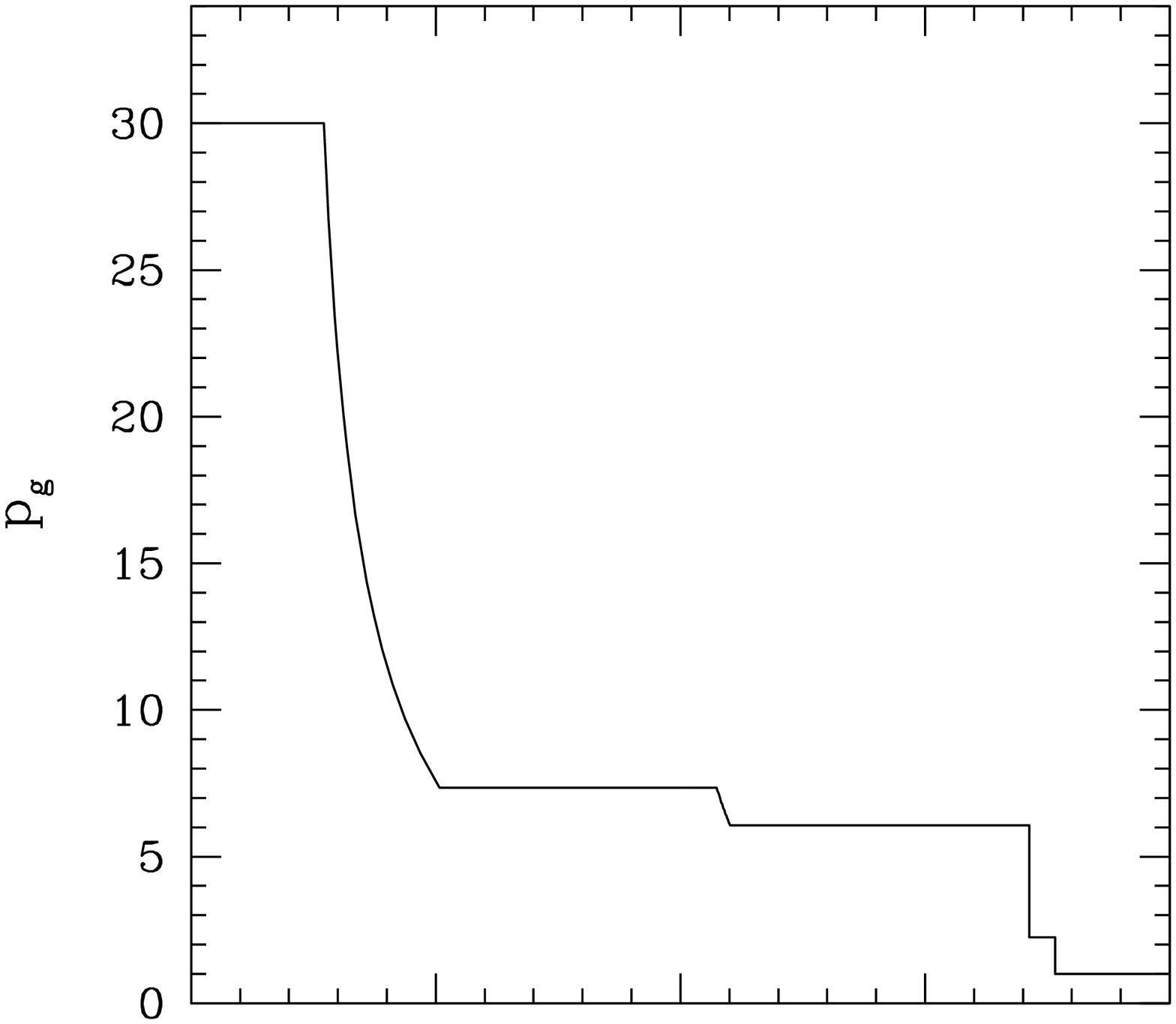}
      \hskip 0.75cm 
      \includegraphics[width=0.45\textwidth]{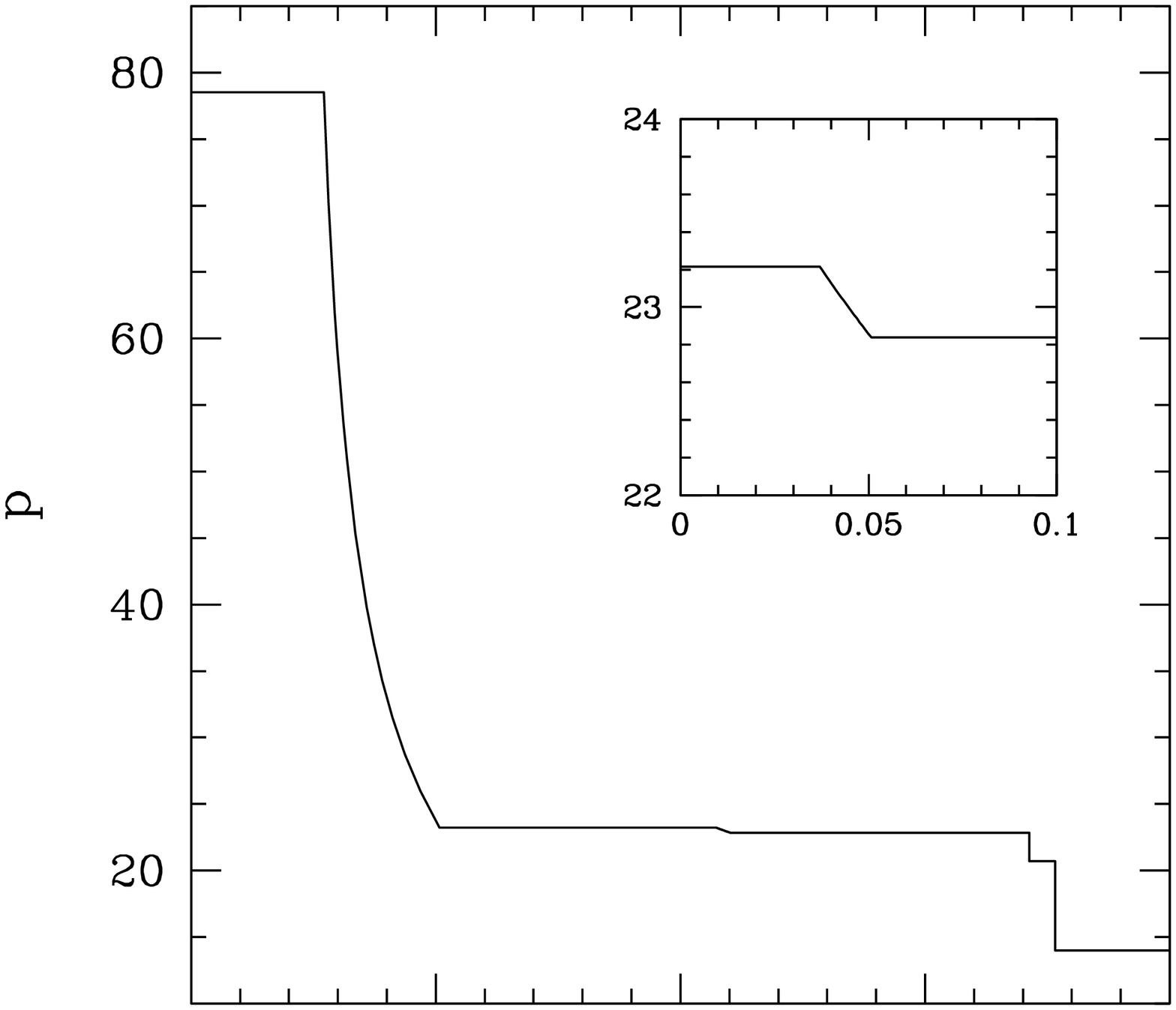}
      \vskip -1.cm 
      \includegraphics[width=0.45\textwidth]{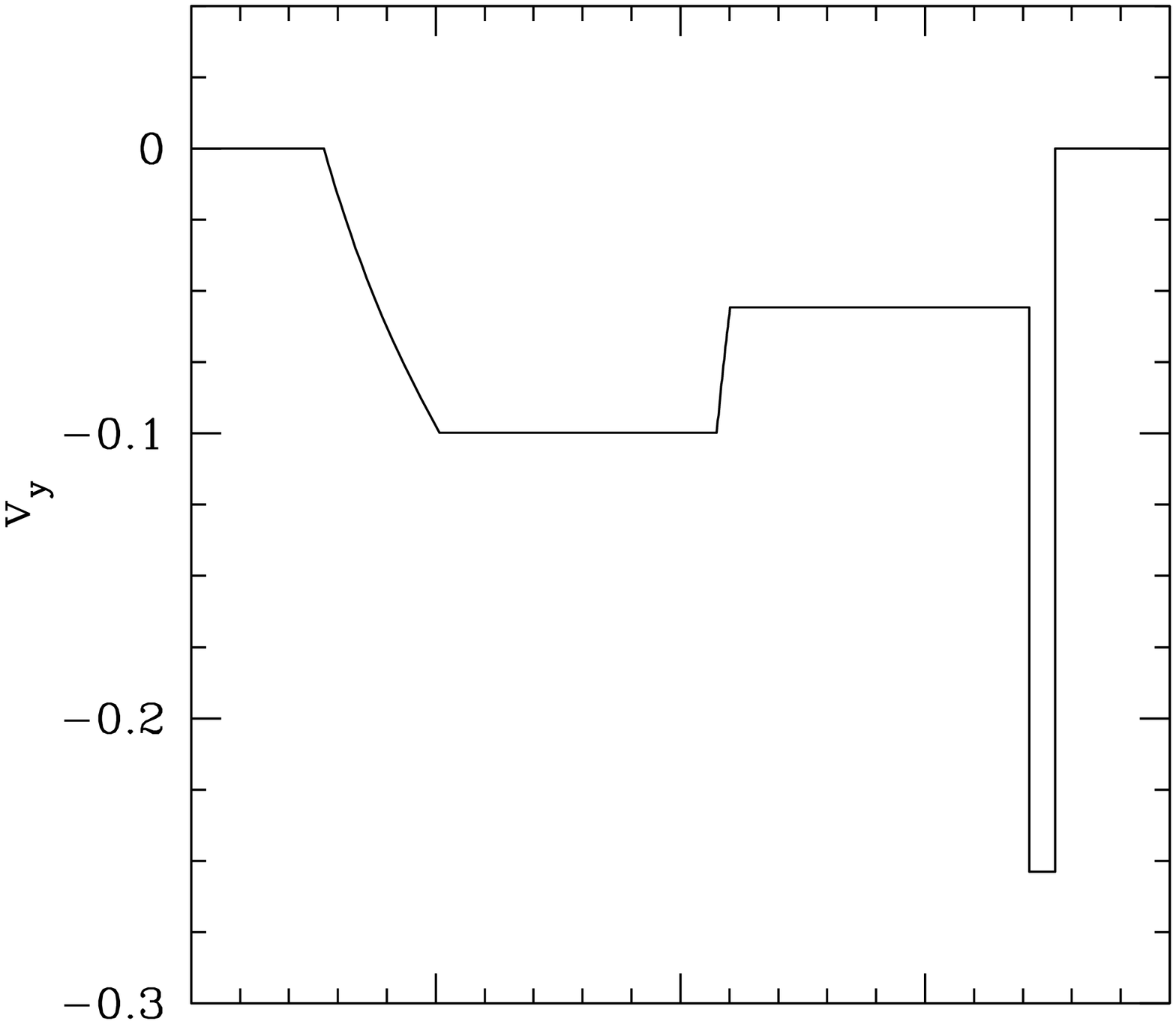}
      \hskip 0.75cm 
      \includegraphics[width=0.45\textwidth]{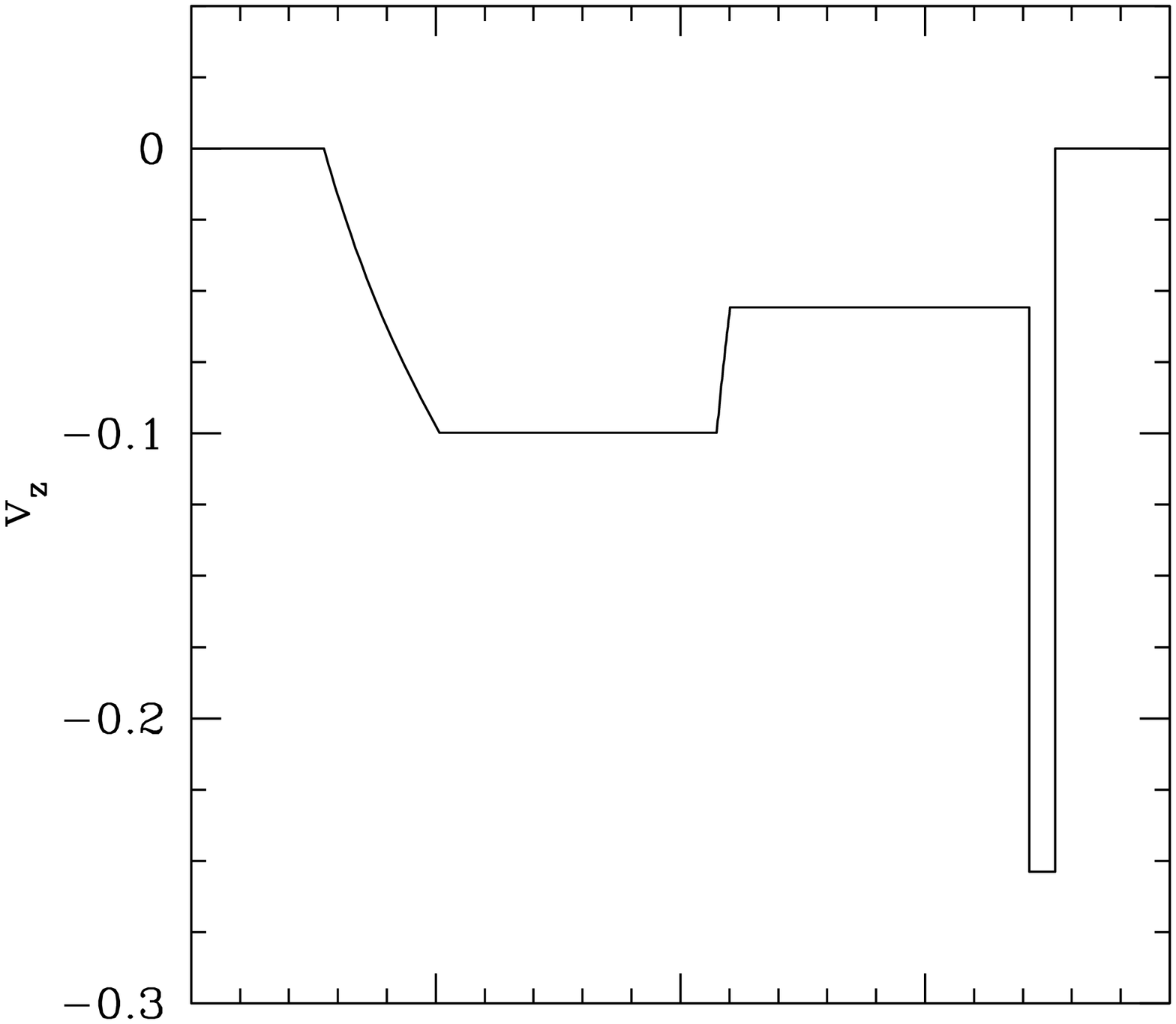}
      \vskip -1.cm 
      \includegraphics[width=0.45\textwidth]{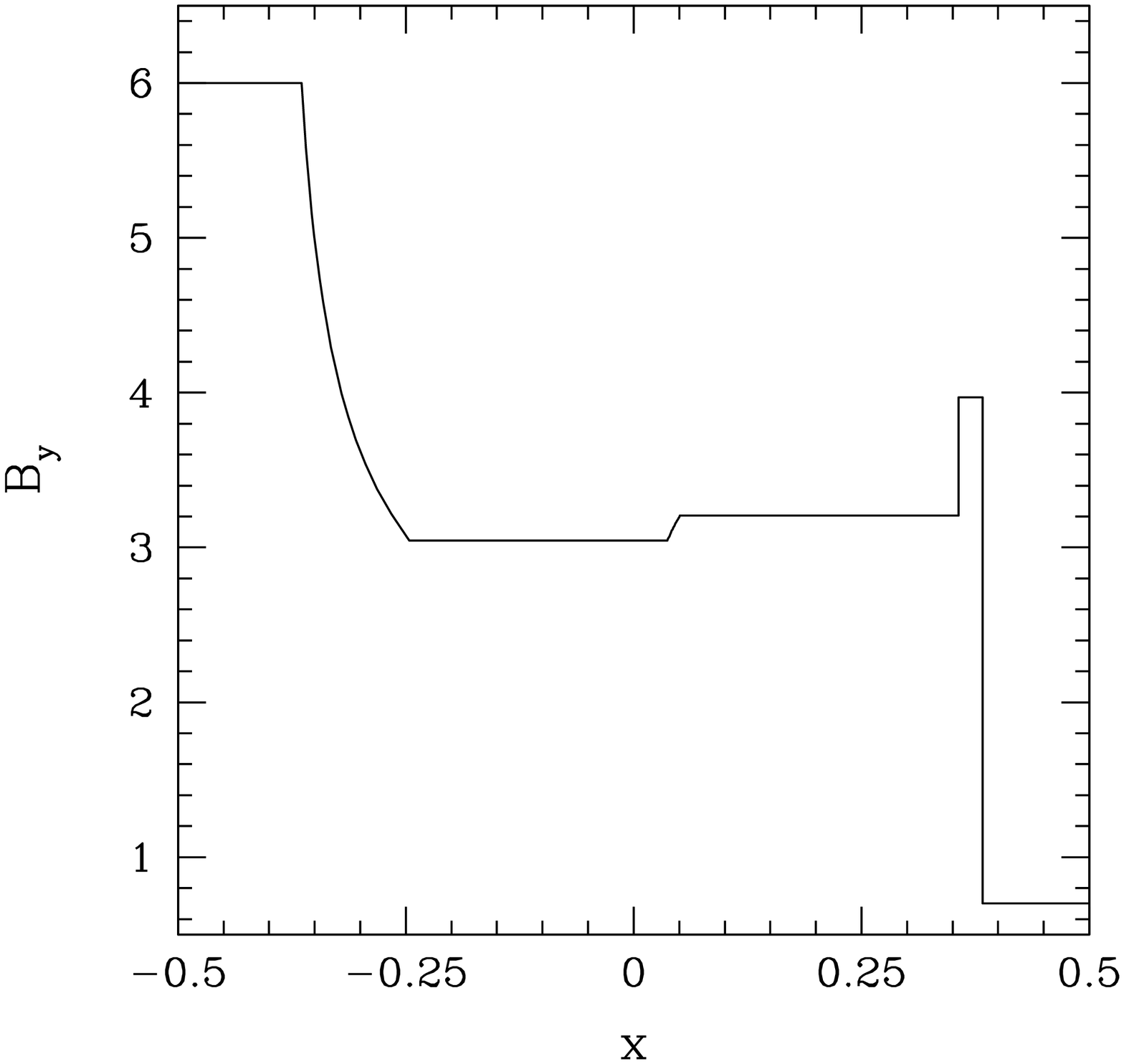}
      \hskip 0.75cm 
      \includegraphics[width=0.45\textwidth]{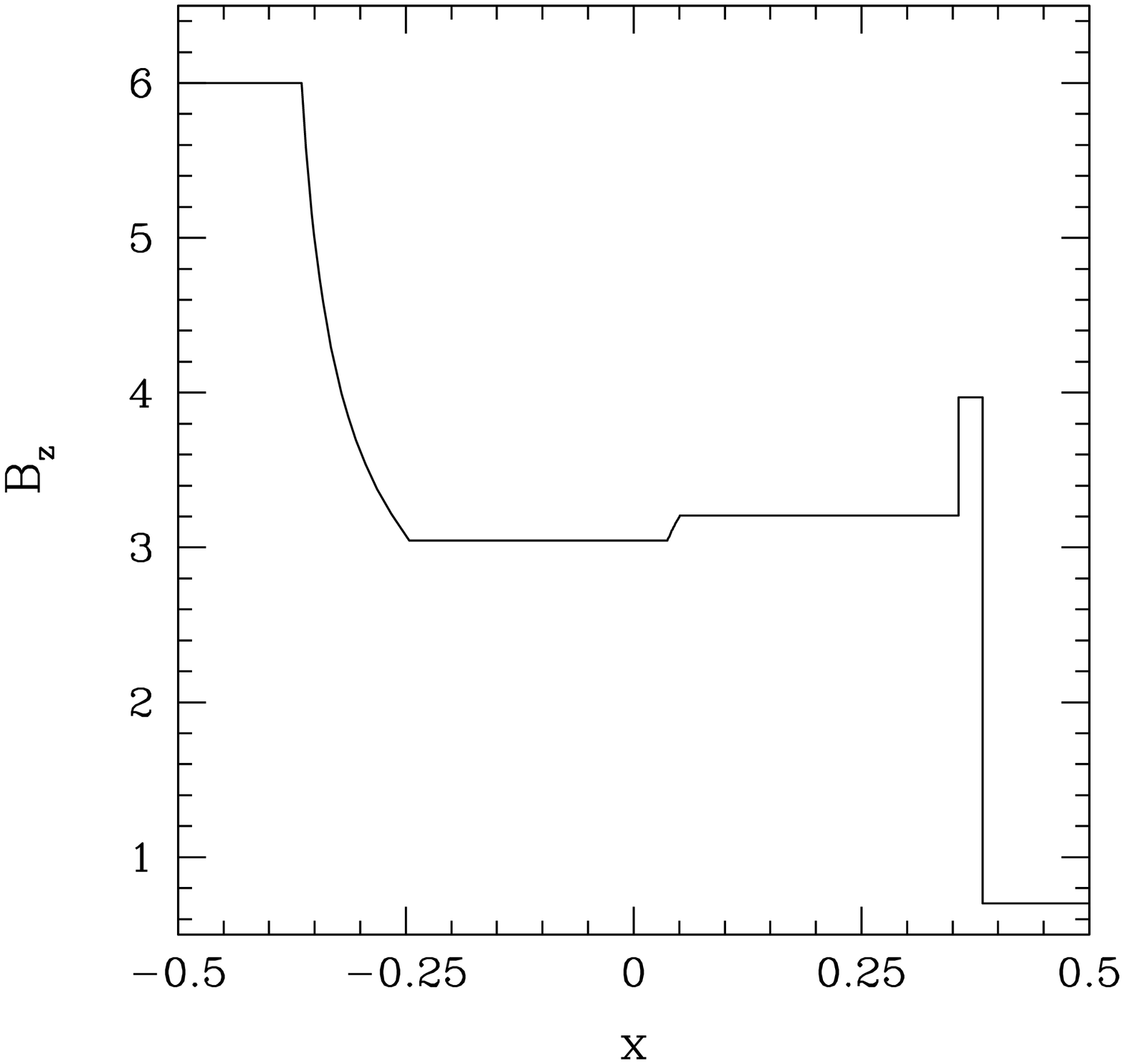}
  \caption{\label{BA2} Exact solution of the test number 2 of Balsara
    (2001) at time $t=0.4$. The solution is composed of two left-going
    fast and slow rarefactions, of a contact discontinuity and of two
    right-going fast and slow shocks.}
  \end{center}
\end{figure}

\begin{figure}
\begin{center}
      \includegraphics[width=0.45\textwidth]{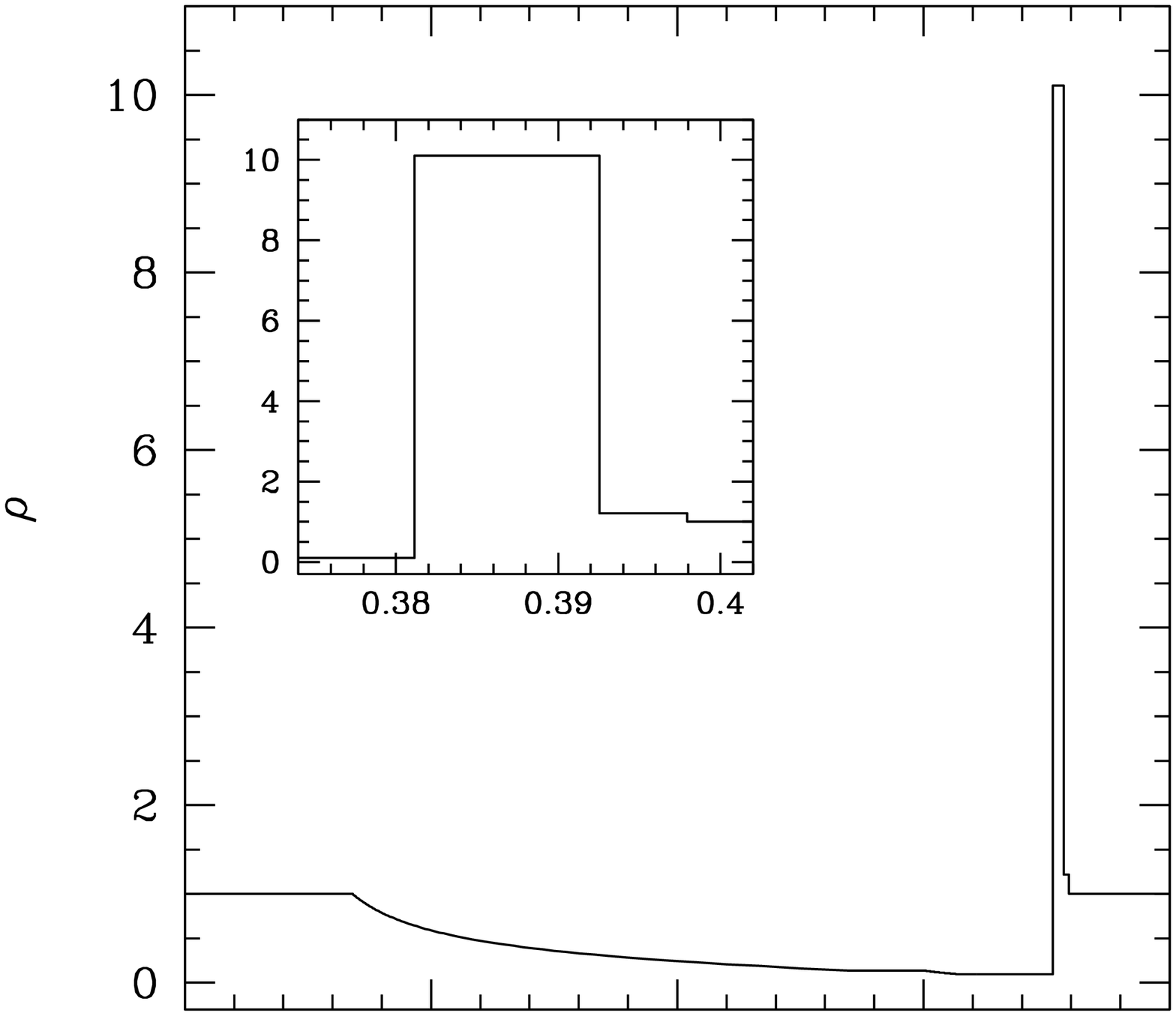}
      \hskip 0.75cm 
      \includegraphics[width=0.45\textwidth]{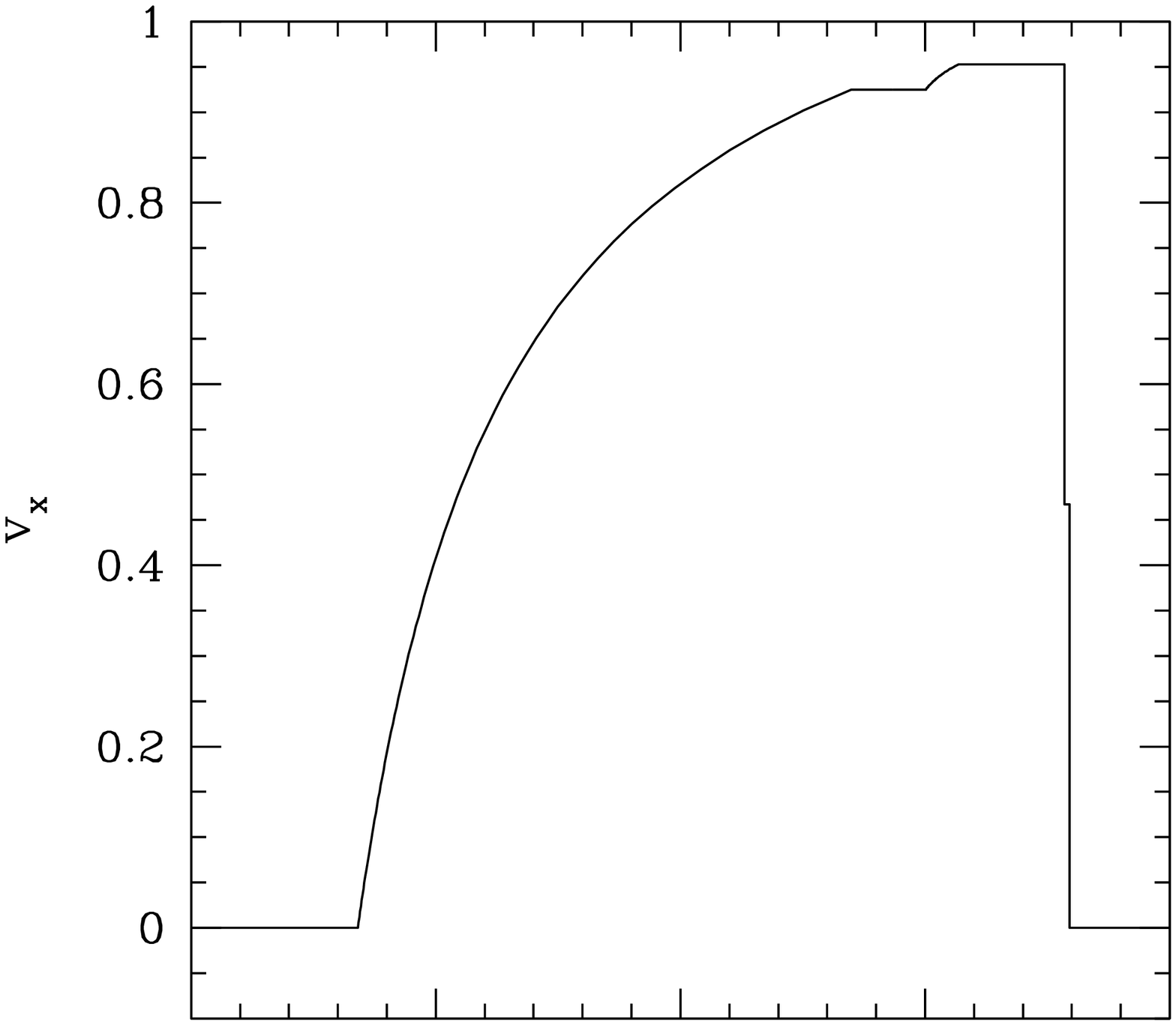}
      \vskip -1.0cm 
      \includegraphics[width=0.45\textwidth]{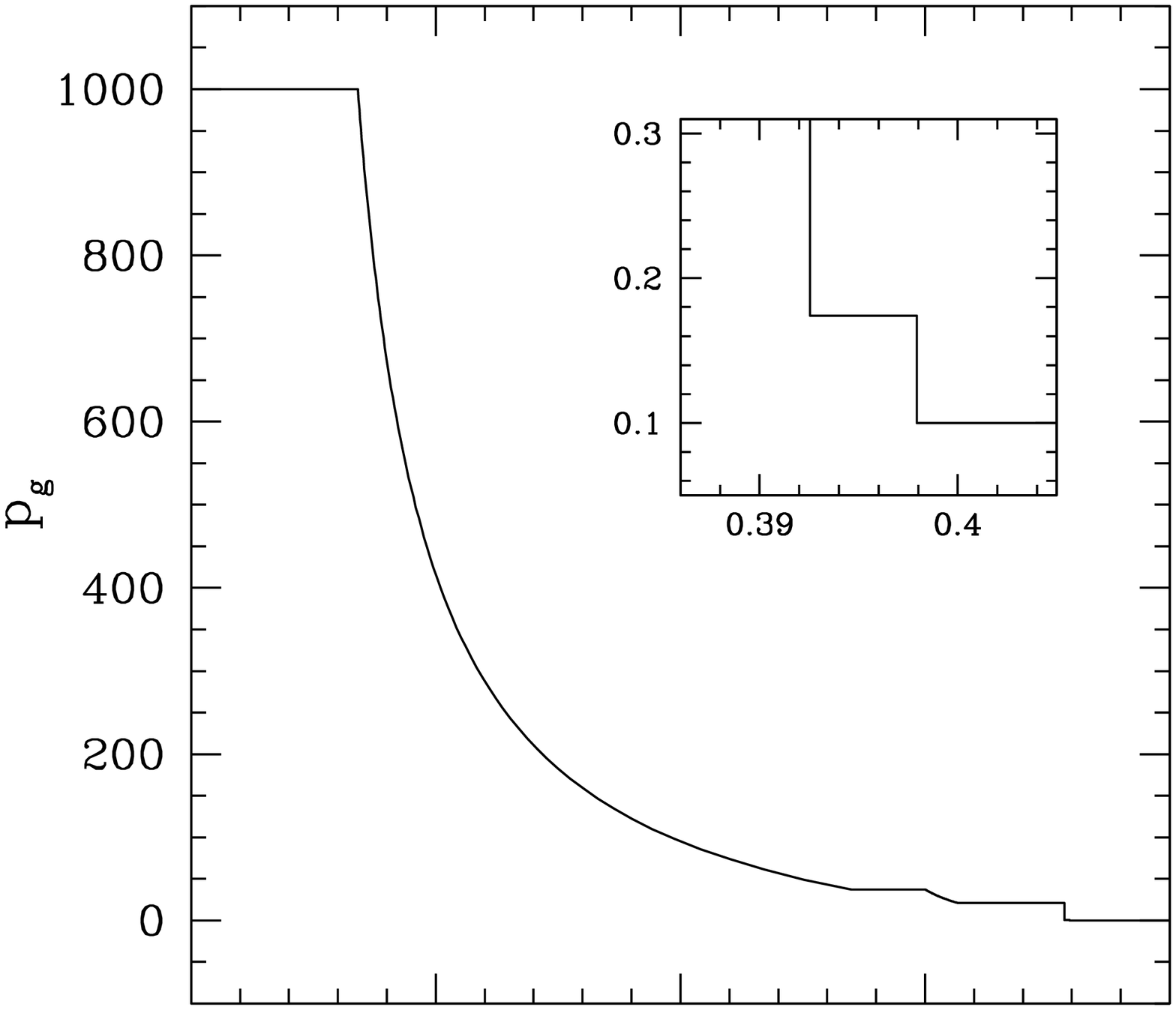}
      \hskip 0.75cm 
      \includegraphics[width=0.45\textwidth]{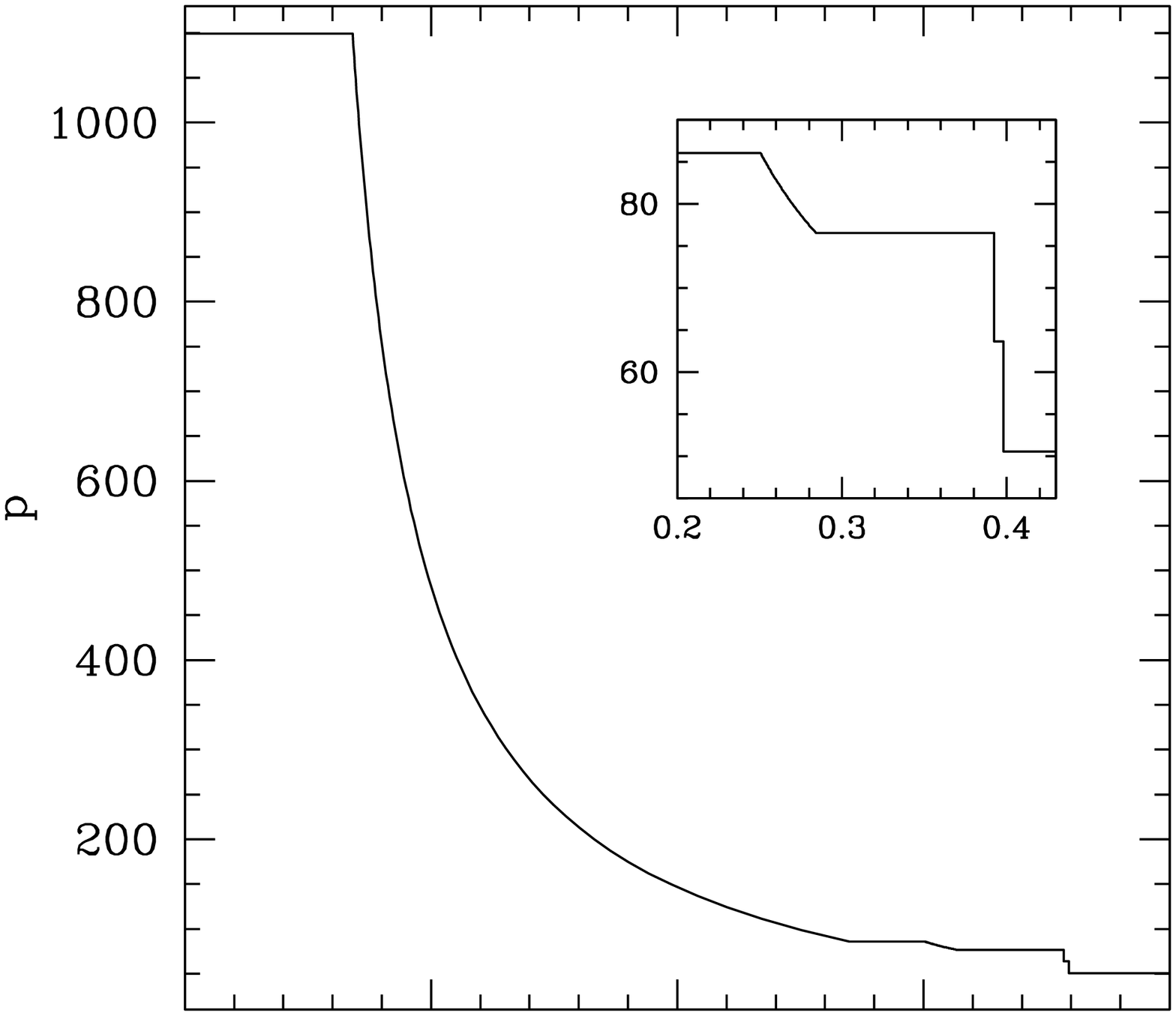}
      \vskip -1.0cm 
      \includegraphics[width=0.45\textwidth]{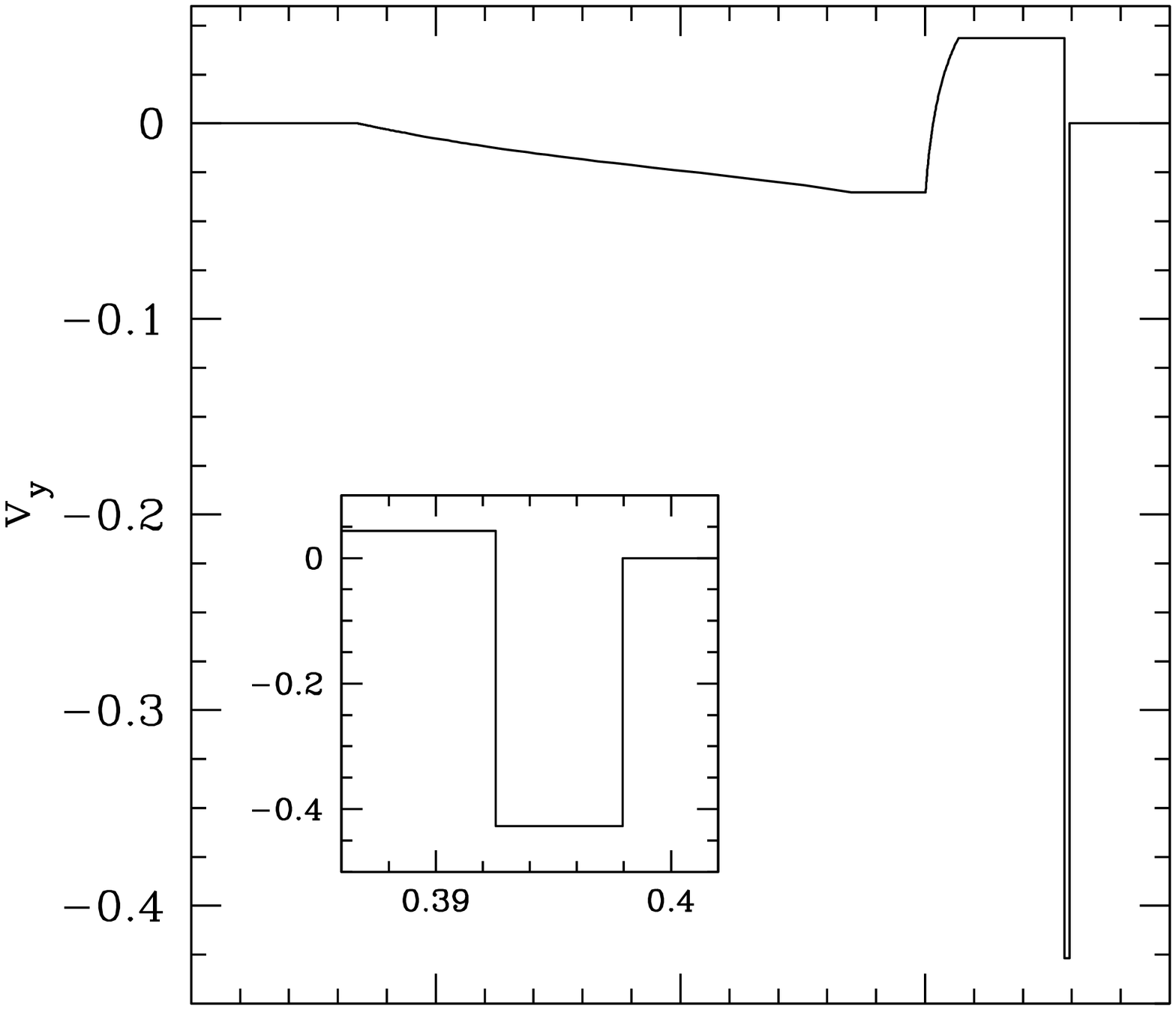}
      \hskip 0.75cm 
      \includegraphics[width=0.45\textwidth]{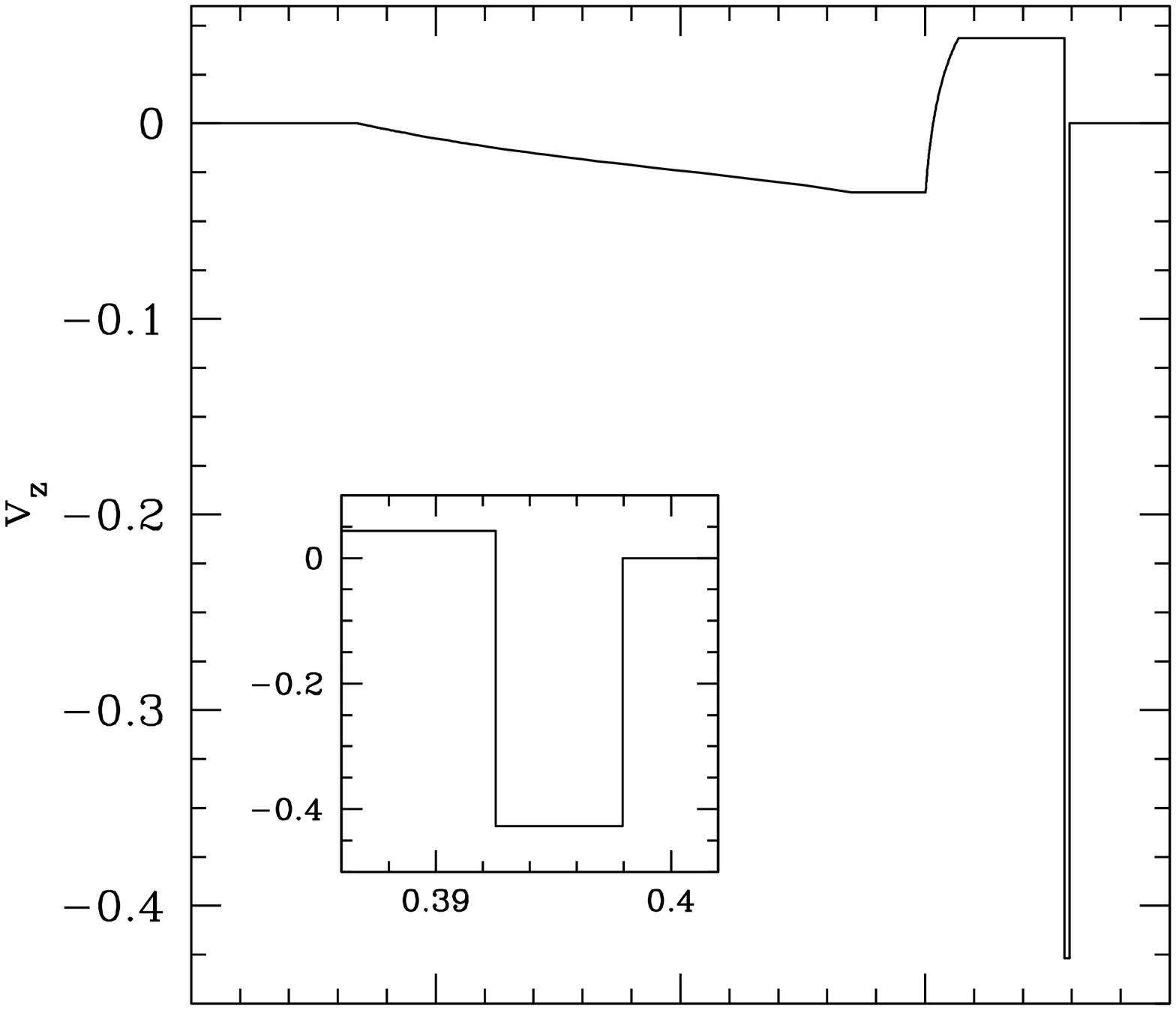}
      \vskip -1.0cm 
      \includegraphics[width=0.45\textwidth]{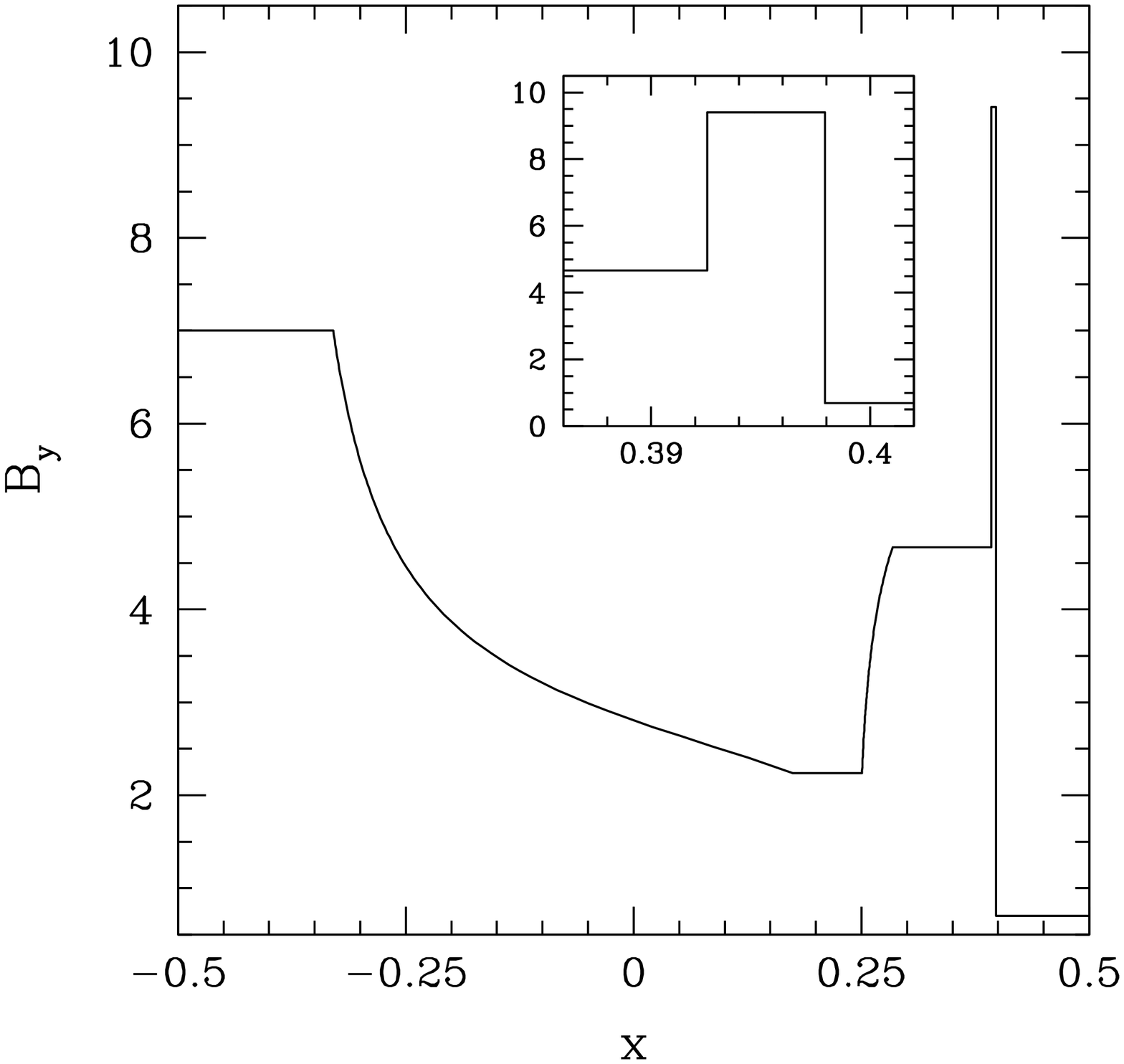}
      \hskip 0.75cm 
      \includegraphics[width=0.45\textwidth]{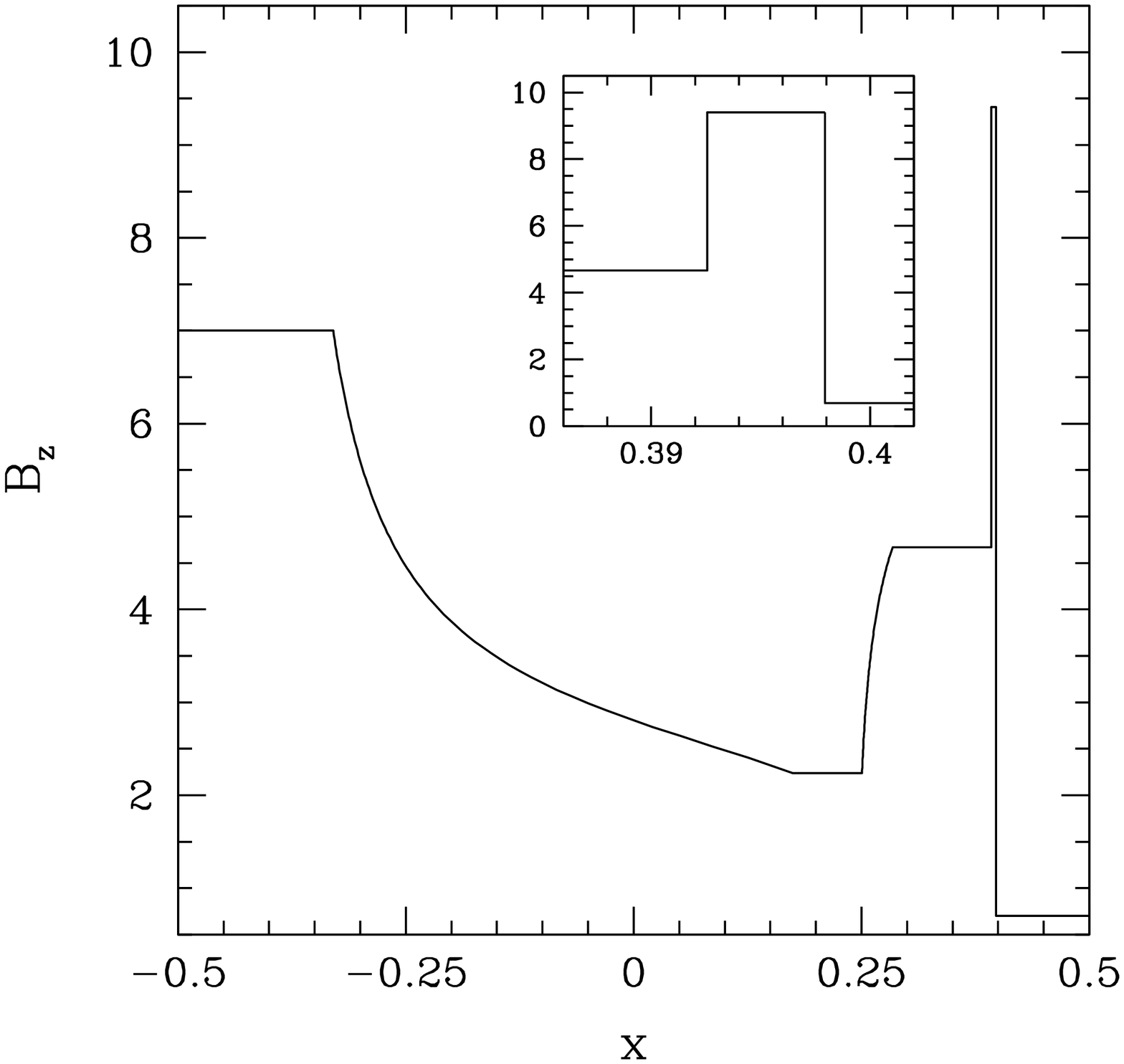}
  \caption{\label{BA3} Exact solution of the test number 3 of Balsara
    (2001) at time $t=0.4$. The solution is composed of two left-going
    fast and slow rarefactions, of a contact discontinuity and of two
    right-going fast and slow shocks.}
  \end{center}
\end{figure}

\begin{figure}
\begin{center}
      \includegraphics[width=0.45\textwidth]{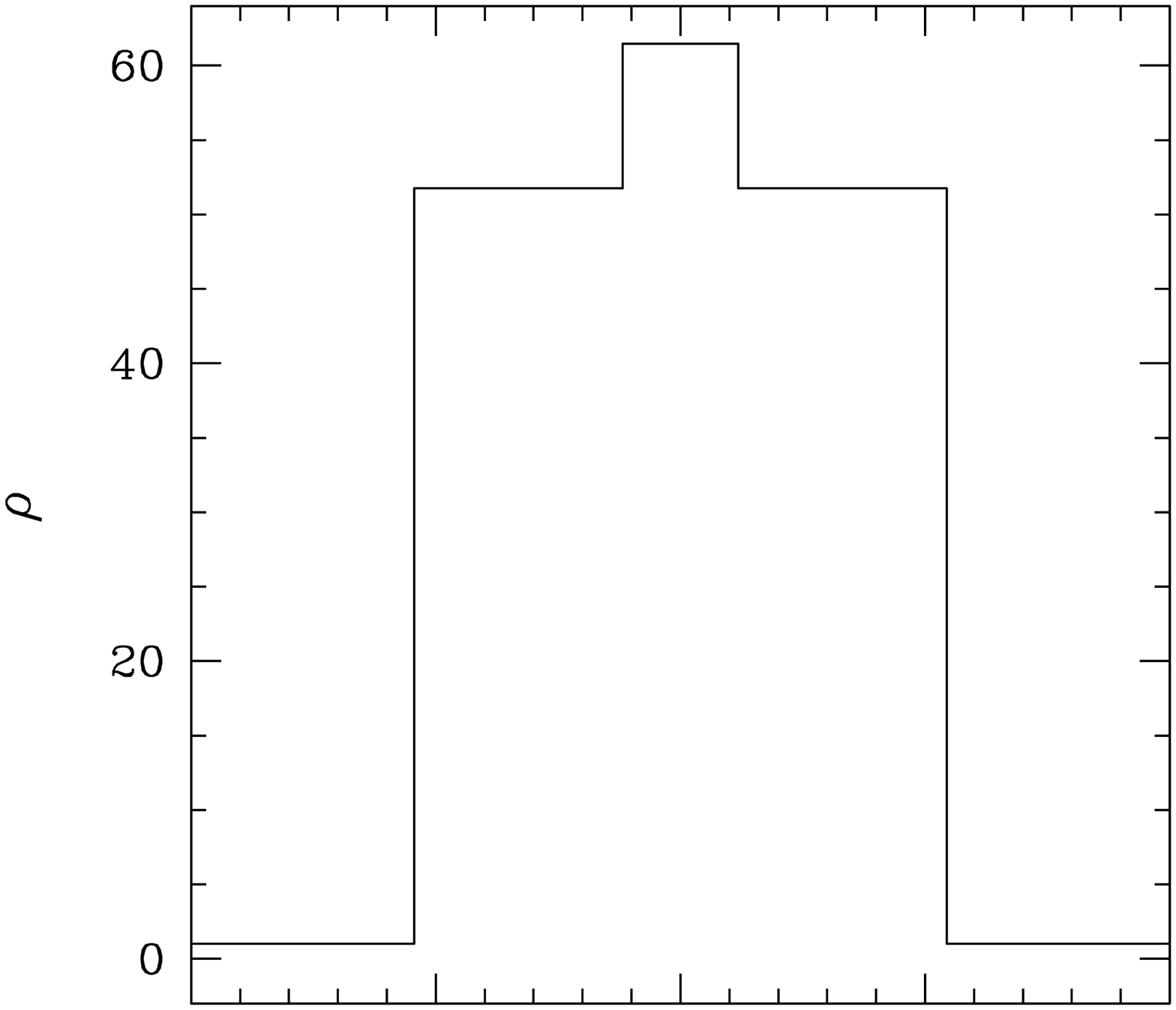}
      \hskip 0.75cm 
      \includegraphics[width=0.45\textwidth]{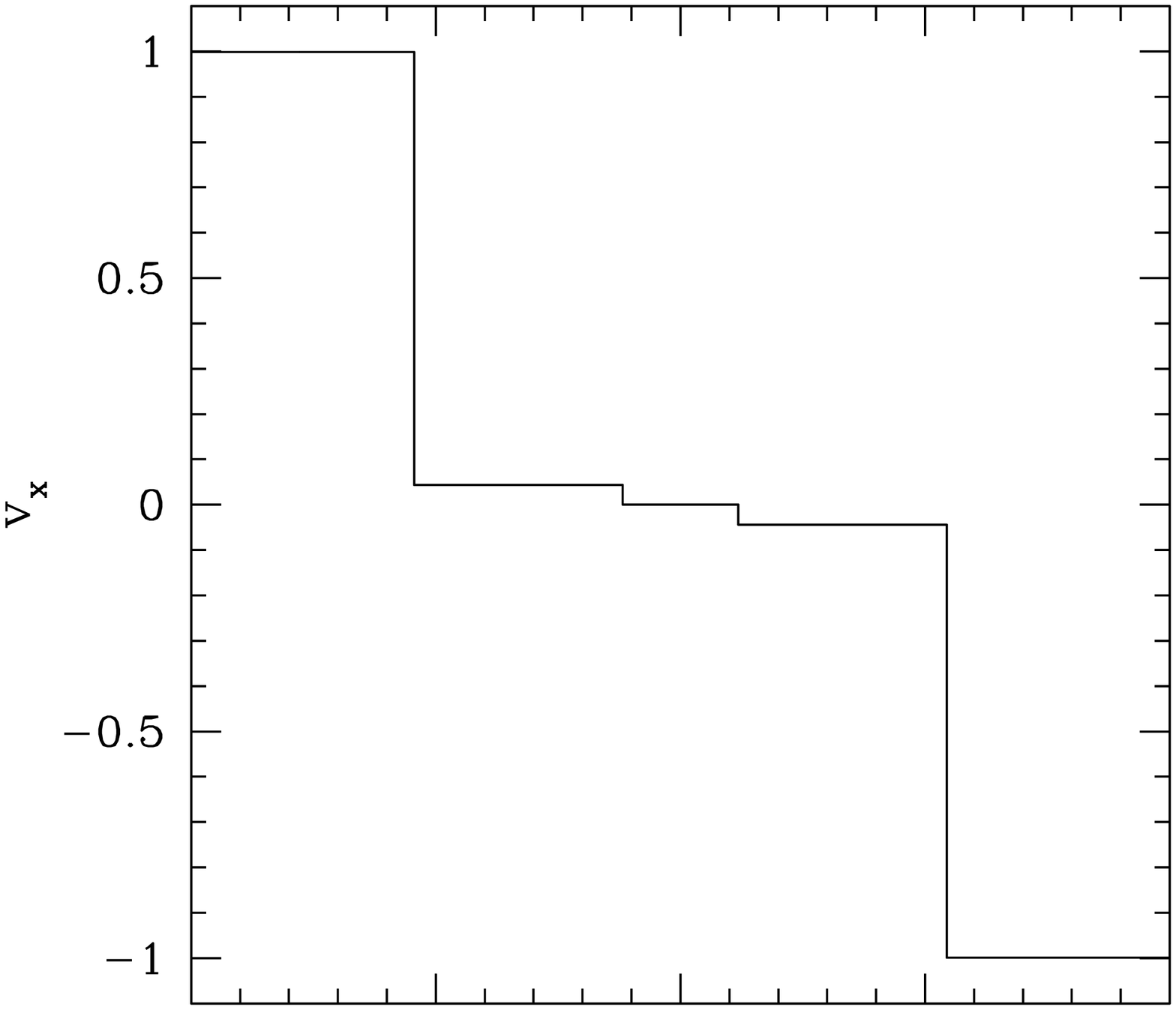}
      \vskip -1.0cm 
      \includegraphics[width=0.45\textwidth]{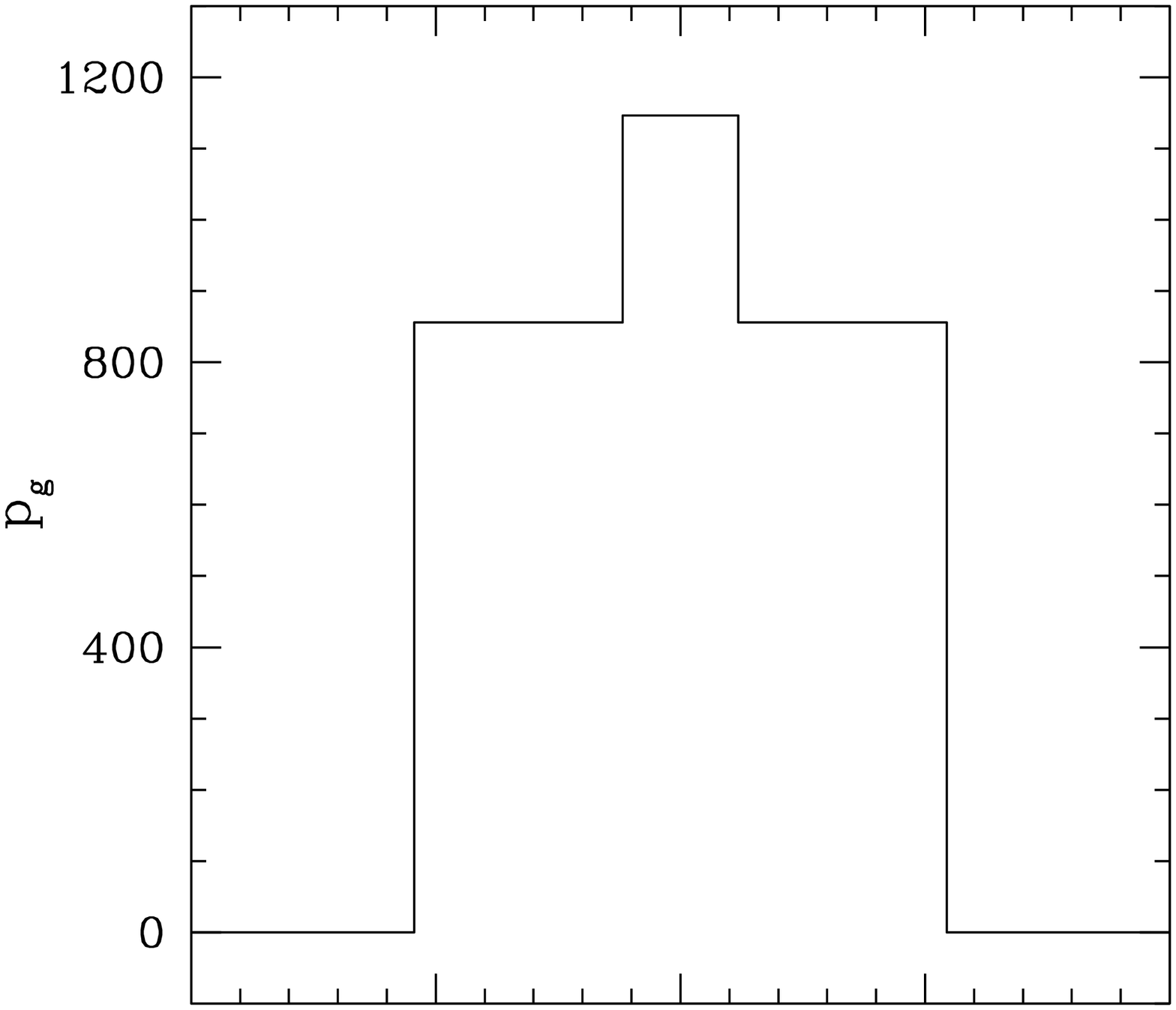}
      \hskip 0.75cm 
      \includegraphics[width=0.45\textwidth]{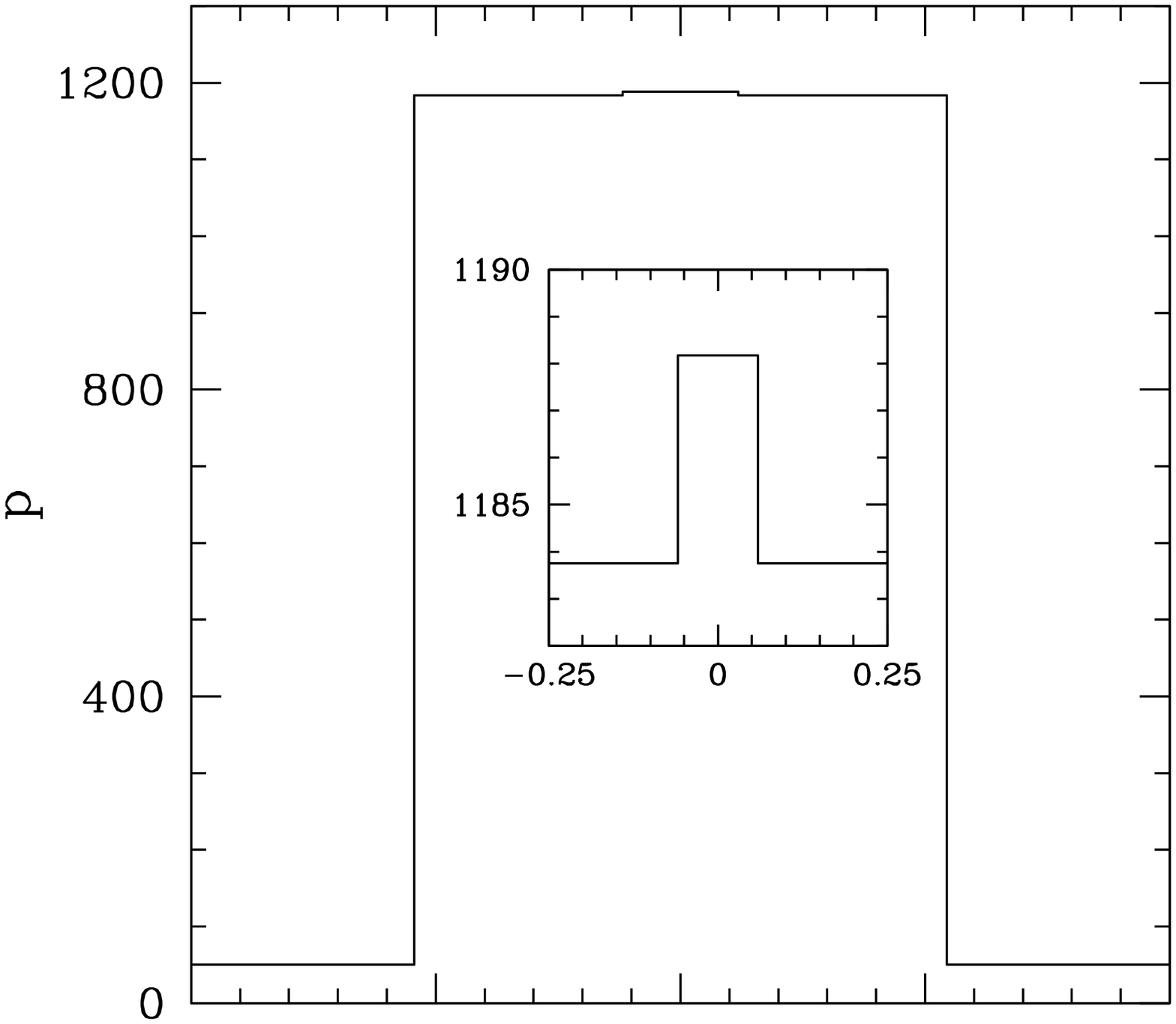}
      \vskip -1.0cm 
      \includegraphics[width=0.45\textwidth]{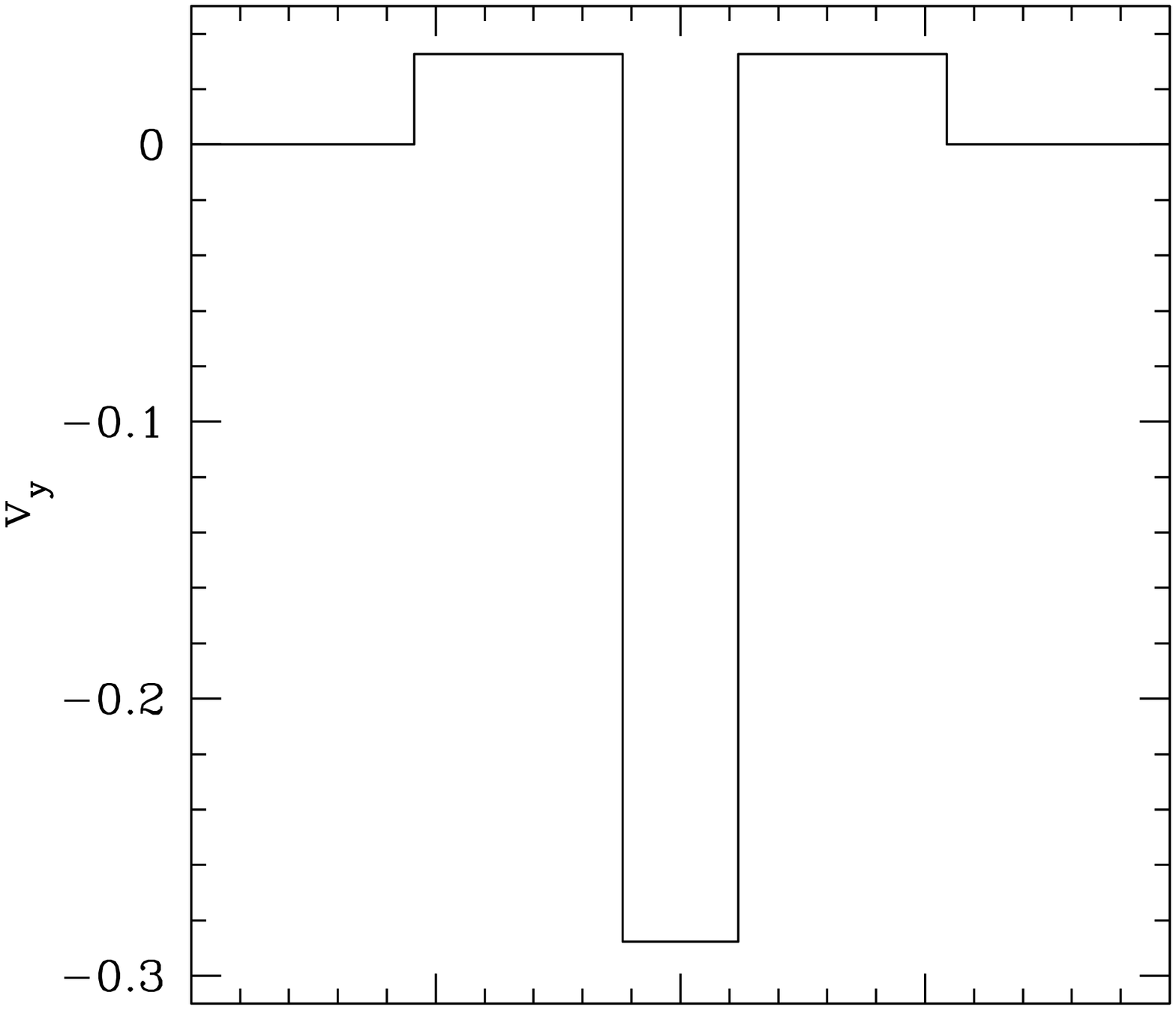}
      \hskip 0.75cm 
      \includegraphics[width=0.45\textwidth]{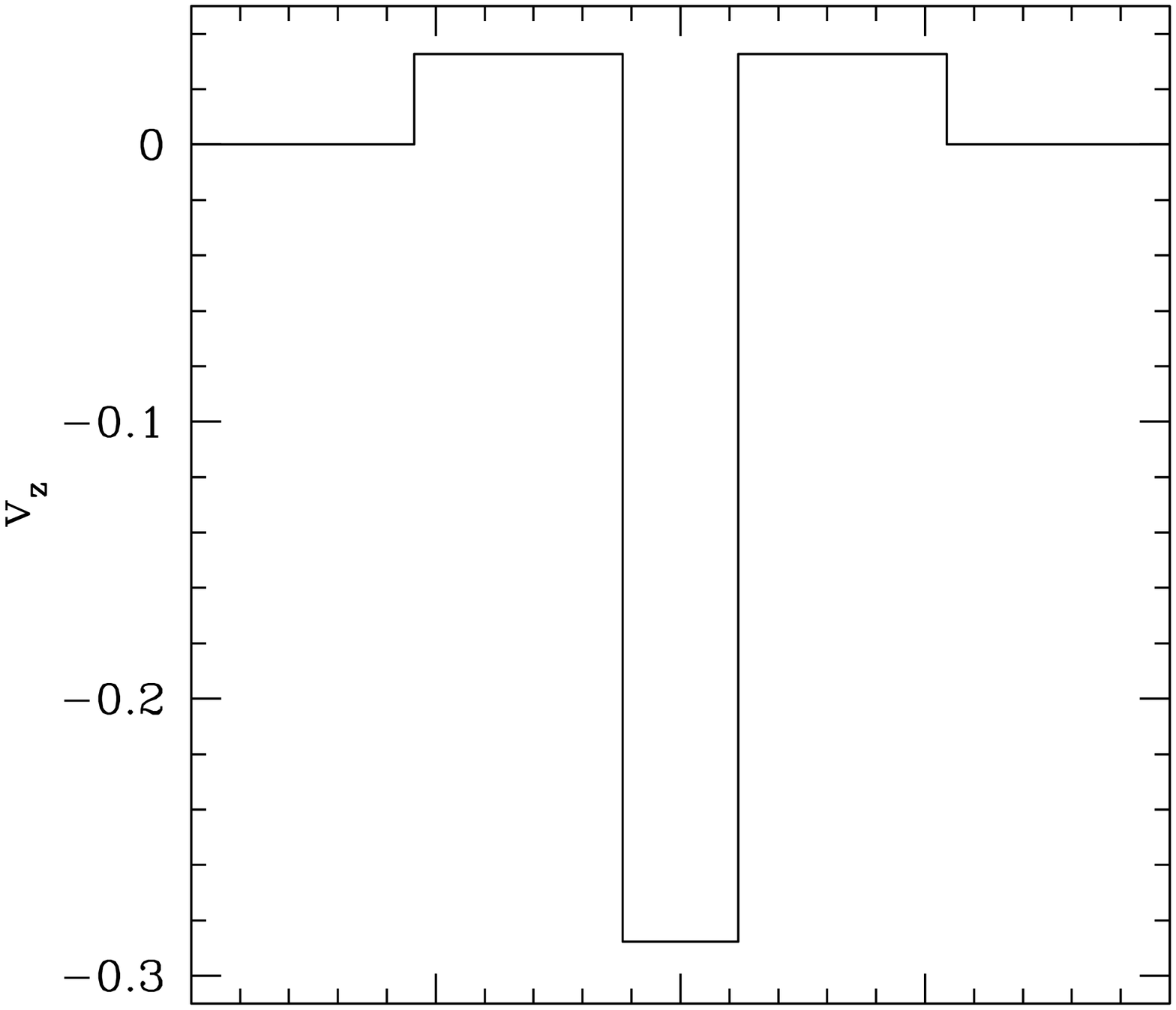}
      \vskip -1.0cm 
      \includegraphics[width=0.45\textwidth]{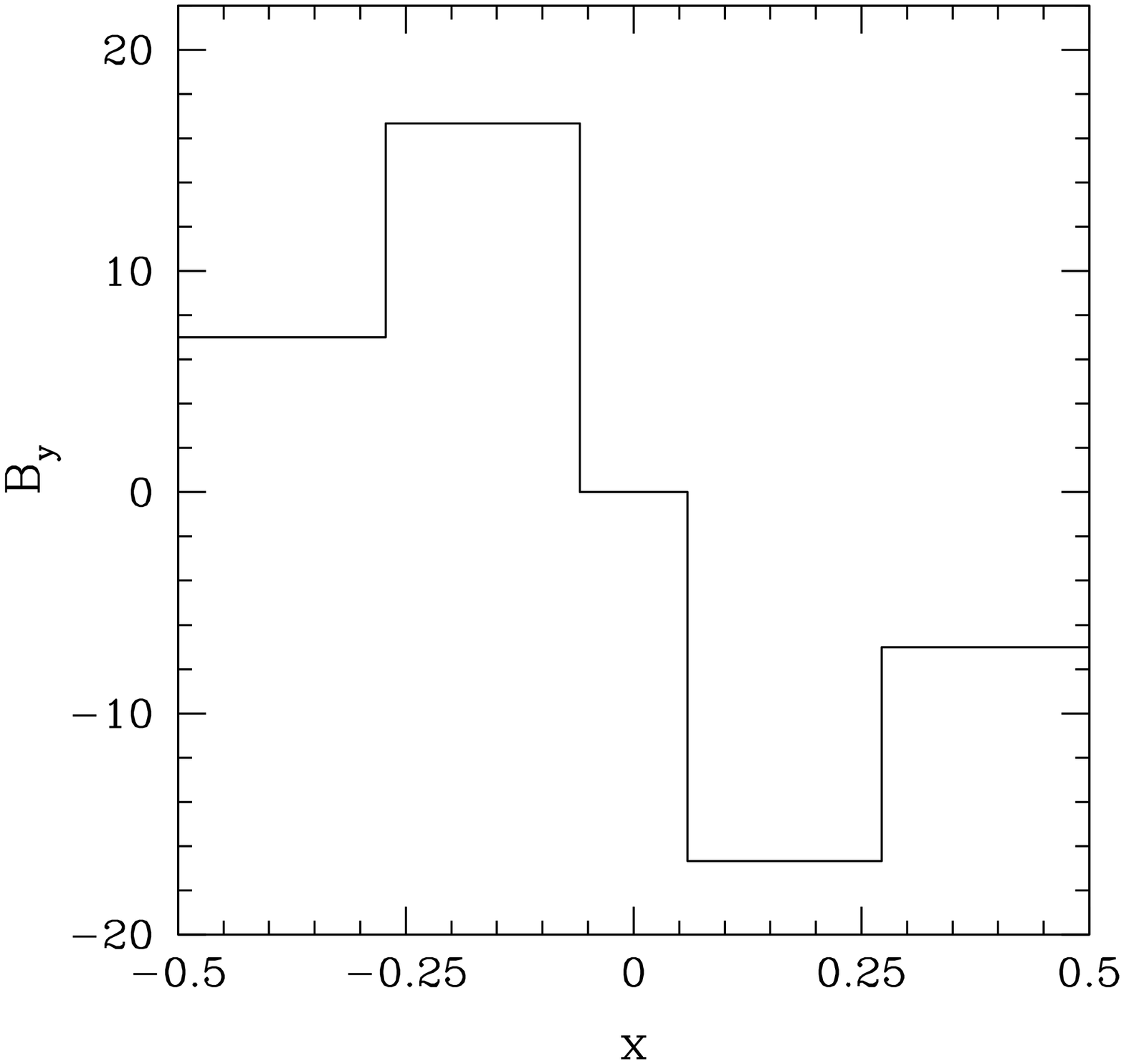}
      \hskip 0.75cm 
      \includegraphics[width=0.45\textwidth]{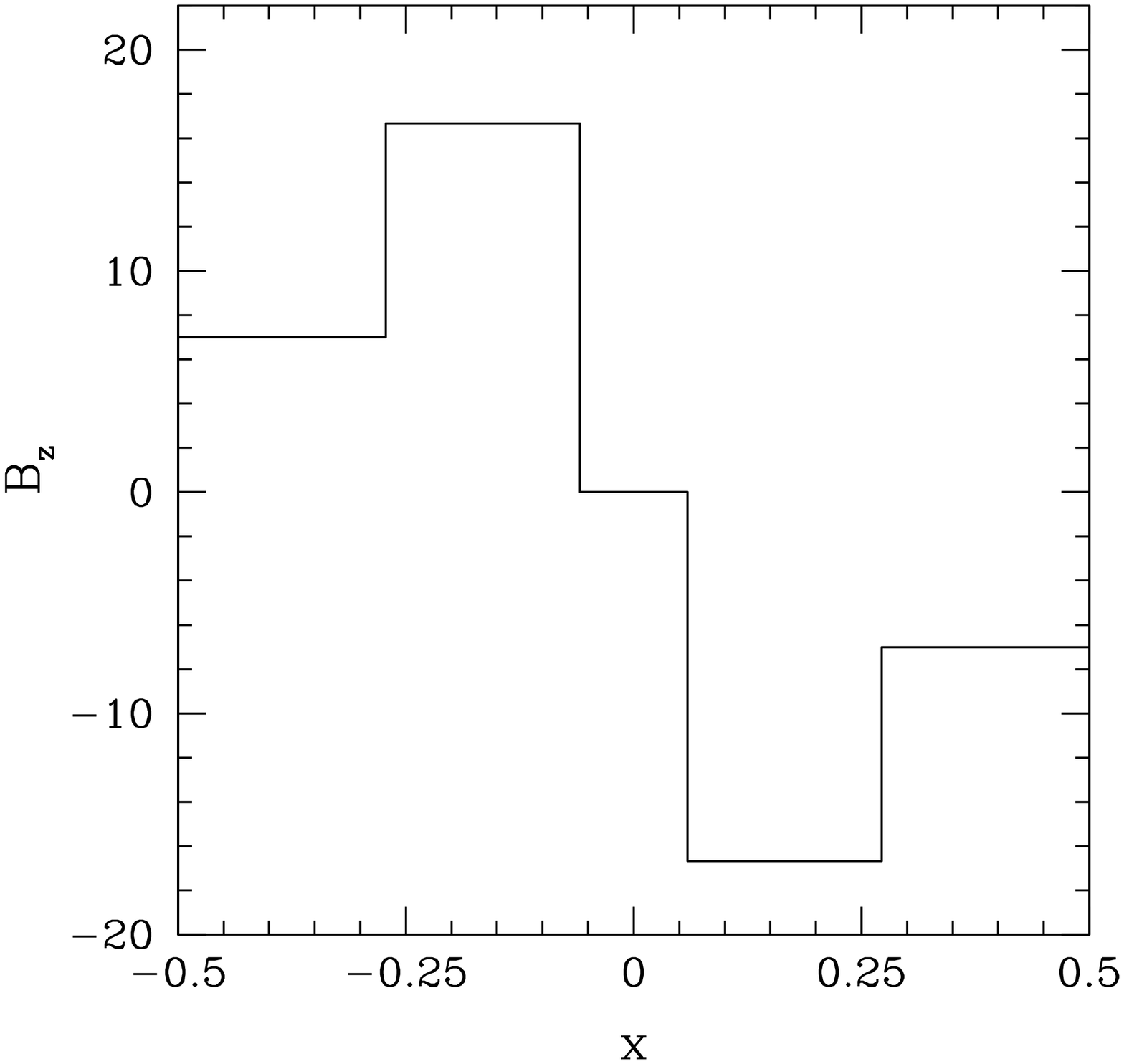}
  \caption{\label{BA4} Exact solution of the test number 4 of Balsara
    (2001) at time $t=0.4$. The solution is composed of two left-going
    fast and slow shocks and of two right-going fast and slow shocks.}
  \end{center}
\end{figure}

\begin{figure}
\begin{center}
      \includegraphics[width=0.45\textwidth]{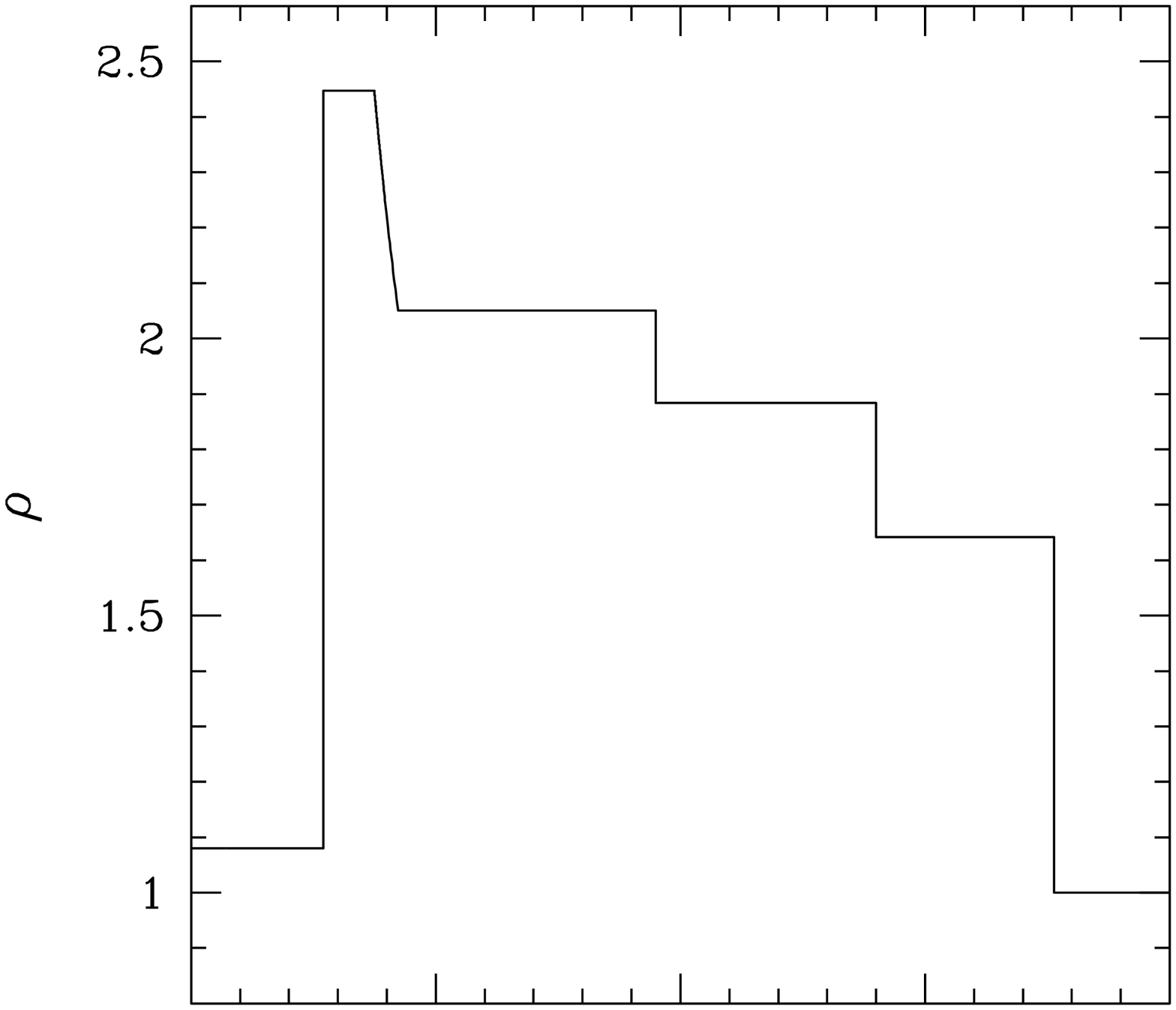}
      \hskip 0.75cm 
      \includegraphics[width=0.45\textwidth]{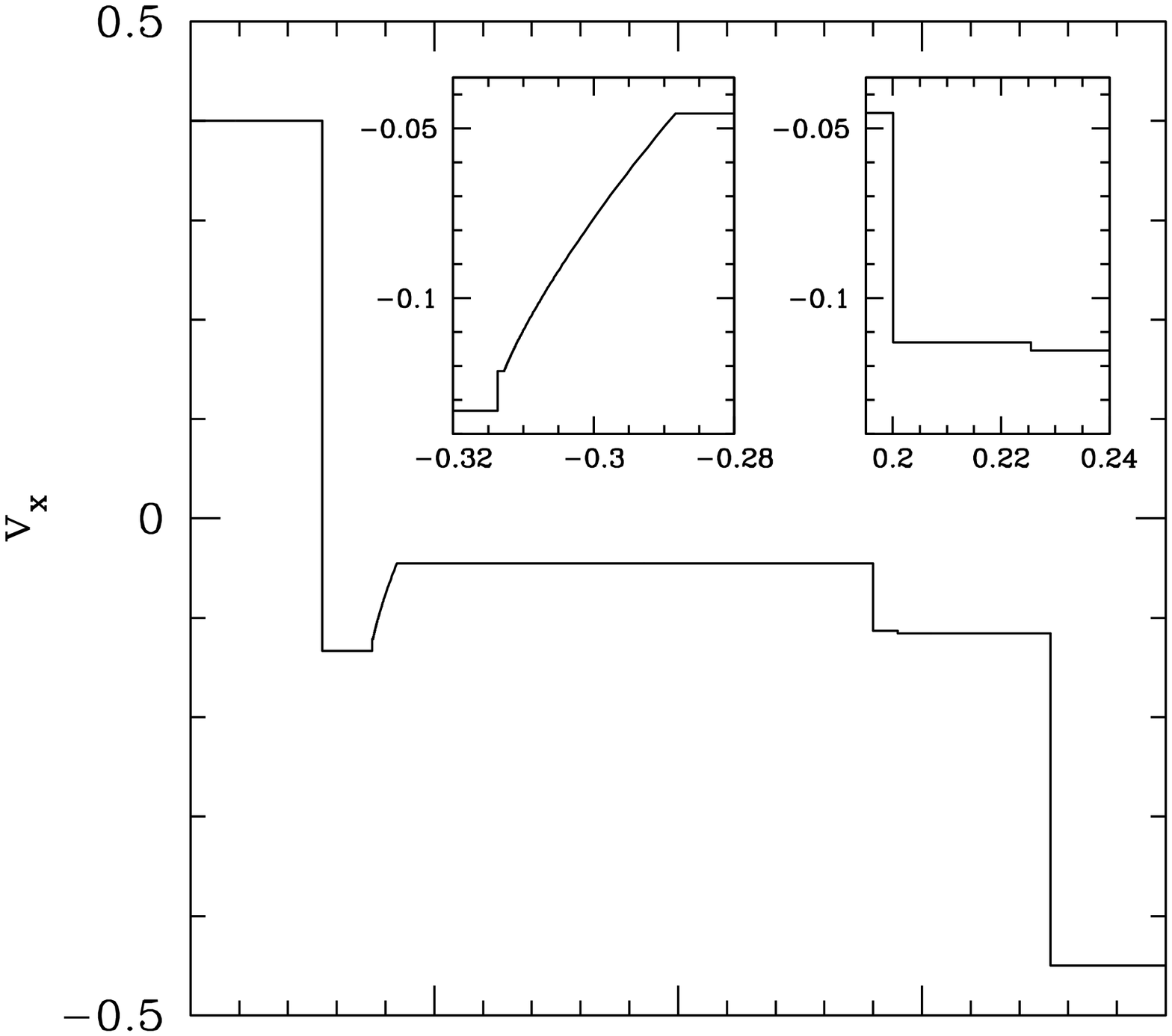}
      \vskip -1.0cm 
      \includegraphics[width=0.45\textwidth]{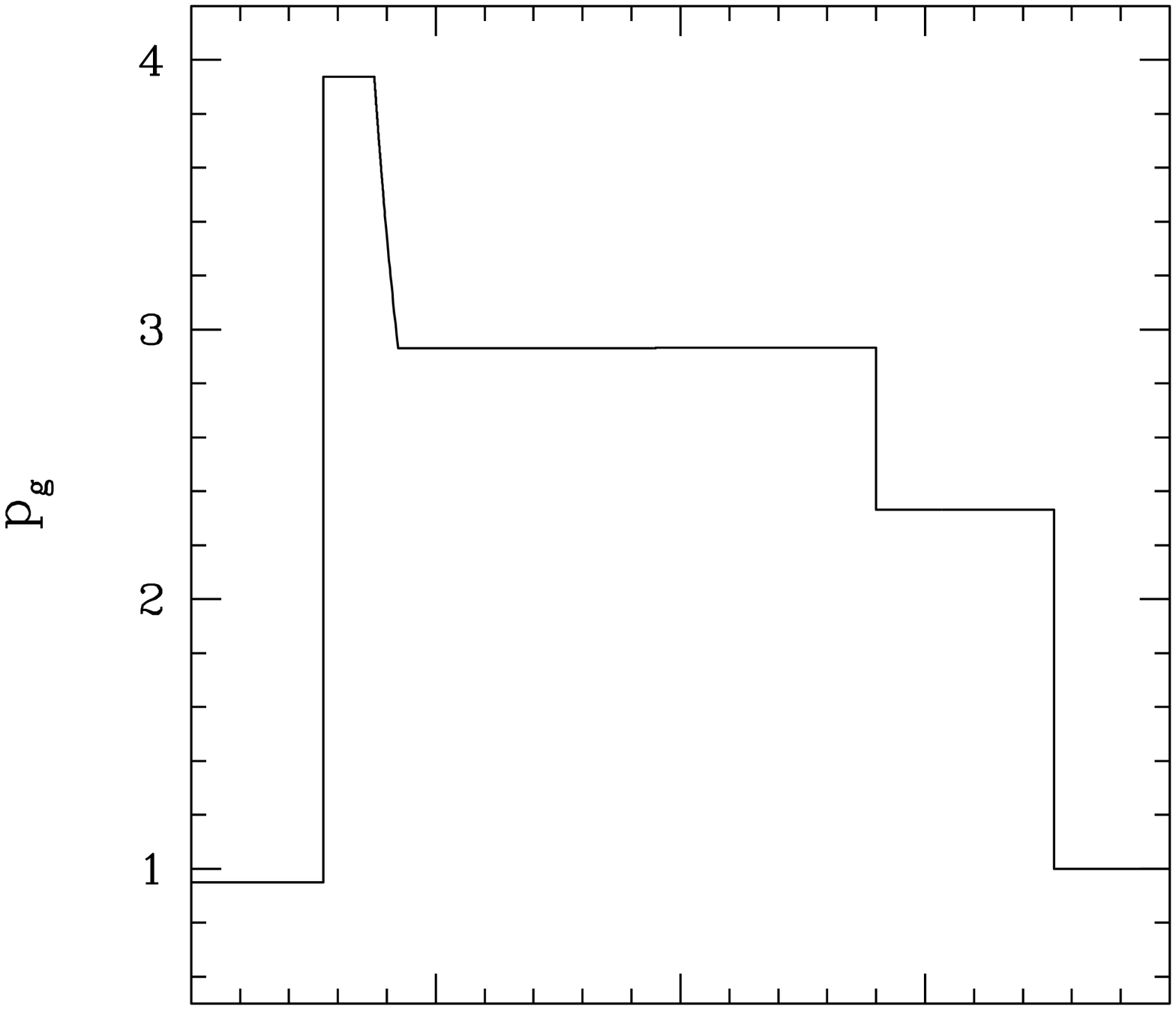}
      \hskip 0.75cm 
      \includegraphics[width=0.45\textwidth]{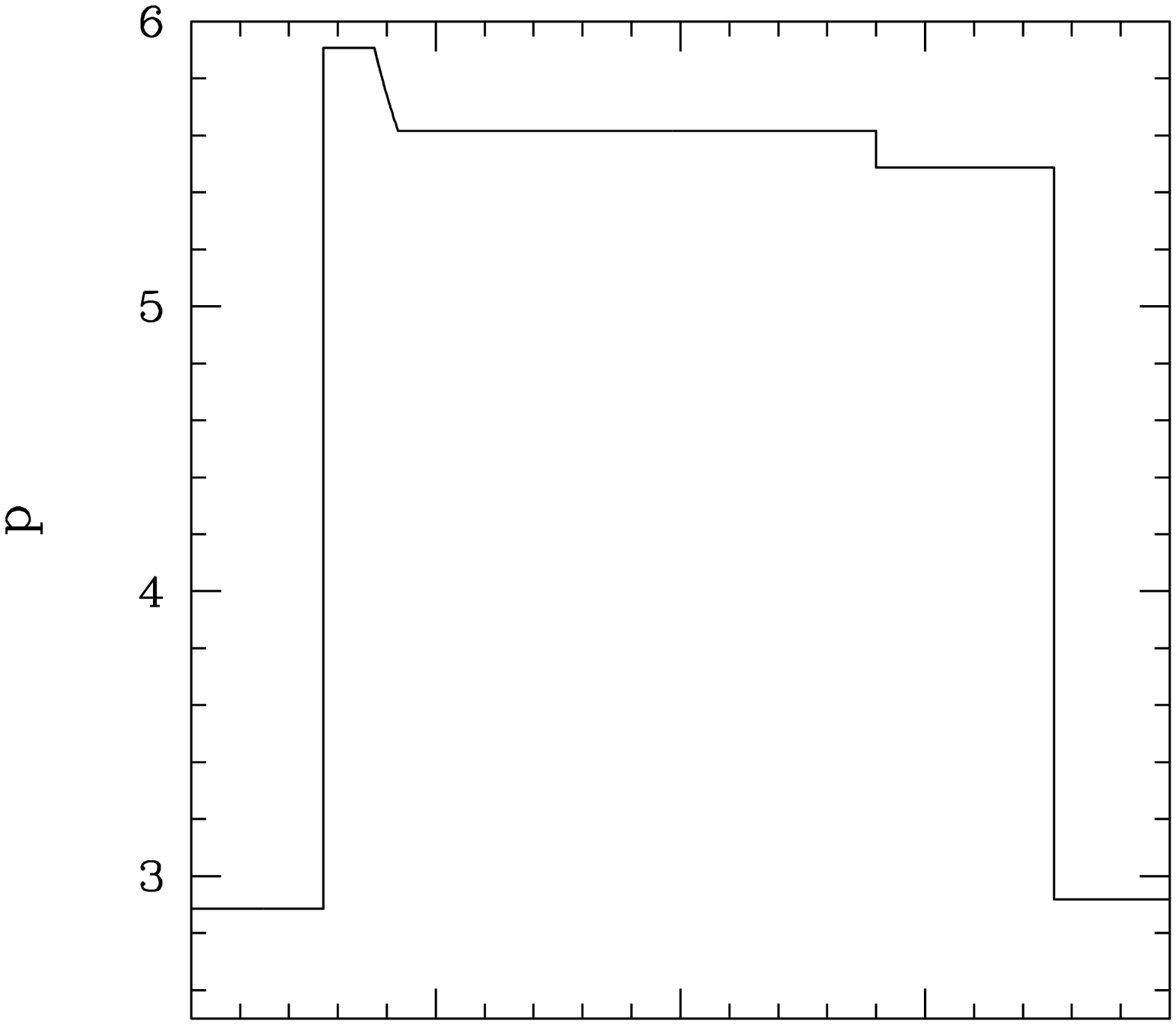}
      \vskip -1.0cm 
      \includegraphics[width=0.45\textwidth]{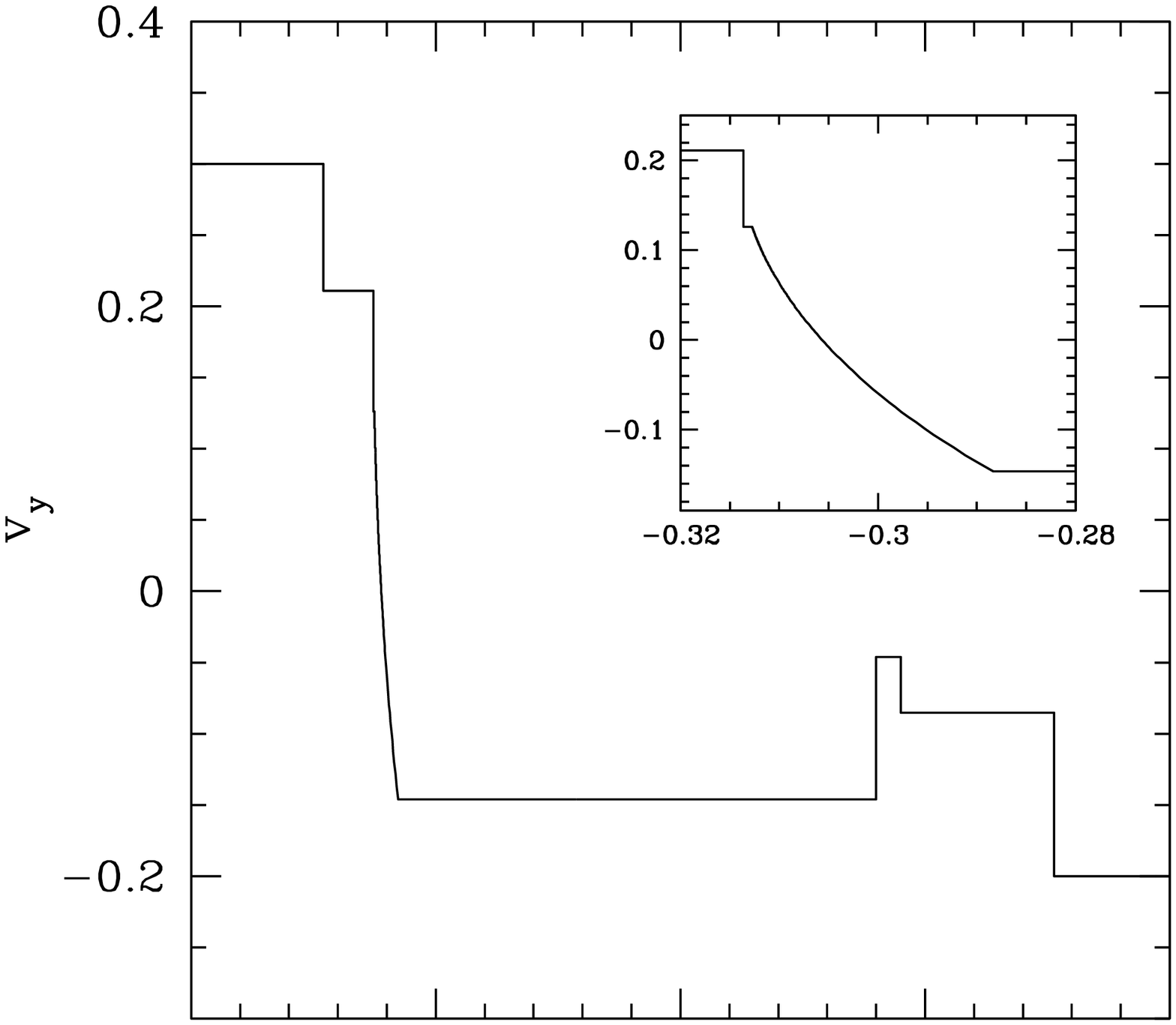}
      \hskip 0.75cm 
      \includegraphics[width=0.45\textwidth]{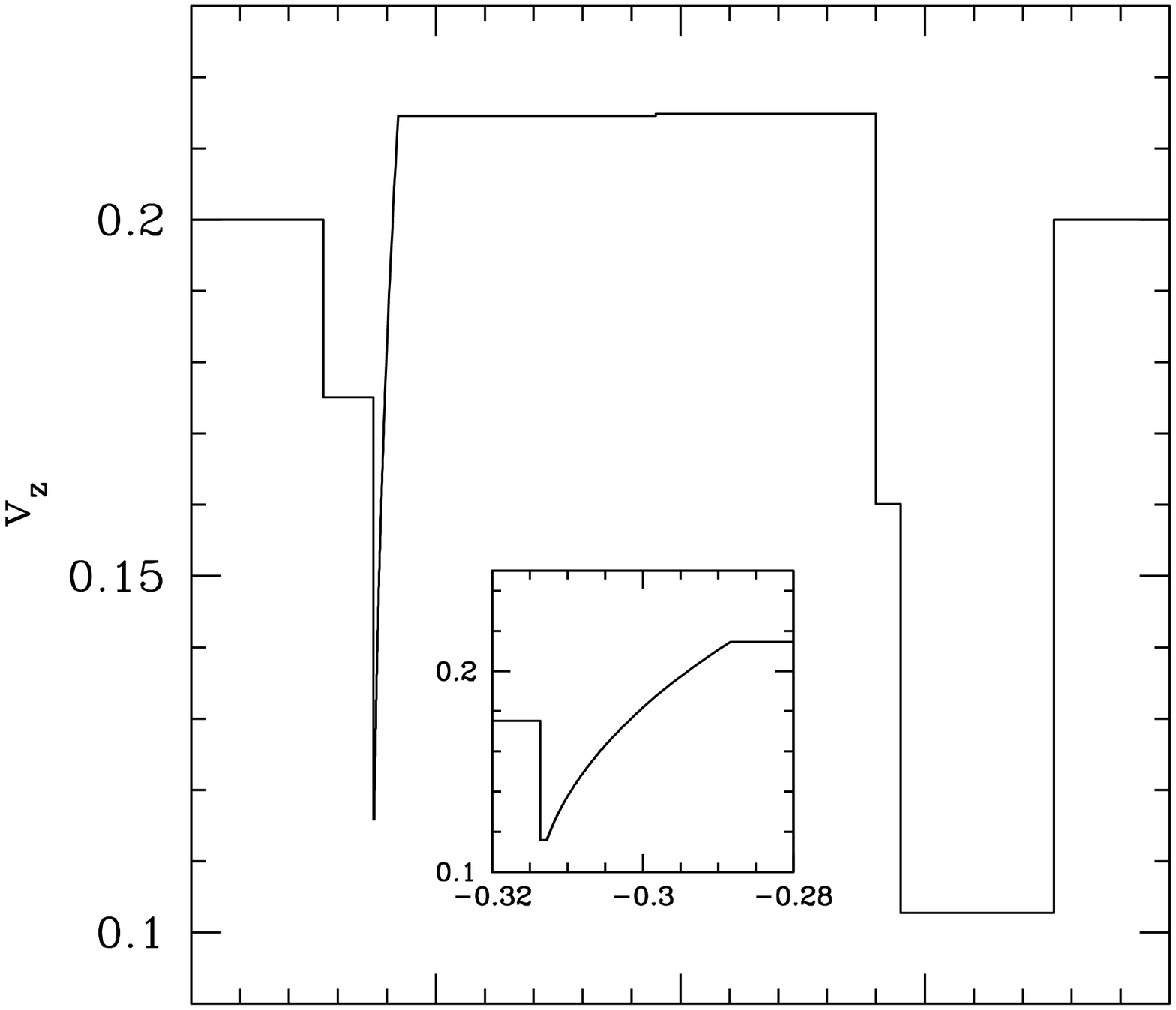}
      \vskip -1.0cm 
      \includegraphics[width=0.45\textwidth]{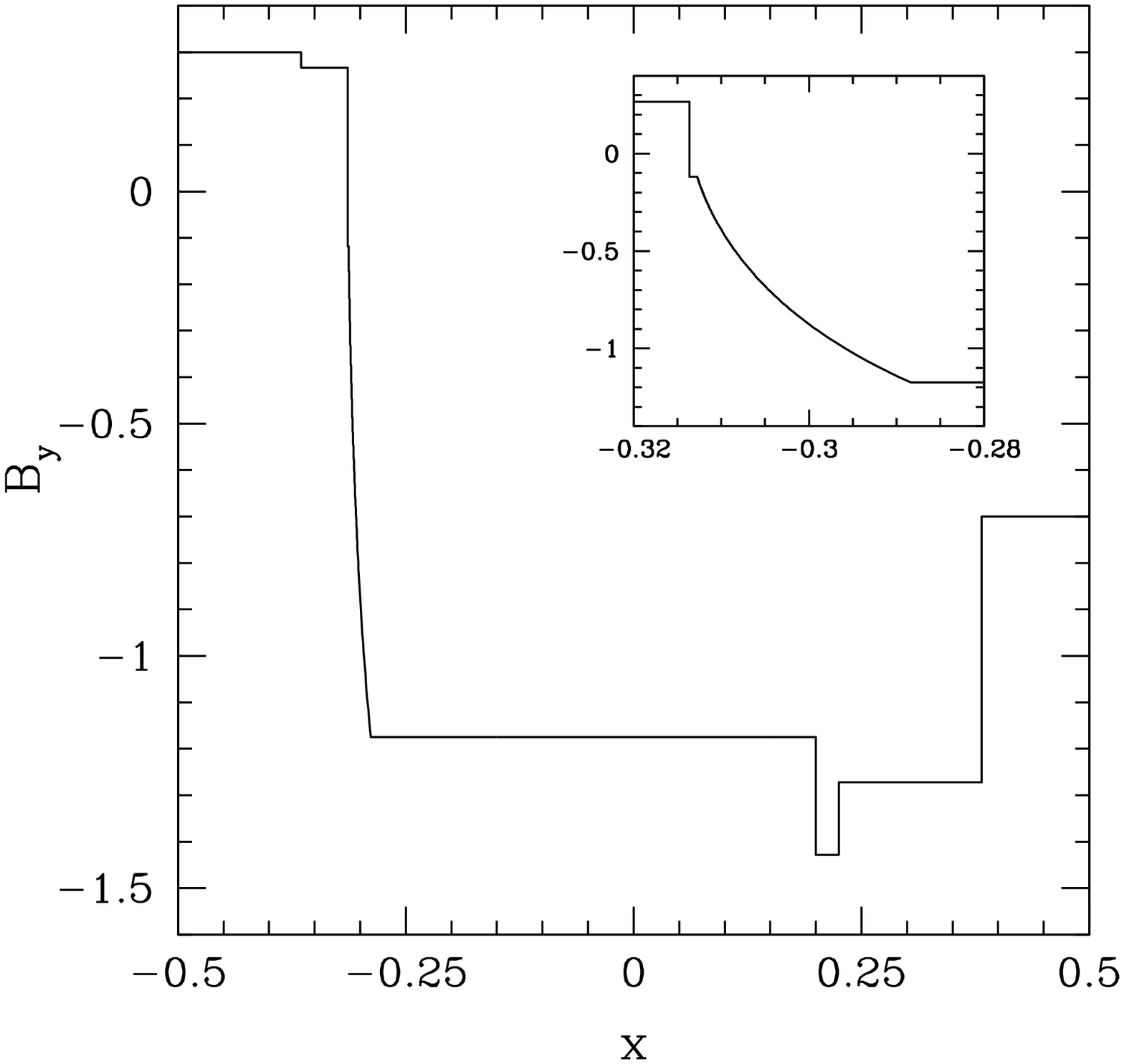}
      \hskip 0.75cm 
      \includegraphics[width=0.45\textwidth]{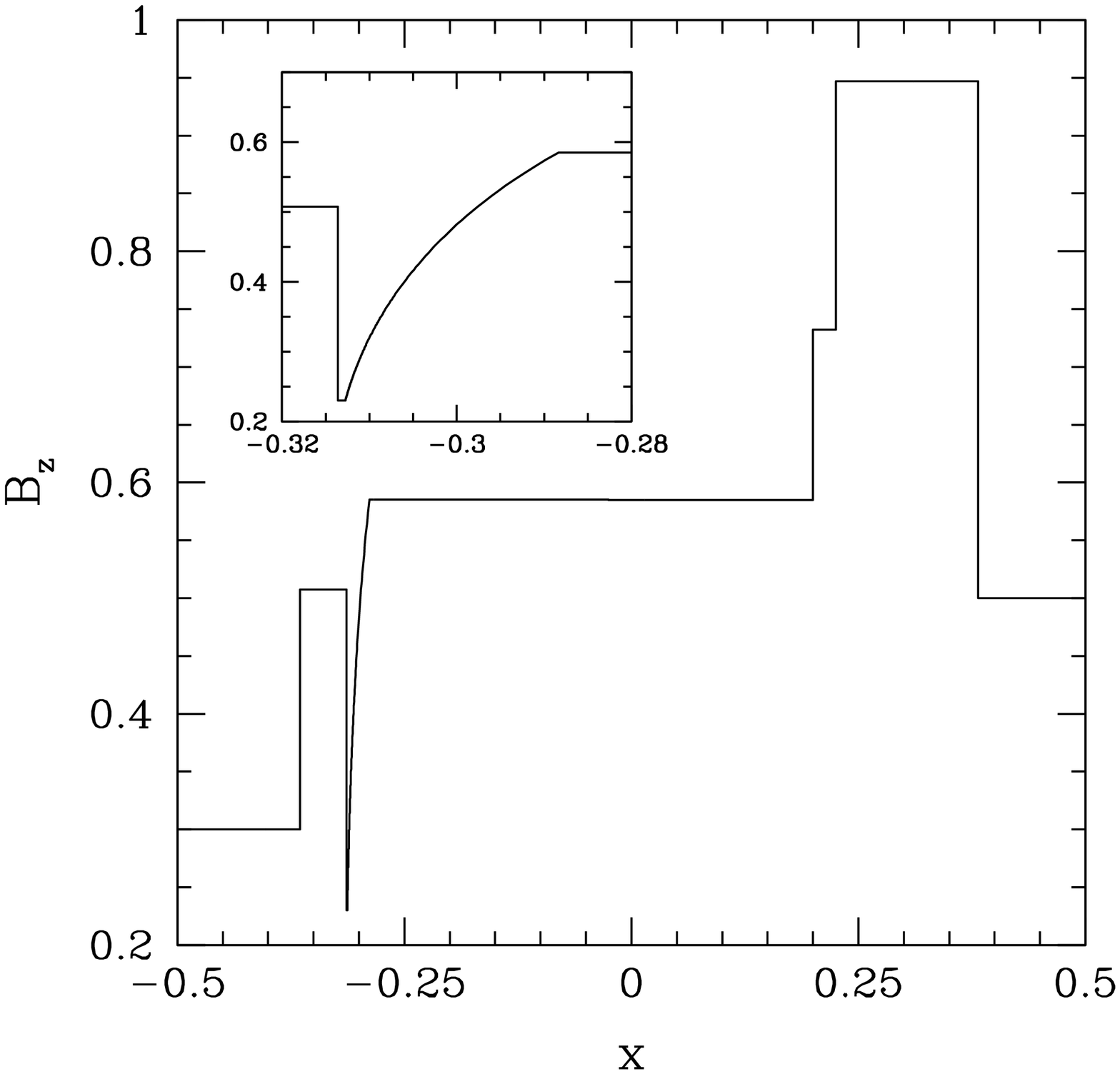}
  \caption{\label{BA5} Exact solution of the test number 5 of Balsara
  (2001) at time $t=0.55$. The solution is composed of a left-going fast
  shock, of a left-going Alfv\`en discontinuity, of a left-going slow
  rarefaction, of a contact discontinuity, of a right-going slow shock,
  of a right-going Alfv\`en discontinuity and of a right-going fast
  shock. Note that the accuracy in this test is only rather low: $3
  \times 10^{-4}$.}
  \end{center}
\end{figure}

\begin{figure}
\begin{center}
      \includegraphics[width=0.45\textwidth]{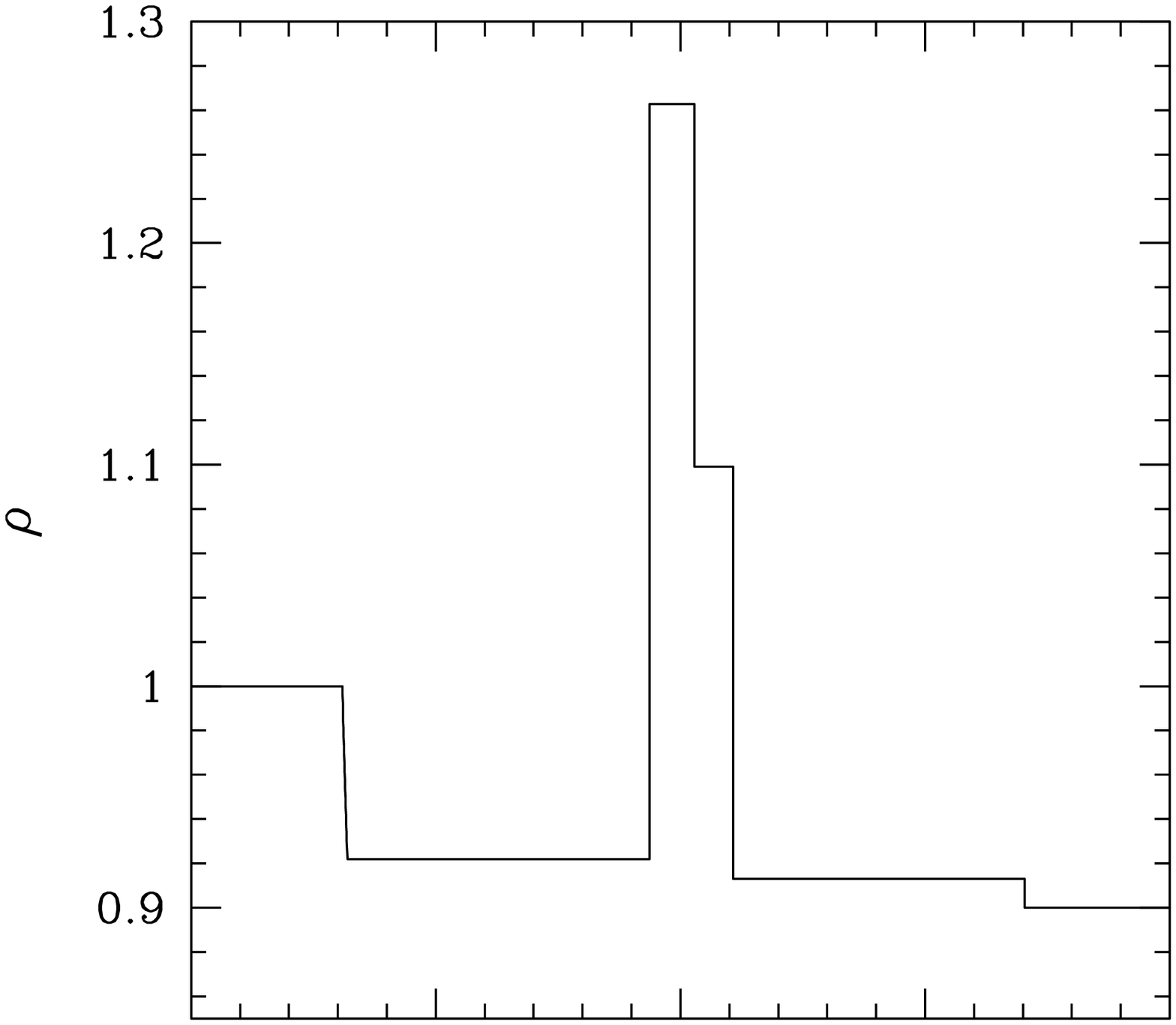}
      \hskip 0.75cm 
      \includegraphics[width=0.45\textwidth]{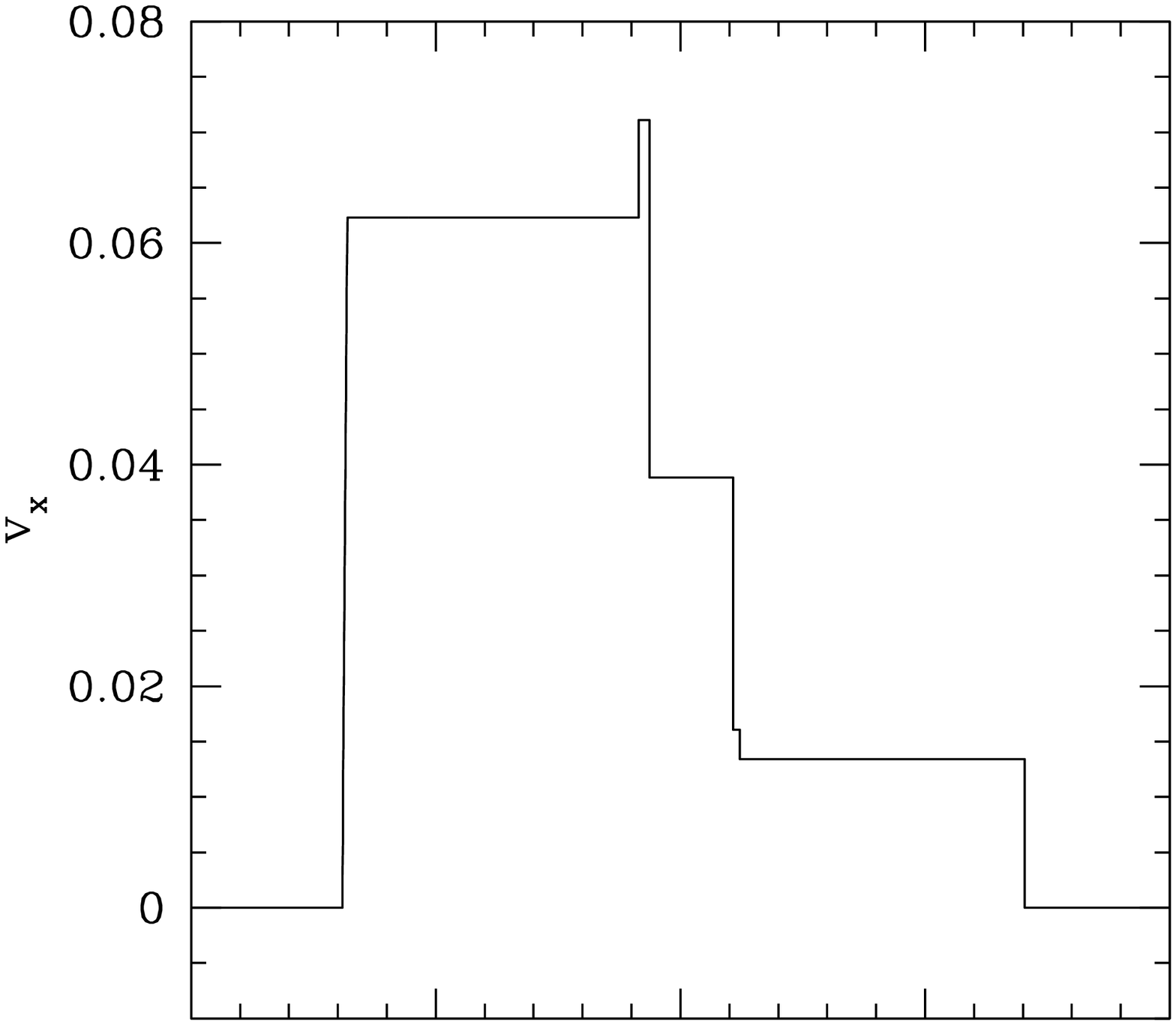}
      \vskip -1.0cm 
      \includegraphics[width=0.45\textwidth]{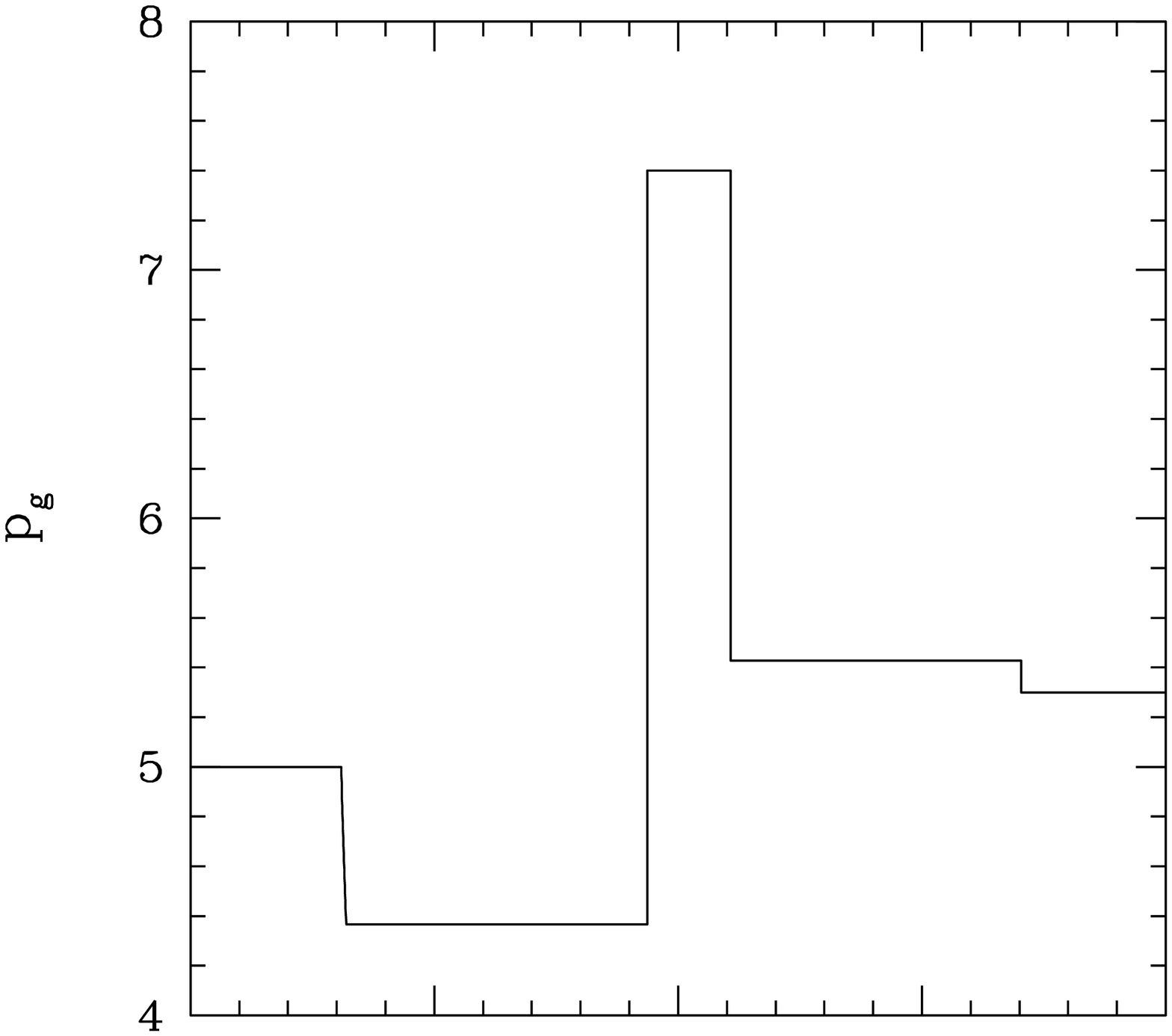}
      \hskip 0.75cm 
      \includegraphics[width=0.45\textwidth]{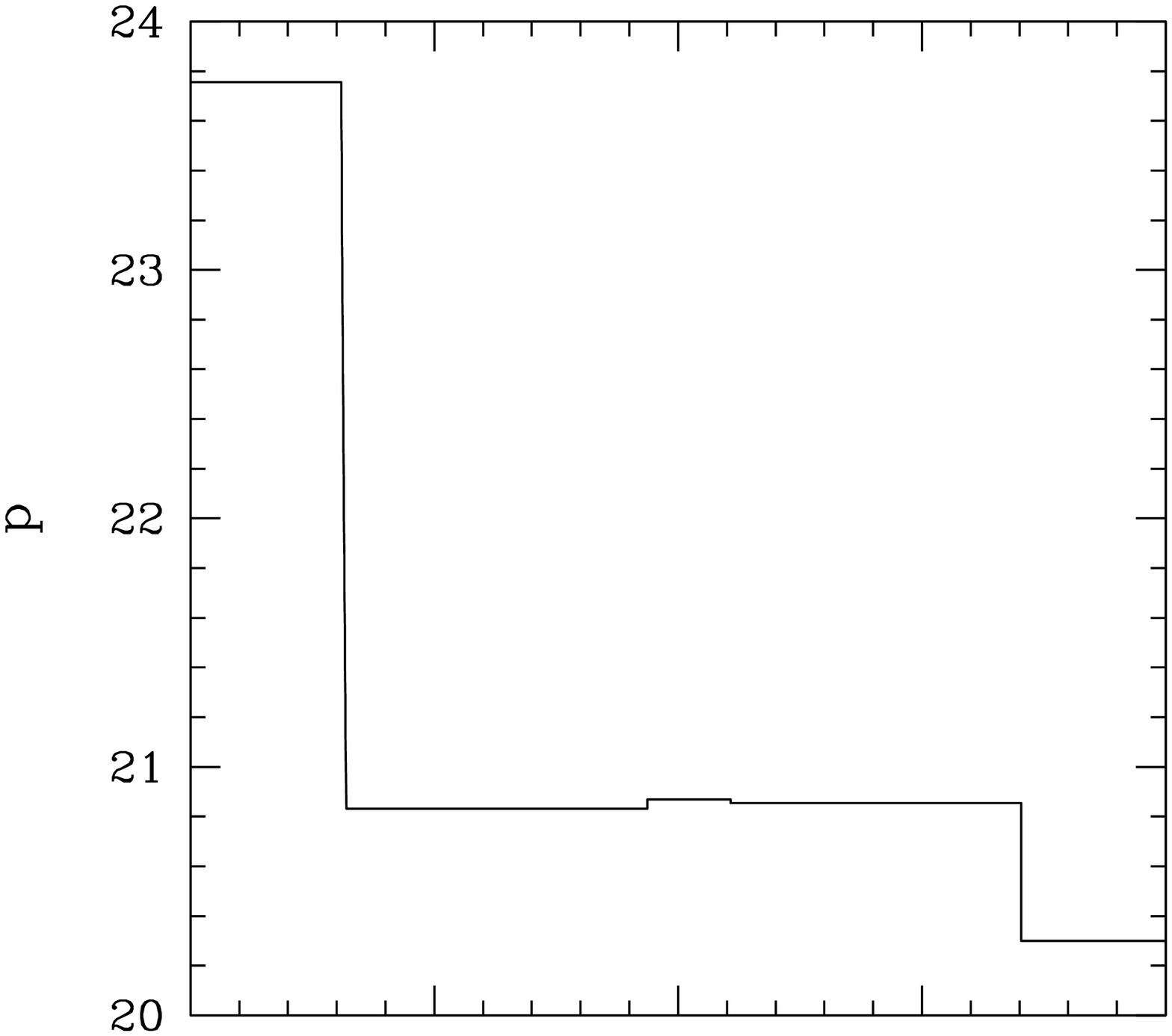}
      \vskip -1.0cm 
      \includegraphics[width=0.45\textwidth]{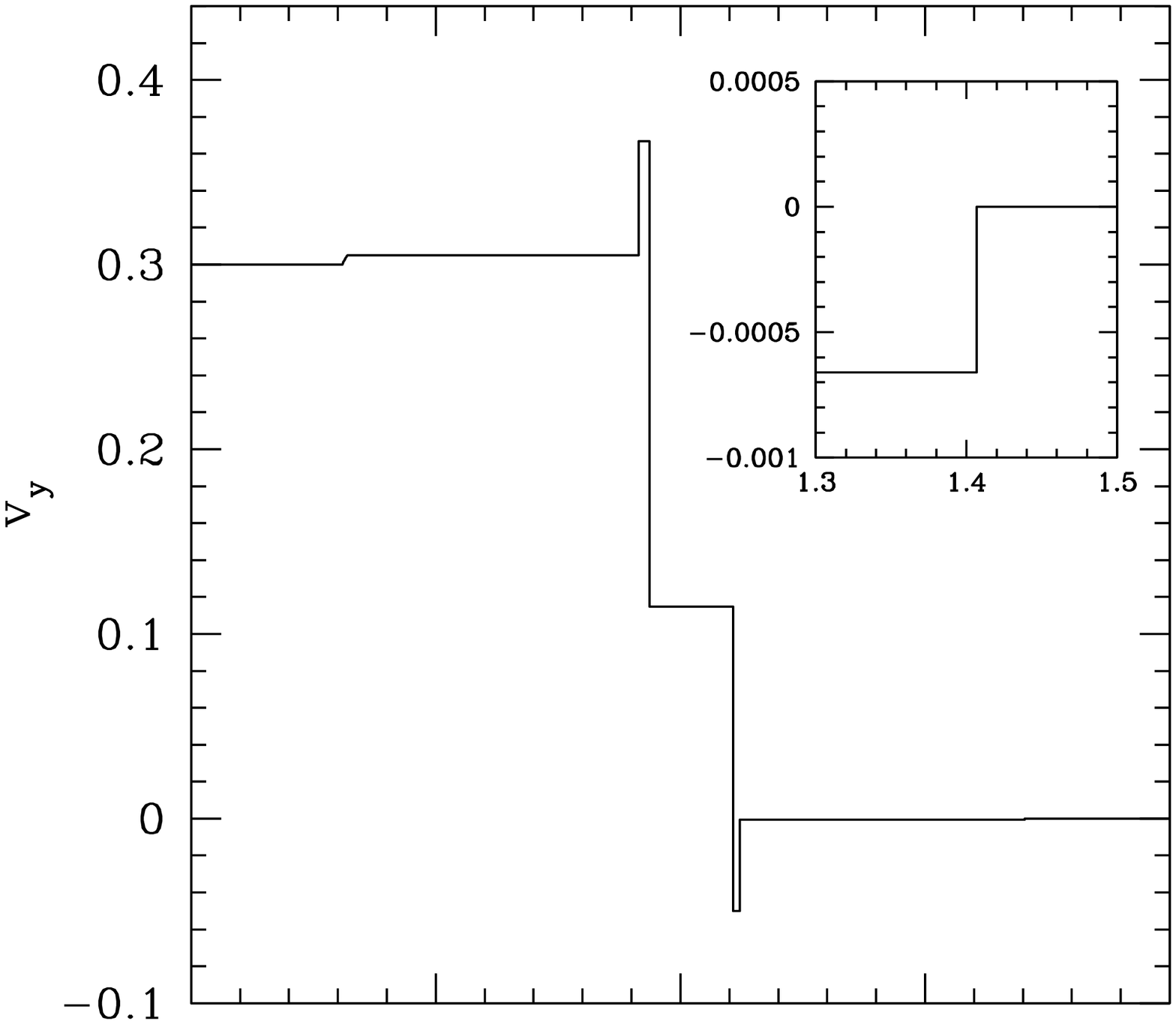}
      \hskip 0.75cm 
      \includegraphics[width=0.45\textwidth]{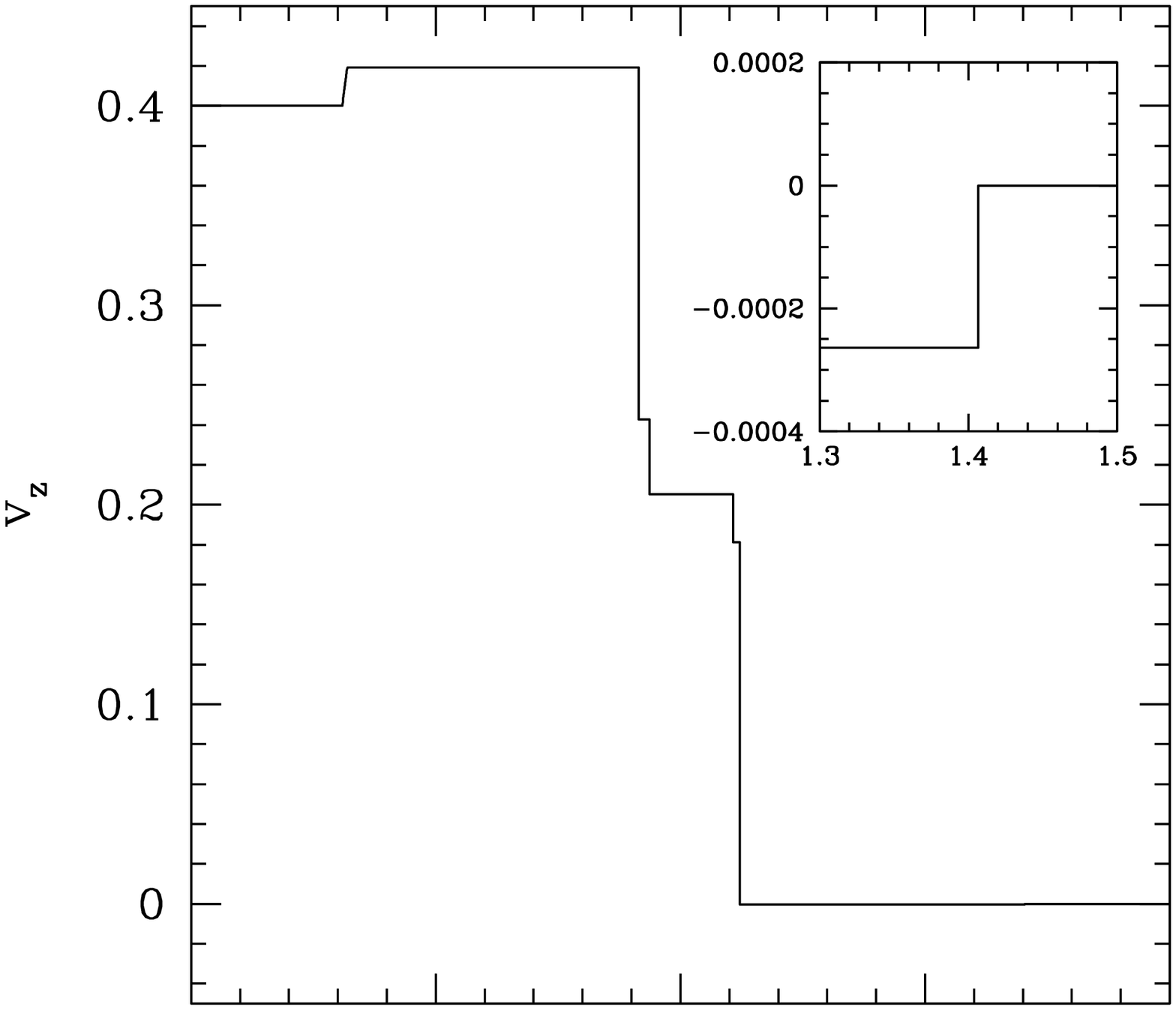}
      \vskip -1.0cm 
      \includegraphics[width=0.45\textwidth]{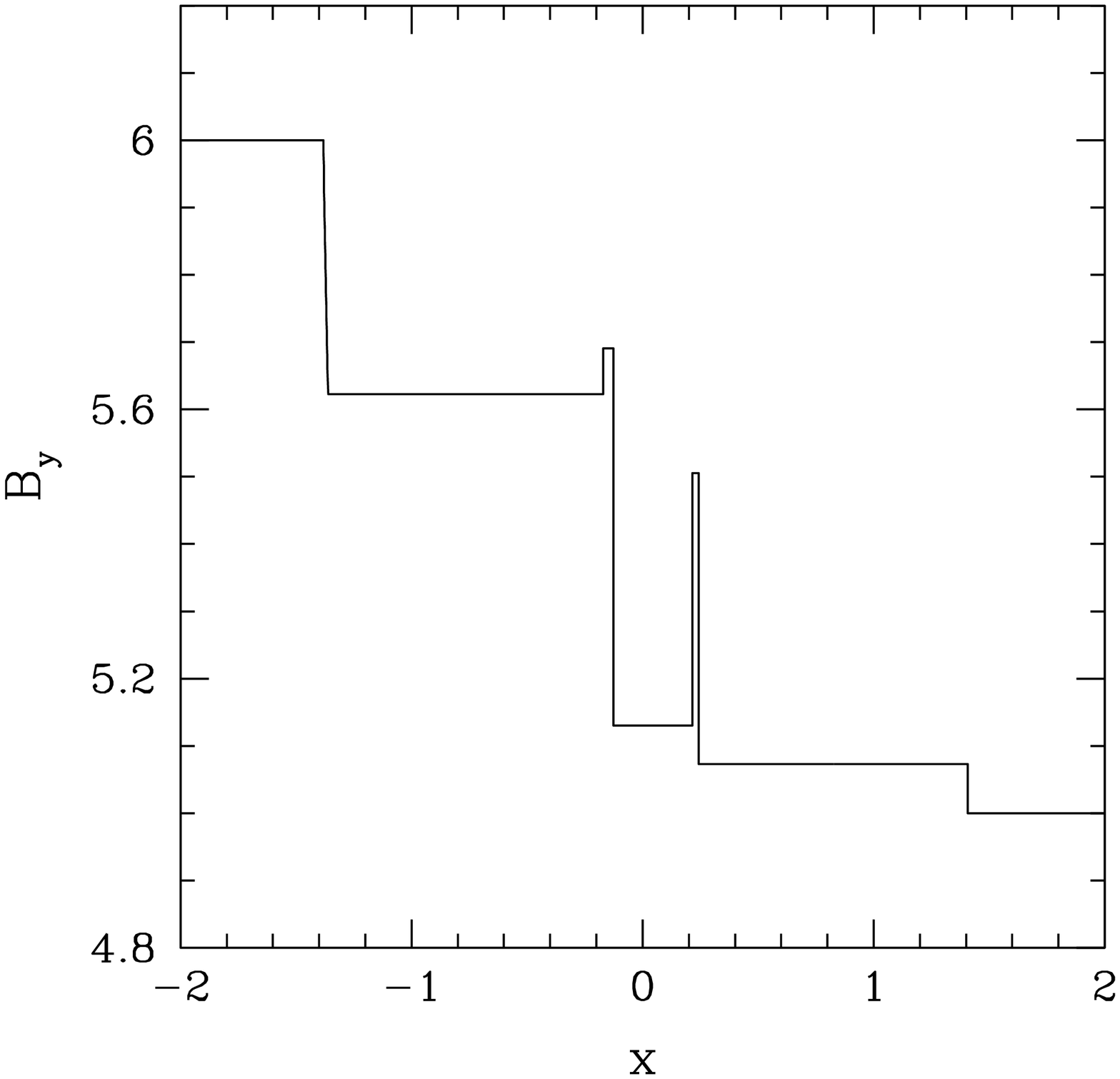}
      \hskip 0.75cm 
      \includegraphics[width=0.45\textwidth]{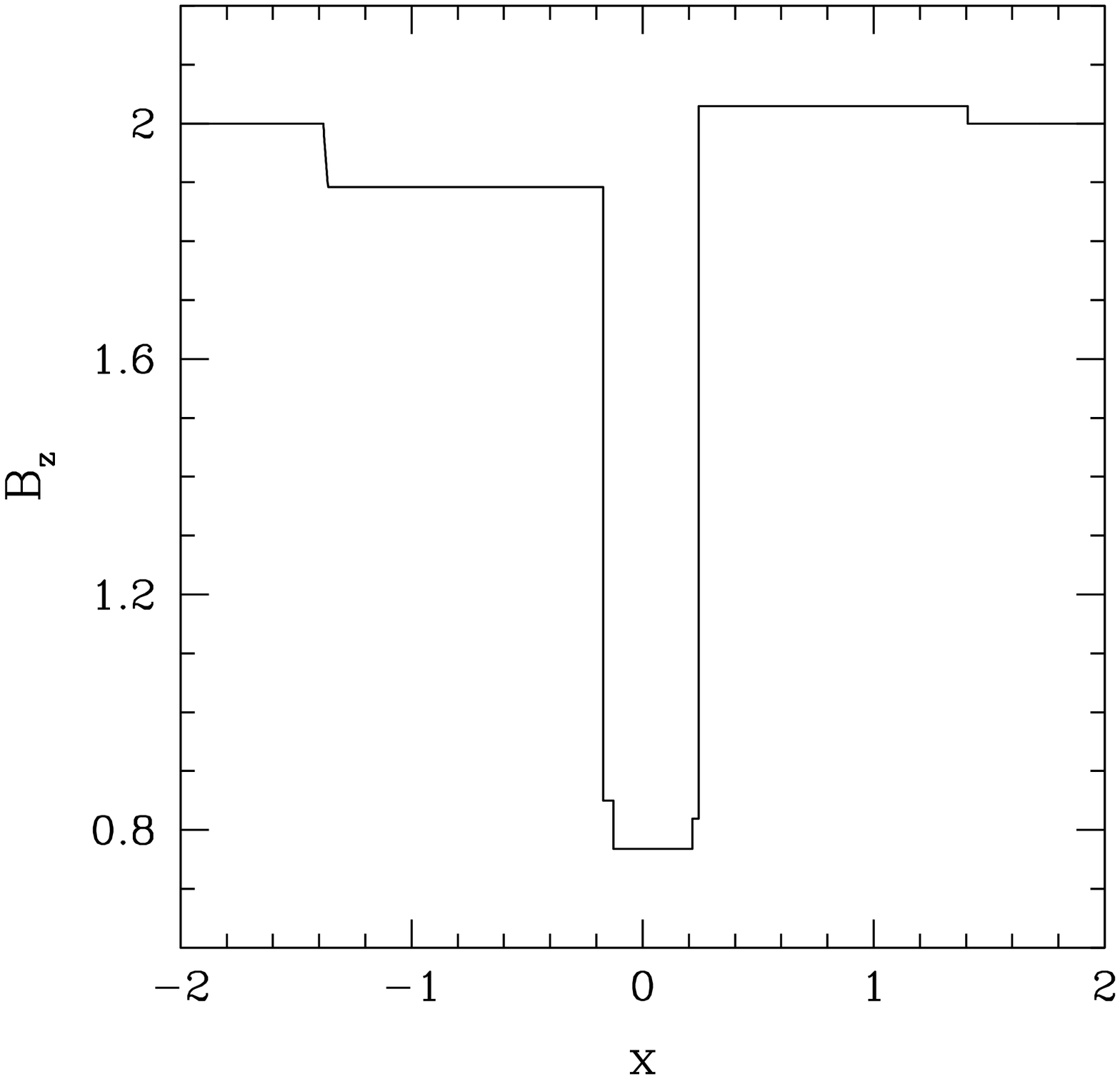}
  \caption{\label{BG_AL1} Exact solution of the generic Alfv\`en test
  at time $t=1.5$. The solution is composed of a left-going fast
  rarefaction, of a left-going Alfv\`en discontinuity, of a left-going
  slow shock, of a contact discontinuity, of a right-going slow shock,
  of a right-going Alfv\`en discontinuity and of a right-going fast
  shock.}
  \end{center}
\end{figure}

\begin{figure}
\begin{center}
      \includegraphics[width=0.45\textwidth]{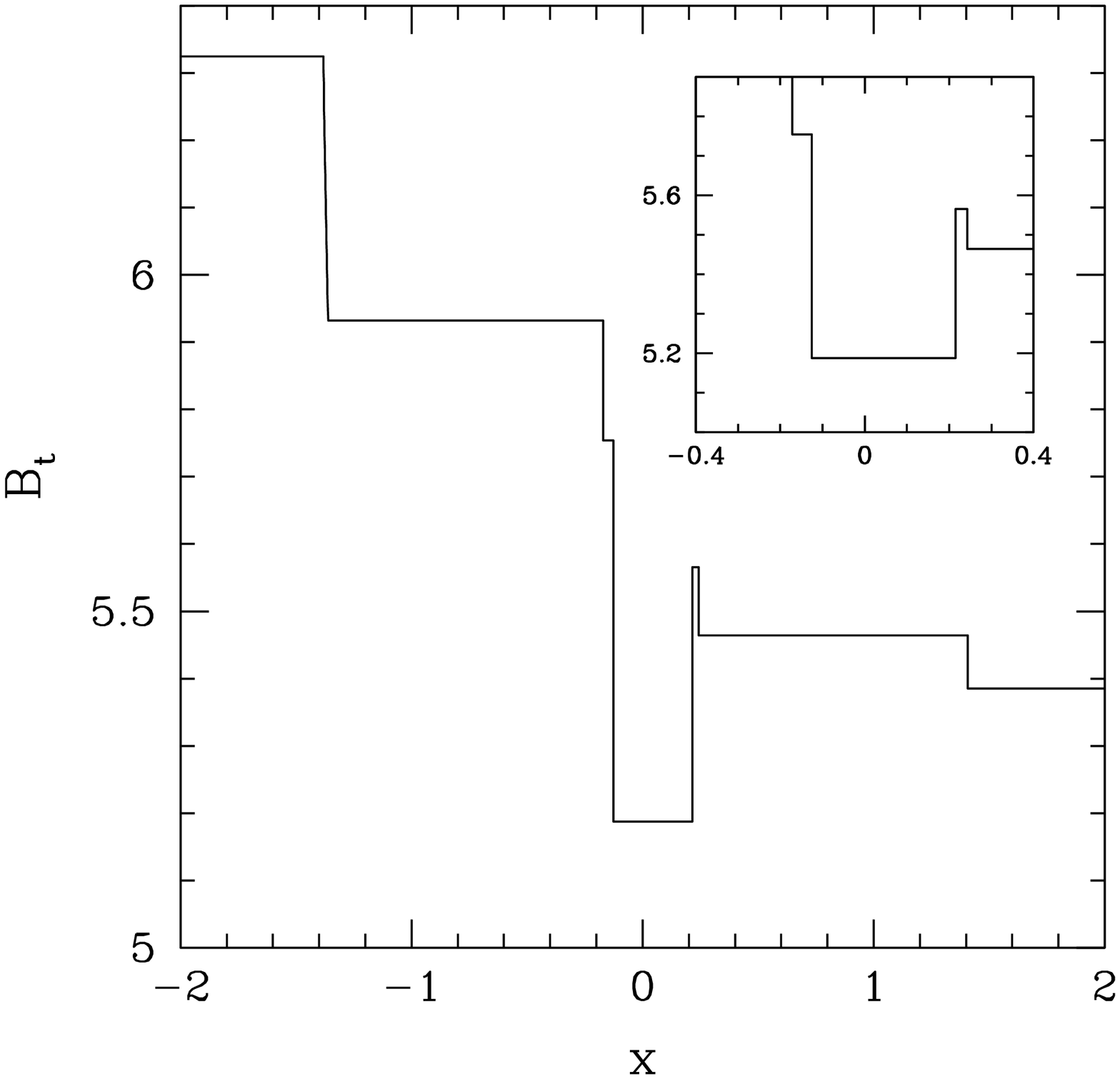}
      \hskip 0.75cm 
      \includegraphics[width=0.45\textwidth]{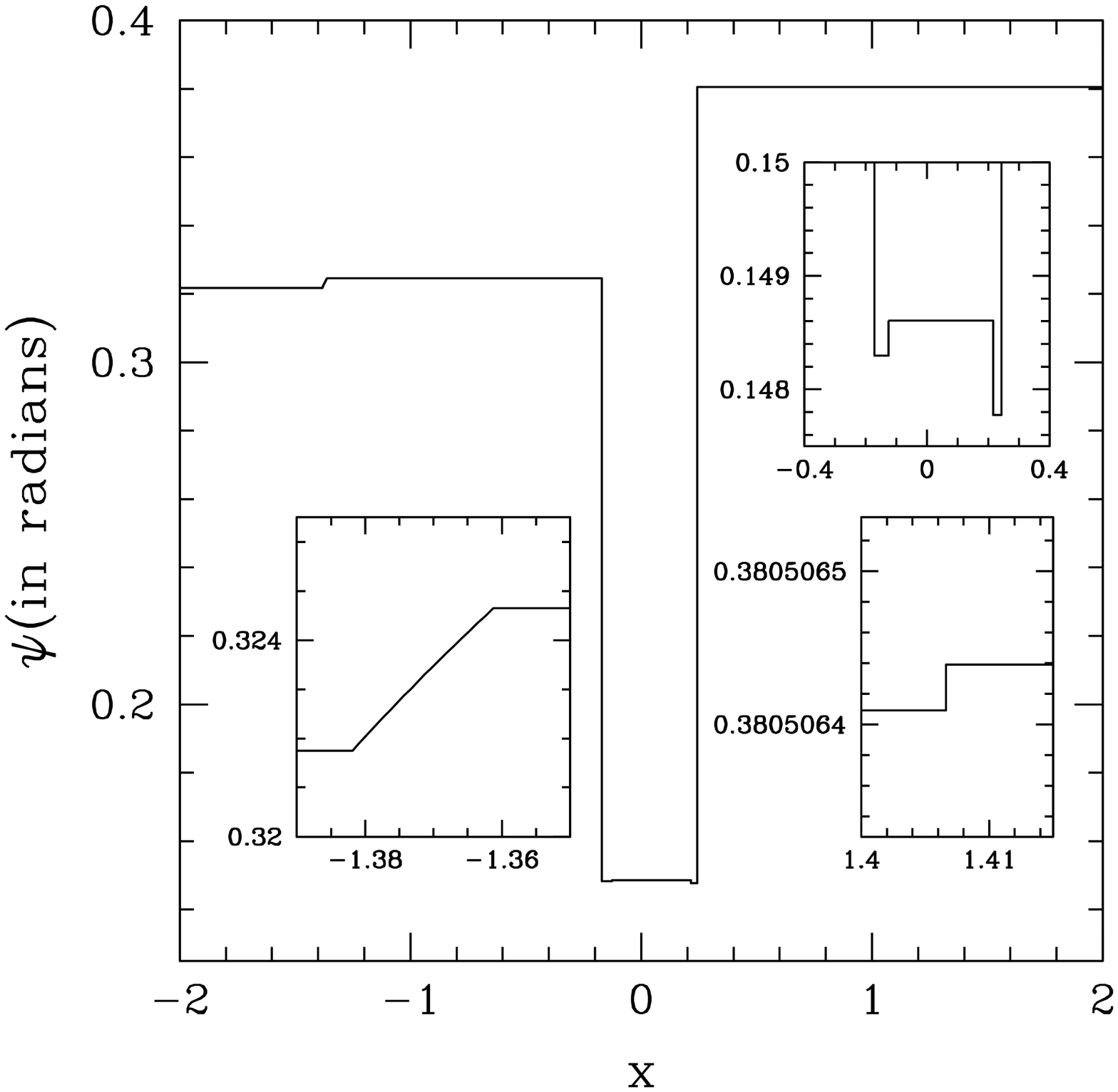}
  \caption{\label{BG_AL1_normb} Exact solution of the generic Alfv\`en
  test at time $t=1.5$. The left panel shows the norm of the tangential
  magnetic field $B_t$, while the right panel the angle
  $\psi\equiv\arctan{(B^z/B^y)}$. Note that both quantities vary across
  all the fast, slow and Alfv\`en waves as a result of a relativistic
  effect.}
  \end{center}
\end{figure}

	Although the numerical code developed for the exact solution
of the Riemann problem in relativistic MHD could in principle be used
for generic initial data, we have used it in particular to calculate
the exact solutions of those less trivial initial states that over the
years have become standard references ({\it e.g.} \cite{KO99}, or
\cite{BA01}). Table \ref{tab:tests} collects the set of initial
conditions used in the tests solved, while we report in
Tables~\ref{tab:KOshosktube1}--\ref{tab:testalfven1} the first
significant digits for the exact solution of the same tests, reporting
in all cases the accuracy obtained (which usually is $\sim 10^{-10}$).
Finally, the full solutions in space of the various Riemann problems
listed in Table \ref{tab:tests} and for the quantities $\rho$, $v^x$,
$p_{\rm g}$, $p$, $v^y$, $v^z$, $B^y$, and $B^z$ are shown in
Figures~\ref{ST1}--\ref{BG_AL1} at the indicated representative
times. In addition, Figure~\ref{BG_AL1_normb} offers a quantitative
view of the changes in the tangential magnetic field $B_t$ and of the
rotation angle $\psi$ across the fast, slow and Alfv\`en waves in the
case of a generic Alfv\`en test.

	In all of the tests reported in Table \ref{tab:tests}, the
HLLE solver with about 800 gridpoints was able to track rather well
the exact solution in all of its waves. The only exception to this has
been test number 1 of Balsara (2001) which represents the relativistic
version of the test proposed by Brio-Wu (1988) in Newtonian
hydrodynamics (\cite{vanputten93}). The approximate numerical solution
of this test, in fact, shows the development of a left-going slow
compound-wave, that is a wave composed by a slow shock adjacent to a
slow rarefaction. Since we assume that a slow or fast-wave can either
be a pure rarefaction or a pure shock, compound structures of this
type cannot be found by construction and thus are not present in the
exact solution found ({\it cf.}  Table~\ref{tab:balsara1} and
Figure~\ref{BA1}). We remark that it is not yet clear whether compound
waves have to be considered acceptable physical solutions of the ideal
MHD equations and a debate on this is still ongoing (see, for
instance,~\cite{myong_roe97a,myong_roe97b},~\cite{falle01},
~\cite{torrilhon03a,torrilhon03b}, \cite{torrilhon04}). We here prefer
to adopt the same standpoint of Ryu and Jones (1995) in the
development of their exact Riemann solver in nonrelativistic
magnetohydrodynamics and not comment further on this until a commonly
accepted view has emerged.

	Another test which deserves a special comment is test number 5 of
Balsara (2001), in which the left-going Alfv\`en discontinuity and the
left-going slow rarefaction wave have very similar propagation
velocities. Indeed they are so close to each other that not even the HLLE
approximate Riemann solver with 40000 gridpoints was able to capture the
precise location of the discontinuity. As a consequence, the initial
guess for the jumps across the left-going Alfv\`en discontinuity was
sufficiently good to yield a convergent solution, but not good enough to
provide an exact solution with a truncation error comparable with the one
reached in all of the other tests ({\it cf.} data in
Table~\ref{tab:balsara5}). In addition, another distinctive feature of
this test and which has not been found in any of the others, is the
rotation of the angle $\psi$ across the left-going slow rarefaction. To
handle this we have followed the procedure discussed in
Section~\ref{sarw} and used equation (\ref{eq:psi}) to compute the
changes in the tangential magnetic field across the rarefaction wave.

%
%
%
%
%
%
%
%

%
\begin{table}
\begin{center}
\begin{tabular}{|c|c|c|c|c|c|c|c|}
\cline{1-8} & & & & & & & \\
& $\rho$ & $p$ & $v^x$ & $v^y$ & $v^z$ & $B^y$ & $B^z$ \\
& & & & & & & \\ \cline{1-8} & & & & & & & \\
R1 & 0.1000E+01 &    0.1001E+04 &    0.0000E+00 &    0.0000E+00 &    0.0000E+00 &    0.0000E+00 &    0.0000E+00 \\
R2 & 0.6984E-01 &    0.2927E+02 &    0.9115E+00 &    0.0000E+00 &    0.0000E+00 &    0.0000E+00 &    0.0000E+00 \\
R3 & 0.6984E-01 &    0.2927E+02 &    0.9115E+00 &    0.0000E+00 &    0.0000E+00 &    0.0000E+00 &    0.0000E+00 \\
R4 & 0.6984E-01 &    0.2927E+02 &    0.9115E+00 &    0.0000E+00 &    0.0000E+00 &    0.0000E+00 &    0.0000E+00 \\
R5 & 0.8846E+00 &    0.2927E+02 &    0.9115E+00 &    0.0000E+00 &    0.0000E+00 &    0.0000E+00 &    0.0000E+00 \\
R6 & 0.8846E+00 &    0.2927E+02 &    0.9115E+00 &    0.0000E+00 &    0.0000E+00 &    0.0000E+00 &    0.0000E+00 \\
R7 & 0.8846E+00 &    0.2927E+02 &    0.9115E+00 &    0.0000E+00 &    0.0000E+00 &    0.0000E+00 &    0.0000E+00 \\
R8 & 0.1000E+00 &    0.1500E+01 &    0.0000E+00 &    0.0000E+00 &    0.0000E+00 &    0.0000E+00 &    0.0000E+00 \\
\cline{1-8} 
\end{tabular}
\end{center}
\smallskip
\caption{First significant digits for the exact solution of the test
shock-tube 1 of Komissarov (1999) computed with an accuracy of
$10^{-10}$. The left column indicates the regions in which the solution
is computed ({\it cf.} Fig.~\ref{mhd_zones}). }
\label{tab:KOshosktube1}
\end{table}
%
\begin{table}
\begin{center}
\begin{tabular}{|c|c|c|c|c|c|c|c|}
\cline{1-8} & & & & & & & \\
& $\rho$ & $p$ & $v^x$ & $v^y$ & $v^z$ & $B^y$ & $B^z$ \\
& & & & & & & \\ \cline{1-8} & & & & & & & \\
R1 & 0.1000E+01 &    0.5292E+02 &    0.9806E+00 &    0.0000E+00 &    0.0000E+00 &    0.1000E+02 &    0.0000E+00 \\
R2 & 0.6331E+01 &    0.2571E+03 &    0.4380E+00 &    0.4069E+00 &    0.0000E+00 &    0.1960E+02 &    0.0000E+00 \\
R3 & 0.6331E+01 &    0.2571E+03 &    0.4380E+00 &    0.4069E+00 &    0.0000E+00 &    0.1960E+02 &    0.0000E+00 \\
R4 & 0.2742E+02 &    0.2819E+03 &    0.2453E-07 &   -0.6811E+00 &    0.0000E+00 &    0.2250E-06 &    0.0000E+00 \\
R5 & 0.2742E+02 &    0.2819E+03 &   -0.2810E-07 &   -0.6811E+00 &    0.0000E+00 &    0.2250E-06 &    0.0000E+00 \\
R6 & 0.6331E+01 &    0.2571E+03 &   -0.4380E+00 &    0.4069E+00 &    0.0000E+00 &   -0.1960E+02 &    0.0000E+00 \\
R7 & 0.6331E+01 &    0.2571E+03 &   -0.4380E+00 &    0.4069E+00 &    0.0000E+00 &   -0.1960E+02 &    0.0000E+00 \\
R8 & 0.1000E+01 &    0.5292E+02 &   -0.9806E+00 &    0.0000E+00 &    0.0000E+00 &   -0.1000E+02 &    0.0000E+00 \\
\cline{1-8} 
\end{tabular}
\end{center}
\smallskip
\caption{The same as Table~\ref{tab:KOshosktube1} but for the exact solution of
the test Collision of Komissarov (1999) computed with an accuracy of
$10^{-6}$.}
\label{tab:KOcollision}
\end{table}
%
\begin{table}
\begin{center}
\begin{tabular}{|c|c|c|c|c|c|c|c|}
\cline{1-8} & & & & & & & \\
& $\rho$ & $p$ & $v^x$ & $v^y$ & $v^z$ & $B^y$ & $B^z$ \\
& & & & & & & \\ \cline{1-8} & & & & & & & \\
R1 & 0.1000E+01 &    0.1625E+01 &    0.0000E+00 &    0.0000E+00 &    0.0000E+00 &    0.1000E+01 &    0.0000E+00 \\
R2 & 0.6257E+00 &    0.6989E+00 &    0.3742E+00 &   -0.3561E-01 &    0.0000E+00 &    0.6594E+00 &    0.0000E+00 \\
R3 & 0.6257E+00 &    0.6989E+00 &    0.3742E+00 &   -0.3561E-01 &    0.0000E+00 &    0.6594E+00 &    0.0000E+00 \\
R4 & 0.7092E+00 &    0.7062E+00 &    0.2555E+00 &   -0.6804E+00 &    0.0000E+00 &   -0.4285E+00 &    0.0000E+00 \\
R5 & 0.2695E+00 &    0.7062E+00 &    0.2555E+00 &   -0.6804E+00 &    0.0000E+00 &   -0.4285E+00 &    0.0000E+00 \\
R6 & 0.1223E+00 &    0.6976E+00 &   -0.2080E-01 &   -0.3460E-02 &    0.0000E+00 &   -0.9769E+00 &    0.0000E+00 \\
R7 & 0.1223E+00 &    0.6976E+00 &   -0.2080E-01 &   -0.3460E-02 &    0.0000E+00 &   -0.9769E+00 &    0.0000E+00 \\
R8 & 0.1250E+00 &    0.7250E+00 &    0.0000E+00 &    0.0000E+00 &    0.0000E+00 &   -0.1000E+01 &    0.0000E+00 \\
\cline{1-8}
\end{tabular}
\end{center}
\smallskip
\caption{The same as Table~\ref{tab:KOshosktube1} but for the exact solution of
the test number 1 of Balsara (2001) computed with an accuracy of
$10^{-10}$. This test represents the relativistic version of the test
proposed by Brio \& Wu (1988)}
\label{tab:balsara1}
\end{table}
%
\begin{table}
\begin{center}
\begin{tabular}{|c|c|c|c|c|c|c|c|}
\cline{1-8} & & & & & & & \\
& $\rho$ & $p$ & $v^x$ & $v^y$ & $v^z$ & $B^y$ & $B^z$ \\
& & & & & & & \\ \cline{1-8} & & & & & & & \\
R1 & 0.1000E+01 &    0.7850E+02 &    0.0000E+00 &    0.0000E+00 &    0.0000E+00 &    0.6000E+01 &    0.6000E+01 \\
R2 & 0.4300E+00 &    0.2321E+02 &    0.6344E+00 &   -0.9981E-01 &   -0.9981E-01 &    0.3045E+01 &    0.3045E+01 \\
R3 & 0.4300E+00 &    0.2321E+02 &    0.6344E+00 &   -0.9981E-01 &   -0.9981E-01 &    0.3045E+01 &    0.3045E+01 \\
R4 & 0.3830E+00 &    0.2284E+02 &    0.6770E+00 &   -0.5566E-01 &   -0.5566E-01 &    0.3205E+01 &    0.3205E+01 \\
R5 & 0.2828E+01 &    0.2284E+02 &    0.6770E+00 &   -0.5566E-01 &   -0.5566E-01 &    0.3205E+01 &    0.3205E+01 \\
R6 & 0.1582E+01 &    0.2072E+02 &    0.4688E+00 &   -0.2538E+00 &   -0.2538E+00 &    0.3971E+01 &    0.3971E+01 \\
R7 & 0.1582E+01 &    0.2072E+02 &    0.4688E+00 &   -0.2538E+00 &   -0.2538E+00 &    0.3971E+01 &    0.3971E+01 \\
R8 & 0.1000E+01 &    0.1399E+02 &    0.0000E+00 &    0.0000E+00 &    0.0000E+00 &    0.7000E+00 &    0.7000E+00 \\
\cline{1-8} 
\end{tabular}
\end{center}
\smallskip
\caption{The same as Table~\ref{tab:KOshosktube1} but for the exact solution of
the test number 2 of Balsara (2001) computed with an accuracy of
$10^{-10}$.}
\label{tab:balsara2}
\end{table}
%
\begin{table}
\begin{center}
\begin{tabular}{|c|c|c|c|c|c|c|c|}
\cline{1-8} & & & & & & & \\
& $\rho$ & $p$ & $v^x$ & $v^y$ & $v^z$ & $B^y$ & $B^z$ \\
& & & & & & & \\ \cline{1-8} & & & & & & & \\
R1 & 0.1000E+01 &    0.1099E+04 &    0.0000E+00 &    0.0000E+00 &    0.0000E+00 &    0.7000E+01 &    0.7000E+01 \\
R2 & 0.1381E+00 &    0.8604E+02 &    0.9246E+00 &   -0.3513E-01 &   -0.3513E-01 &    0.2238E+01 &    0.2238E+01 \\
R3 & 0.1381E+00 &    0.8604E+02 &    0.9246E+00 &   -0.3513E-01 &   -0.3513E-01 &    0.2238E+01 &    0.2238E+01 \\
R4 & 0.9798E-01 &    0.7653E+02 &    0.9529E+00 &    0.4366E-01 &    0.4366E-01 &    0.4670E+01 &    0.4670E+01 \\
R5 & 0.1010E+02 &    0.7653E+02 &    0.9529E+00 &    0.4366E-01 &    0.4366E-01 &    0.4670E+01 &    0.4670E+01 \\
R6 & 0.1218E+01 &    0.6363E+02 &    0.4670E+00 &   -0.4270E+00 &   -0.4270E+00 &    0.9408E+01 &    0.9408E+01 \\
R7 & 0.1218E+01 &    0.6363E+02 &    0.4670E+00 &   -0.4270E+00 &   -0.4270E+00 &    0.9408E+01 &    0.9408E+01 \\
R8 & 0.1000E+01 &    0.5059E+02 &    0.0000E+00 &    0.0000E+00 &    0.0000E+00 &    0.7000E+00 &    0.7000E+00 \\
\cline{1-8} 
\end{tabular}
\end{center}
\smallskip
\caption{The same as Table~\ref{tab:KOshosktube1} but for the exact solution of
the test number 3 of Balsara (2001) computed with an accuracy of
$10^{-10}$.}
\label{tab:balsara3}
\end{table}
%
\begin{table}
\begin{center}
\begin{tabular}{|c|c|c|c|c|c|c|c|}
\cline{1-8} & & & & & & & \\
& $\rho$ & $p$ & $v^x$ & $v^y$ & $v^z$ & $B^y$ & $B^z$ \\
& & & & & & & \\ \cline{1-8} & & & & & & & \\
R1 & 0.1000E+01 &    0.5020E+02 &    0.9990E+00 &    0.0000E+00 &    0.0000E+00 &    0.7000E+01 &    0.7000E+01 \\
R2 & 0.5175E+02 &    0.1184E+04 &    0.4408E-01 &    0.3263E-01 &    0.3263E-01 &    0.1668E+02 &    0.1668E+02 \\
R3 & 0.5175E+02 &    0.1184E+04 &    0.4408E-01 &    0.3263E-01 &    0.3263E-01 &    0.1668E+02 &    0.1668E+02 \\
R4 & 0.6148E+02 &    0.1188E+04 &    0.1086E-07 &   -0.2877E+00 &   -0.2877E+00 &    0.8042E-09 &    0.8036E-09 \\
R5 & 0.6148E+02 &    0.1188E+04 &   -0.1089E-07 &   -0.2877E+00 &   -0.2877E+00 &    0.8042E-09 &    0.8036E-09 \\
R6 & 0.5175E+02 &    0.1184E+04 &   -0.4408E-01 &    0.3263E-01 &    0.3263E-01 &   -0.1668E+02 &   -0.1668E+02 \\
R7 & 0.5175E+02 &    0.1184E+04 &   -0.4408E-01 &    0.3263E-01 &    0.3263E-01 &   -0.1668E+02 &   -0.1668E+02 \\
R8 & 0.1000E+01 &    0.5020E+02 &   -0.9990E+00 &    0.0000E+00 &    0.0000E+00 &   -0.7000E+01 &   -0.7000E+01 \\
\cline{1-8} 
\end{tabular}
\end{center}
\smallskip
\caption{The same as Table~\ref{tab:KOshosktube1} but for the exact solution of
the test number 4 of Balsara (2001) computed with an accuracy of
$10^{-7}$.}
\label{tab:balsara4}
\end{table}
%
\begin{table}
\begin{center}
\begin{tabular}{|c|c|c|c|c|c|c|c|}
\cline{1-8} & & & & & & & \\
& $\rho$ & $p$ & $v^x$ & $v^y$ & $v^z$ & $B^y$ & $B^z$ \\
& & & & & & & \\ \cline{1-8} & & & & & & & \\
R1 & 0.1080E+01 &    0.2885E+01 &    0.4000E+00 &    0.3000E+00 &    0.2000E+00 &    0.3000E+00 &    0.3000E+00 \\
R2 & 0.2447E+01 &    0.5908E+01 &   -0.1331E+00 &    0.2111E+00 &    0.1751E+00 &    0.2662E+00 &    0.5076E+00 \\
R3 & 0.2447E+01 &    0.5908E+01 &   -0.1215E+00 &    0.1264E+00 &    0.1158E+00 &   -0.1182E+00 &    0.2302E+00 \\
R4 & 0.2050E+01 &    0.5616E+01 &   -0.4547E-01 &   -0.1463E+00 &    0.2146E+00 &   -0.1175E+01 &    0.5852E+00 \\
R5 & 0.1884E+01 &    0.5616E+01 &   -0.4543E-01 &   -0.1462E+00 &    0.2149E+00 &   -0.1175E+01 &    0.5850E+00 \\
R6 & 0.1642E+01 &    0.5488E+01 &   -0.1129E+00 &   -0.4606E-01 &    0.1601E+00 &   -0.1429E+01 &    0.7320E+00 \\
R7 & 0.1642E+01 &    0.5488E+01 &   -0.1155E+00 &   -0.8536E-01 &    0.1027E+00 &   -0.1272E+01 &    0.9468E+00 \\
R8 & 0.1000E+01 &    0.2918E+01 &   -0.4500E+00 &   -0.2000E+00 &    0.2000E+00 &   -0.7000E+00 &    0.5000E+00 \\
\cline{1-8} 
\end{tabular}
\end{center}
\smallskip
\caption{The same as Table~\ref{tab:KOshosktube1} but for the exact solution of
the test number 5 of Balsara (2001) computed with an accuracy of $3
\times 10^{-4}$.}
\label{tab:balsara5}
\end{table}
%
\begin{table}
\begin{center}
\begin{tabular}{|c|c|c|c|c|c|c|c|}
\cline{1-8}  & & & & & & & \\
& $\rho$ & $p$ & $v^x$ & $v^y$ & $v^z$ & $B^y$ & $B^z$ \\
& & & & & & & \\ \cline{1-8} & & & & & & & \\
R1 & 0.1000E+01 & 0.2376E+02 & 0.0000E+00 &  0.3000E+00 & 0.4000E+00 & 0.6000E+01 & 0.2000E+01 \\
R2 & 0.9219E+00 & 0.2083E+02 & 0.6232E-01 &  0.3050E+00 &  0.4193E+00 & 0.5622E+01 & 0.1892E+01 \\
R3 & 0.9219E+00 & 0.2083E+02 & 0.7109E-01 &  0.3669E+00 &  0.2429E+00 & 0.5691E+01 & 0.8502E+00 \\
R4 & 0.1263E+01 & 0.2087E+02 & 0.3886E-01 &  0.1147E+00 &  0.2054E+00 & 0.5130E+01 & 0.7680E+00 \\
R5 & 0.1099E+01 & 0.2087E+02 & 0.3886E-01 &  0.1147E+00 &  0.2054E+00 & 0.5130E+01 & 0.7680E+00 \\
R6 & 0.9130E+00 & 0.2085E+02 & 0.1607E-01 & -0.5009E-01 &  0.1813E+00 & 0.5505E+01 & 0.8195E+00 \\
R7 & 0.9130E+00 & 0.2085E+02 & 0.1341E-01 & -0.6599E-03 & -0.2640E-03 & 0.5073E+01 & 0.2029E+01 \\
R8 & 0.9000E+00 & 0.2030E+02 & 0.0000E+00 &  0.0000E+00 &  0.0000E+00 & 0.5000E+01 & 0.2000E+01 \\
\cline{1-8} 
\end{tabular}
\end{center}
\smallskip
\caption{The same as Table~\ref{tab:KOshosktube1} but for the exact
solution of the generic Alfv\`en test computed with an accuracy of
$10^{-10}$.}
\label{tab:testalfven1}
\end{table}

\section{Conclusions}
\label{conclusions}

	We have presented the procedure for the solution of the exact
Riemann problem in special relativistic MHD. Special care has been paid
in treating both degenerate initial states ({\it i.e.} with zero normal
magnetic field) leading to a set of only three waves analogous to the
ones in relativistic hydrodynamics, as well as generic initial states
({\it i.e.} with nonzero normal magnetic field) leading to the full set
of seven MHD waves.

	The approach discussed for the numerical solution of the exact
Riemann problem reflects this distinction and different sets of equations
are used according to the values of the normal magnetic field. In
particular, when $B^x=0$, all of the equations needed for the solution of
the Riemann problem are written as a function of the total pressure, thus
following a procedure which is logically equivalent to the one adopted in
relativistic hydrodynamics (we have referred to this as to the
$p$-method). When $B^x \ne 0$, on the other hand, an hybrid approach is
adopted in which the solution across fast-waves and Alfv\`en
discontinuities is still computed using the $p$-method, but the one
across slow-waves and the contact discontinuity is computed using
equations which are written in terms of the tangential components of the
magnetic field (we have referred to this as to the $B_t$-method). The use
of a combined approach for the general case of $B^x \ne 0$ has turned out
to be crucial for a successful solution of the problem.

	Because of its generality, the solution presented here could
serve as a useful if not indispensable test for those numerical codes
that solve the MHD equations in relativistic regimes. As the astronomical
observations become increasingly more accurate, such numerical codes will
become increasingly more important to explain and describe in detail the
complex physics of astrophysical compact objects.

	As a final remark we note that despite the considerable
improvements in the performance of modern computers, the exact solution
of the Riemann problem at each grid interface is still computationally
too expensive to be used routinely in sophisticated multidimensional
numerical codes solving the equations of relativistic hydrodynamics or
MHD in either stationary or dynamical spacetimes (see, for instance,
Baiotti et al. 2005, Duez et al. 2005). While a numerical code based on
exact Riemann solvers may represent at least in principle the most
accurate approach to the solution of the hydrodynamics and MHD equations,
considerable work is still required to make it competitive with less
accurate but more computationally efficient methods. A first step in this
direction would be, for instance, the search for an analytic solution for
the shock velocity and this will be the subject of future work. {Another
important problem deserving equal attention is that of the uniqueness of
the solution.  While a global consensus on this issue still needs to be
reached, it will remain essential in order to construct a complete and
consistent picture of the exact solution of the Riemann problem in
relativistic MHD.}

	The numerical codes computing the exact solution both when
$B^x=0$ and when $B^x\ne 0$ are available from the authors upon
request. Users of the codes can give credit by mentioning the source and
citing this paper.

\bigskip
\acknowledgements

It is a pleasure to thank Jos\'e M$^{\underline{\mbox{a}}}$. Mart\'i,
Jos\'e A. Pons and Olindo Zanotti for useful discussions and comments.

\appendix
\newpage
\section{}
\label{appendixa}

	We here report the expressions for the tangential components of
the velocity behind the shock ({\it i.e.} $v_b^y, v_b^z$) when
expressed as function of post-shock $p$ and $J$. First, we
consider $v^y_b$ as function of $p_b$, $J$ and $v^z_b$.

\begin{eqnarray*}
v^y_b &=& -\frac{1}{N_y}\Bigl\{ 
J^3 W_a^4 \Bigl[{v^y_a} (D_a + p_a + \tau_a - \eta_a^2 W_a^2)
              (D_a + p_b + \tau_a -B^2_x - \eta_a^2 W_a^2) +\\ 
&& {B^y_a} \Bigl({B_x} (p_a + \tau_a)
              {v^x_a} + {B^z_a} (p_b + \tau_a) {v^z_b} 
		-\eta_a (p_b + \tau_a) - 
	      \\ 
&& D_a (\eta_a - B_x v^x_a - B^z_a v^z_b)  + \eta_a^3 W_a^2 -
              	\eta_a^2 ({B_x} {v^x_a} + 
		{B^z_a} {v^z_b}) W_a^2\Bigr)\Bigr] -\\
&& J^2 D_a W_a^2 \Bigl[ 
		\Bigl(2B^3_x \eta_a {v^y_a} + B^2_x {v^x_a} 
	[(D_a + p_a + \tau_a) {v^y_a}-2 {B^y_a} \eta_a] + \\ 
&&{v^x_a} (p_a - p_b) [2 (D_a + p_a + \tau_a) {v^y_a} + {B^y_a} 
	({B^z_a} {v^z_b} - 2 \eta_a  )] - 
{B_x} \{\eta_a (3 D_a + 2 p_a + p_b + 3 \tau_a) {v^y_a} - 
\\ && 
	{B^z_a} (p_a - p_b)
              {v^y_a} {v^z_b}  +
{B^y_a} [(p_a + \tau_a) ({v^x_a}^2 - 1) -3 \eta_a^2 +  
	      2 {B^z_a} \eta_a {v^z_b} + (p_b +
              \tau_a) {v^z_a} {v^z_b} - M D_a ]\}\Bigr) W_a^2- \\ 
&&  {B_x} {B^y_a} (D_a + p_b + \tau_a)+\eta_a^ 2 [2 (p_b -p_a ) {v^x_a} {v^y_a} - B^2_x {v^x_a}{v^y_a} +
	{B_x} (3 \eta_a {v^y_a} - M {B^y_a} )] W_a^4\Bigr] W_{ s} - \\ 
&& J D_a^2 W_a^2 \Bigl[B^3_x {v^x_a} ({B^y_a} + 
	2 \eta_a {v^y_a} W_a^2) -B^4_x {v^y_a} +\\ 
&& {B_x} {v^x_a} (p_a - p_b) [2 {B^y_a} + (2
              \eta_a {v^y_a} + {B^z_a} {v^y_a} {v^z_b} + {B^y_a}
              {v^z_a} {v^z_b}) W_a^2] + \\ 
&& Q [(D_a + p_a + \tau_a) {v^y_a} - \eta_a ({B^y_a} +
              \eta_a {v^y_a} W_a^2)]  +\\
&& B^2_x \Bigl({B^y_a} ({B^z_a} {v^z_b} -3 \eta_a + 2 M \eta_a W_a^2) +
	{v^y_a} [D_a + p_a + \tau_a - (3 \eta_a^2 +
	      (p_b - p_a ) {v^z_a} {v^z_b}) W_a^2]\Bigr)\Bigr] W_{ s}^2 
	      +\\
&& {B_x} D_a^3 \Bigl(B^3_x {v^x_a} {v^y_a} W_a^2 -
              {B_x} (p_a - p_b) {v^x_a} {v^y_a} {v^z_a} {v^z_b} W_a^4
              - Q ({B^y_a} + \eta_a
              {v^y_a} W_a^2) -\\
&& B^2_x [{B^y_a} + (\eta_a {v^y_a} -
              M {B^y_a} ) W_a^2]\Bigr)
              W_{ s}^3
	      \Bigr\} \ , 
\end{eqnarray*}
where we have defined $M \equiv (1 - {v^x_a}^2 - {v^z_a} {v^z_b})$,
$Q \equiv W_a^2 (p_a - p_b) ({v^x_a}^2 - 1)$  and
\begin{eqnarray*}
N_y&\equiv& W_a ^ 2 \Bigl\{ 
	{B_x} D_a W_a^2 W_{ s} \Bigl(J ^ 2 [2 {B^y_a}^2 \eta_a -
\\ &&  
	3\eta_a (D_a + p_b + \tau_a) + 3 \eta_a^3 W_a^2 + {B^y_a}
            {v^y_a} (D_a - p_a + 2 p_b + \tau_a - \eta_a^2 W_a^2)] +
\\ && 
	2 D_a J (p_a - p_b) {v^x_a} (2 \eta_a - {B^y_a} {v^y_a})
            W_{ s} + D_a^2 \eta_a (p_a - p_b) ({v^x_a}^2 -1)
            W_{ s}^2\Bigr) -
\\&& 
	B^4_x D_a^2 W_{ s}^2 (J + 
	D_a {v^x_a} W_{s}) - W_a^2 [J (D_a + p_b + \tau_a - \eta_a^2
            W_a^2) +
\\&&
         D_a (p_b - p_a) {v^x_a} W_{ s}] [
	J^2 (D_a + p_b + \tau_a - \eta_a^2 W_a^2) + 2 D_a J(p_b - p_a) {v^x_a} W_{ s} -
\\ && 
            {B^y_a}^2
            J^2 + 
	D_a^2 (p_b - p_a) 
	    ({v^x_a}^2 - 1) W_{ s}^2] + B^3_x D_a W_{ s} [2 \eta_a
            J^2 W_a^2 + 
\\ && 
	2 D_a \eta_a J {v^x_a} W_a^2 W_{ s} +
	D_a^2 (\eta_a - {B^y_a} {v^y_a}) W_{ s}^2] + B^2_x \Bigl[J^3
            W_a^2 (D_a + p_b + \tau_a - \eta_a^2 W_a^2) + 
\\ && 
	D_a J^2 {v^x_a} W_a^2 (D_a - 2 p_a + 3 p_b + \tau_a - \eta_a^2
            W_a^2) W_{ s} +
\\ && 
	D_a^2 J \Bigl(D_a + p_b -{B^y_a}^2 + 
            \tau_a + [2 {B^y_a} \eta_a {v^y_a} - 3 \eta_a^2 - (p_a -
            p_b) (3 {v^x_a}^2 - 1 + 
	{v^y_a}^2)] W_a^2\Bigr) W_{ s}^2 -
\\ && 
            D_a^3 {v^x_a} {v^z_a}^2 W_a^2 (p_b - p_a)W_{ s}^3\Bigr]\Bigr\} \ .
\end{eqnarray*}

	We next consider the expression of $v_b^z$ as function of
post-shock $p$ and $J$
\begin{eqnarray*}
v^z_b &=& -\frac{1}{N_z}\Bigl\{
J^3 W_a ^ 4 \Bigl[{v^z_a} (D_a + p_a + \tau_a - \eta_a^2 W_a^2)
              (D_a-{B^x}^2 - {B^y_a}^2 +
\\ && 
p_b + \tau_a - \eta_a^2
              W_a^2) + {B^z_a} \Bigl((p_a +
              \tau_a) ({B^x} {v^x_a} + {B^y_a} {v^y_a})-\eta_a (p_b + \tau_a) +
\\ && 
D_a ({B^x} {v^x_a} + {B^y_a} {v^y_a}-\eta_a) + \eta_a^3 W_a^2 -
              \eta_a^2 ({B^x} {v^x_a} + {B^y_a} {v^y_a})
	      W_a^2\Bigr)\Bigr] + 
\\ && 
D_a J^2 W_a^2 \Bigl[{B^x} {B^z_a} (D_a + p_b + \tau_a) +
              \Bigl({B^x} {B^z_a} [ 2 {B^y_a} \eta_a {v^y_a}-3 \eta_a^2  +
\\ && 
D_a ({v^x_a}^2 + {v^y_a}^2 - 1) + (p_a + \tau_a) ({v^x_a}^2 + 
	      {v^y_a}^2 - 1)] -
\\ && 
2 {B^x}^3 \eta_a {v^z_a} -
              {B^x} [2 {B^y_a}^2 \eta_a - \eta_a (3 D_a + 2 p_a + p_b
              + 3 \tau_a) +
\\&& 
{B^y_a} (D_a + p_a + \tau_a) {v^y_a}]
              {v^z_a} -{B^x}^2 {v^x_a} [(D_a +
              p_a + \tau_a) {v^z_a}-2 {B^z_a} \eta_a] -
\\&& 
2 (p_a - p_b) {v^x_a}
              [(D_a + p_a + \tau_a) {v^z_a}-{B^z_a} \eta_a]\Bigr)
              W_a^2 +
\\&& 
\eta_a ^ 2 \Bigl({B^x} {B^z_a} ({v^x_a}^2 +
              {v^y_a}^2 - 1) -
({B^x}^2 {v^x_a}-3 {B^x} \eta_a + 2 p_a
              {v^x_a} - 2 p_b {v^x_a} + {B^x} {B^y_a} {v^y_a})
              {v^z_a}\Bigr) W_a^4\Bigr] W_{ s} - 
\\ && 
D_a^2 J W_a^2 \Bigl(
	      2 {B^x} (p_a - p_b) {v^x_a} ({B^z_a} + \eta_a
              {v^z_a} W_a^2)-{B^x}^4
              {v^z_a} +
{B^x}^3 {v^x_a} ({B^z_a} + 2 \eta_a
              {v^z_a} W_a^2) +
\\&& 
{B^x}^2 \{{v^z_a} [D_a-{B^y_a}^2 +
              p_a + \tau_a - \eta_a (3 \eta_a - 2 {B^y_a} {v^y_a})
              W_a^2] +
\\ && 
{B^z_a} [{B^y_a} {v^y_a} -3 \eta_a - 2 \eta_a
              ({v^x_a}^2 + {v^y_a}^2 - 1) W_a^2]\} +
\\ && 
(p_a - p_b) ({v^x_a}^2 - 1) 
	      W_a^2 [(D_a + p_a + \tau_a) {v^z_a} - \eta_a
              ({B^z_a} + \eta_a {v^z_a} W_a^2)]\Bigr) W_{ s}^2 +
\\ && 
{B^x} D_a ^ 3 \Bigl[B_x^3 {v^x_a} {v^z_a} W_a^2 - (p_a - p_b)
              ({v^x_a}^2 - 1) W_a^2 ({B^z_a} + \eta_a {v^z_a} W_a^2) -
\\ &&
              B_x^2 \Bigl({B^z_a} + [{B^z_a} ({v^x_a}^2 + {v^y_a}^2 - 1)
              + (\eta_a - {B^y_a} {v^y_a}) {v^z_a}] W_a^2\Bigr)\Bigr] W_{
              s}^3\Bigr\} \ ,
\end{eqnarray*}
where
\begin{eqnarray*}
N_z &\equiv&
W_a ^ 2 \Bigl\{{B^x} D_a W_a^2 W_{ s} \{J ^ 2 [2 {B^y_a}^2 \eta_a +
2 {B^z_a}^2 \eta_a - 3 \eta_a (D_a + p_b + \tau_a) + 3
            \eta_a^3 W_a^2 +
\\&& 
{B^y_a} {v^y_a} (D_a - p_a + 2 p_b +
            \tau_a - \eta_a^2 W_a^2) +
{B^z_a} {v^z_a} (D_a - p_a + 2
            p_b + \tau_a - \eta_a^2 W_a^2)] +
\\ && 
2 D_a J (p_a - p_b)
            {v^x_a} (2 \eta_a - {B^y_a} {v^y_a} - {B^z_a} {v^z_a})
            W_{ s} +
D_a^2 \eta_a (p_a - p_b) ( {v^x_a}^2 - 1 )
            W_{ s}^2\} -
\\ && 
B_x^4 D_a^2 W_{ s}^2 (J + D_a {v^x_a} W_{ s})  -
W_a^2 [J (D_a + p_b + \tau_a - \eta_a^2
            W_a^2) - 
\\&& 
D_a (p_a - p_b) {v^x_a} W_{ s}] [J ^ 2
            (D_a -{B^y_a}^2 - {B^z_a}^2 + p_b + \tau_a - \eta_a^2
            W_a^2) -
\\&& 
2 D_a J (p_a - p_b) {v^x_a} W_{ s} - D_a^2
            (p_a - p_b) ({v^x_a}^2 - 1) W_{ s}^2] +
\\ && 
B_x^3 D_a
            W_{ s} [2 \eta_a J^2 W_a^2 + 2 D_a \eta_a J {v^x_a}
            W_a^2 W_{ s} + D_a^2 (\eta_a - {B^y_a} {v^y_a} -
            {B^z_a} {v^z_a}) W_{ s}^2] +
\\ && 
B_x^2 \Bigl(J^3 W_a^2 (D_a +
            p_b + \tau_a - \eta_a^2 W_a^2) +
D_a J^2 {v^x_a} W_a^2
            (D_a - 2 p_a + 3 p_b + \tau_a - \eta_a^2 W_a^2) W_{ s}
            + 
\\&& 
D_a^2 J \{D_a -{B^y_a}^2 - {B^z_a}^2 + p_b + \tau_a +
            [2 \eta_a ({B^y_a} {v^y_a} + {B^z_a}
            {v^z_a}) -3 \eta_a^2 - 
\\ && 
(p_a - p_b) (3 {v^x_a}^2 + {v^y_a}^2 +
            {v^z_a}^2 - 1)] W_a^2\} W_{ s}^2 \Bigr)\Bigr\} \ .
\end{eqnarray*}

\section{}
\label{appendixb}

	The explicit form for the system of ODEs to be solved numerically
to determine the solution across a rarefaction wave within the $p$-method
is given by the following set of equations in which the total pressure
$p$ plays the role of the self-similar variable
\begin{eqnarray}
\frac{\mathrm{d}\rho}{\mathrm{d}p} &=& -\rho\left(W^2
v_x+\frac{1}{v_x-\xi}\right)\frac{\mathrm{d}v_x}{\mathrm{d}p}-\rho W^2
v_y \frac{\mathrm{d}v_y}{\mathrm{d}p}-\rho W^2 v_z
\frac{\mathrm{d}v_z}{\mathrm{d}p} \ ,\\
\nonumber \\
\frac{\mathrm{d}v_x}{\mathrm{d}p}
&=& R \Bigl\{
(\rho h_g W^2+ B_x^2)(\xi-v_x)(v_x\xi-1) +
B_x^2\frac{\xi v_x-1}{W^2(v_x-\xi)}+
B_x^2\xi(v_y^2+ v_z^2) + 
\nonumber\\&&  
B_x[\eta(\xi^2-1)-B_xv_x(1-2v_x\xi+\xi^2)]\Bigr\} \ ,\\
\nonumber\\
\frac{\mathrm{d}v_y}{\mathrm{d}p} &=&
R \Biggl\{
2 B_x v_y(\eta-B_zv_z)\xi- B_x^2 v_y\xi(\xi+v_x)+\nonumber\\ && v_y[
B_z^2+W^2(\eta^2-w)](v_x-\xi)\xi + B_y^2v_y(v_x\xi - 1) + B_yB_zv_z(\xi^2
- 1)+\nonumber\\ && B_xB_y\frac{(v_y^2+
v_z^2-1)+(v_x-2v_xv_y^2)\xi+(1+v_y^2- v_z^2)\xi^2-v_x\xi^3}{(v_x-\xi)}
\Biggr\}\ ,\\
\nonumber \\
\frac{\mathrm{d}v_z}{\mathrm{d}p} &=&
R \Biggl\{
2B_x (\eta-B_yv_y)v_z\xi- B_x^2 v_z\xi(v_x+\xi)+\nonumber\\ && v_z[
B_y^2+W^2(\eta^2-w)](v_x-\xi)\xi+B_yB_zv_y(\xi^2-1)+v_z
B_z^2(v_x\xi-1)+\nonumber\\ && B_xB_z\frac{(v_y^2+ v_z^2-1)+(v_x-2v_x
v_z^2)\xi+(1-v_y^2+ v_z^2)\xi^2-v_x\xi^3}{(v_x-\xi)} \Biggr\}\ ,\\
\nonumber \\
\frac{\mathrm{d}B_y}{\mathrm{d}p} &=&-\frac{W^2 (B_y-B_y v_x \xi+B_x
v_y \xi)}{ B_x^2+2 B_x \eta W^2 (v_x-\xi)+W^4 (\eta^2-w)
(v_x-\xi)^2}\ ,\\
\nonumber \\
\frac{\mathrm{d}B_z}{\mathrm{d}p} &=& -\frac{W^2 (B_z-B_z v_x \xi+B_x
  v_z \xi)}{ B_x^2+2 B_x \eta W^2 (v_x-\xi)+W^4 (\eta^2-w)
  (v_x-\xi)^2} \ ,
\end{eqnarray}
where we have defined
\begin{equation}
\label{singular}
R \equiv \frac{1}{\rho h_gW^4(\eta^2-w)(V_A^{+} -\xi)(V_A^{-} -\xi)} \ ,
\end{equation}
with
\begin{equation}
V_A^{\pm} \equiv v_x+\frac{B_x}{W^2(\eta\pm\sqrt{w})} \ ,
\end{equation}
being the Alfv\`en velocities in the two directions. Note that the set of
ODEs has a singular point if the characteristic velocity of the slow or
fast magnetosonic waves is equal to the Alf\'ven velocity [{\it cf.}
eq. (\ref{singular})] and cannot be solved in this case without a proper
regularization. This procedure is not included in the numerical code made
available upon request.


\end{document}